%% file: main.tex
\documentclass[twocolumn]{aastex701}
\usepackage{xspace}
\usepackage{xtab}
\usepackage[encapsulated]{CJK}  
\usepackage{textcomp,gensymb}
\usepackage{multirow}
\usepackage{booktabs}
\usepackage{tabularx}
\usepackage{longtable}
\usepackage{amsmath}

\graphicspath{{figures/}}

\newcommand{\hii}{\ion{H}{2}}
\newcommand{\nii}{\ion{N}{2}}
\newcommand{\sii}{\ion{S}{2}}
\newcommand{\oiii}{\ion{O}{3}}
\newcommand{\ha}{H$\alpha$\xspace}
\newcommand{\hb}{H$\beta$\xspace}

\newcommand{\stromgren}{Str{\"o}mgren }

\newcommand{\rf}{$R_{\text{50}}$\xspace}
\newcommand{\rn}{$R_{\text{90}}$\xspace}
\newcommand{\riso}{$R_{\text{iso}}$\xspace}
\newcommand{\rlum}{$R_{\text{lum}}$\xspace}
\newcommand{\rmom}{$R_{\text{mom}}$\xspace}

\newcommand{\nregions}{2492~}

\begin{document}

\title{Resolved \hii\ regions in NGC\,253: Ionized gas structure and suggestions of a universal density-surface brightness relation}

\correspondingauthor{Rebecca L. McClain}
\email{mcclain.378@osu.edu}


\input{affils}
\input{authors}

\begin{abstract}
We use the full-disk VLT-MUSE mosaic of NGC\,253 to identify \nregions \hii\ regions and study their resolved structure. With an average physical resolution of $17$~pc, this is one of the largest samples of highly resolved spectrally mapped extragalactic \hii\ regions. Regions of all luminosities exhibit a characteristic emission profile described by a double Gaussian with a marginally resolved or unresolved core with radius $<10$~pc surrounded by a more extended halo of emission with radius $20{-}30$~pc. Approximately $80\%$ of the emission of a region originates from the halo component. As a result of this compact structure, the luminosity-radius relations for core and effective radii of \hii\ regions depend sensitively on the adopted methodology. Only the isophotal radius yields a robust relationship in NGC\,253, but this measurement has an ambiguous physical meaning. We invert the measured emission profiles to infer density profiles and find central densities of $n_e \approx 10{-}100$~cm$^{-3}$. In the brightest regions, these agree well with densities inferred from the [\sii]6716,30 doublet. The central density of \hii\ regions correlates well with the surface brightness within the effective radius. We show that this same scaling relation applies to the recent MUSE+HST catalog for 19 nearby galaxies. We also discuss potential limitations, including completeness, impacts of background subtraction and spatial resolution, and the generality of our results when applied to other galaxies. 
\end{abstract}

\keywords{\ion{H}{2} regions(694) --- Interstellar medium(847) --- Warm ionized medium (1788) --- Disk galaxies (405) --- Extragalactic astronomy (506)}

\section{Introduction}
\label{sec:intro}

\hii\ regions surrounding young massive stars produce bright optical line emission. They act as tracers of massive star formation \citep{kennicutt2012} and offer access to the chemical abundances and physical conditions of the gas \citep[e.g.,][]{kewley2019}. The \hii\ regions themselves are the sites of stellar feedback \citep[e.g.,][]{mcleod2020,barnes2021,pathak2025}, and their structure encodes information on the previous effect and current location where this feedback occurs \citep[e.g.,][]{lopez2014,barnes2022,hannon2022,pedrini2024}. 

The resolved ($\sim10$~pc) structure of ionized gas emission within and around \hii\ regions traces the location of key modes of stellar feedback, including photoionization heating and radiation pressure. Spatially resolved measurements also provide information on the density, and in turn the mass, of the ionized gas that makes up these regions, which is immediately impacted by stellar feedback and may represent the remnants of the molecular clouds from which the young star cluster powering the region formed. A concrete understanding of this resolved structure is also crucial for modeling emission from unresolved regions and for understanding the interface between \hii\ regions and the surrounding massive, low-density diffuse ionized gas (DIG). 

The structure of individual \hii\ regions in the Milky Way (MW) has been studied in detail for many years \citep[e.g.,][]{odell2001, mesadelgado2011, mcleod2016, kreckel2024}. These studies provide the foundation of our understanding of the physics and structure of these regions \citep[][]{osterbrock2006,draine2011}. Unfortunately, Galactic \hii\ regions beyond the Solar Neighborhood remain challenging to study at optical wavelengths due to high extinction, unfavorable geometry, and uncertain distances. While significant progress has been made tracing large populations of Galactic regions via infrared \citep[e.g.,][]{churchwell2006,anderson2014} and radio emission \citep[e.g.,][]{armentrout2021}, achieving a complete view of \hii\ region populations using optical line emission requires surveying other galaxies \citep[e.g.,][]{kennicutt1984,pellegrini2011,drory2024}. 

Studies of optical hydrogen recombination line emission from nearby galaxies thus form the basis of our view of populations of \hii\ regions. They have yielded measurements of the \hii\ region luminosity function \citep[e.g.,][]{kennicutt1989,thilker2002,santoro2022}, the fraction of emission associated with the DIG as opposed to \hii\ regions \citep[e.g.,][]{thilker2000,belfiore2022}, and the location and rate of star formation \citep[e.g.,][]{kennicutt2009}. However, the majority of \hii\ regions are small, with radii $\sim 5{-}15$~pc in Milky Way and \textit{Hubble} Space Telescope studies \citep[e.g.,][]{anderson2014,barnes2022,barnes2025}. 
As a result, seeing-limited $\sim 1''$ resolution H$\alpha$ observations in galaxies at $\gtrsim 10$~Mpc (where $1'' \gtrsim 50$~pc) fail to resolve the structure of most \hii\ regions. 

In this paper, we catalog and measure the resolved structure for a complete sample of $\sim2500$ \hii\ regions identified as \ha peaks in the \citet{congiu2025} VLT/MUSE mosaic of the very nearby spiral galaxy NGC\,253 \citep[D~$\approx$~3.5~Mpc so that $1'' \approx 17$~pc;][]{okamoto2024}. This is, to our knowledge, the largest set of resolved structural measurements for \hii\ regions in any nearby galaxy. For each region, we measure resolved profiles of extinction-corrected H$\alpha$ and other line emission. From these profiles, we make a series of fundamental structural measurements addressing:
\begin{enumerate}
\item What is the characteristic shape of the \ha\ emission profile about the \ha\ peak? 
\item What implications do our measured structures have for the commonly invoked luminosity-size relationship for \hii\ regions?
\item What is the density of the ionized gas in \hii\ regions, and does this correlate with other observed properties of the region?
\item How do our results fit in the context of previous work with differing methodology and resolution?
\end{enumerate} 
Because VLT-MUSE \citep[][]{bacon2010} achieves outstanding sensitivity with a large field of view, the \citet{congiu2025} survey of NGC 253 detects \ha\ emission even from regions powered by individual late-type O stars over the whole galaxy. MUSE also covers both \ha and \hb and classic BPT diagnostics, allowing for robust extinction correction and spectroscopic validation of our target \hii\ regions \citep[e.g.,][]{kewley2019}. 

This work forms part of a new era of highly resolved \hii\ region studies, including studies of other very nearby galaxies such as NGC\,300 (D~$\approx2$~Mpc), NGC\,5236/M83 (D~$\approx5$~Mpc), and NGC\,7793 (D~$\approx3.4$~Mpc) \citep{mcleod2020,dellabruna2020,dellabruna2022}. Compared to these studies, our work emphasizes the structure of individual regions \citep[e.g., similar to][in M\,33, D~$\approx840$~kpc]{relano2009} rather than a stellar feedback inventory or population statistics. Our emphasis on the resolved structure of regions is similar to recent \textit{Hubble} and \textit{JWST} work by \citet{barnes2022}, \citet{chandar2025}, and \citet{pedrini2024}. Our work complements those studies by leveraging the excellent surface brightness sensitivity and spectroscopic coverage of VLT-MUSE to achieve a more complete view of the \hii\ regions over the full area of an individual spiral galaxy. Our goal is to offer a more physical picture to feedback studies \citep[e.g.,][]{lopez2014,mcleod2020,barnes2021,pathak2025} and a highly resolved view of \hii\ regions that can inform measurements of their properties from lower resolution data, extending even to high redshift data \citep[e.g.,][]{wisnioski2012,cosens2018}.

In this study, we focus only on ionized gas. Following standard practice, we identify \hii\ regions as bright concentrations of photoionized gas. In actuality, \hii\ regions are structured around massive stars and reflect the interface of ionizing photons from these stars with the surrounding interstellar medium \citep[][]{whitmore2011,hannon2022}. NGC\,253 is the target of extensive surveys that capture optical and UV emission from the massive stellar populations \citep[][HST GO 17809]{dalcanton2009,hassani2024} as well as dust, atomic, and molecular gas \citep[][JWST GO 2987]{leroy2021}. Combining these observations with the \citet{congiu2025} VLT-MUSE view represents a natural next step, but here the reader should view our results as characterizing the structure of $\approx 17$~pc-scale concentrations of photoionized gas.

\input{fig-peakmapcutouts}

\section{Data and Methods} \label{data}

NGC\,253 is nearby \citep[D$=3.50 \pm$ 0.22 Mpc,][]{rekola2005,radburnsmith2011,okamoto2024,congiu2025}, massive \citep[$M_\star \sim 4 \times 10^{10}$ M$_\odot$][]{leroy2021}, and actively forming stars \citep[SFR $\sim 5$ M$_{\odot}$~yr$^{-1}$,][]{leroy2019}. Other galactic properties are listed in Table~\ref{tab:ngc253_props}. At this distance, $1'' = 17$~pc. The galaxy has a well-studied nuclear starburst that produces a strong outflow \citep[e.g.,][]{cronin2025}, though we largely omit this region from our analysis. In the disk of the galaxy, individual spiral arms and star-forming regions are easily distinguished despite the galaxy's inclination \citep[$i \approx 76^\circ$,][see Appendix~\ref{ap:inclination} for discussion]{mccormick2013}. Together, these properties make NGC\,253 one of the best targets for studying the highly resolved ISM in the disk of a typical star-forming galaxy.

\subsection{VLT-MUSE Survey} 
\label{sec:musedata}

NGC\,253 was observed with MUSE on the Very Large Telescope as part of ESO programs 108.2289 (P.I. Congiu, the full disk mosaic) and 0102.B-0078 (P.I. Zschaechner, two central pointings). \citet{congiu2025} present this combined dataset and describe the data reduction. Briefly, the observations consist of 103 individual pointings that together form a $\approx 20' \times 5'$ area that covers the disk of the galaxy (Figure~\ref{fig:peaks}). The data have a spaxel size of $0.2''$ and seeing-limited resolution (see Appendix~\ref{ap:psf}). The total exposure per field was $\approx 844$~s (1920~s for the central pointings).

MUSE observes the wavelength range between 4600 and 9300 \AA\ (WFM-NOAO-E mode). We use the emission line maps produced by the PHANGS-MUSE Data Analysis Pipeline (DAP)\footnote[1]{\url{https://gitlab.com/francbelf/ifu-pipeline}} \citep{emsellem2022} applied to the NGC\,253 MUSE data as described in \citet{congiu2025}. This procedure subtracts the stellar continuum and produces emission-line intensities with associated uncertainties. To maximize the detected signal, we use a version of the data where the cube was convolved with a small $0.6''$ ($\approx 3\times3$ pixel) kernel before running the DAP. This has a modest effect on the PSF, increasing it to $\approx1''$ (see Appendix~\ref{ap:psf}) but a large effect on the S/N, which improves by a factor of $\approx 3$. The DAP produces two sets of line data products, one based on a single Gaussian fit to each line and another ``moment~0'' based on integrating the part of the continuum-subtracted spectrum where the line is expected to occur. To avoid the S/N-based clipping implicit in the Gaussian fits, our analysis uses the moment~0 maps. Both the line fitting and the moment~0 become less reliable in the galaxy center and near the starburst-driven galactic wind, which often exhibit multiple distinct velocity components and are heavily affected by extinction \citep[][]{cronin2025}. For this reason, we exclude a 600~pc by 600~pc box centered at the galactic center from our analysis (black box in the top panel of Figure~\ref{fig:peaks}).

A typical $1\sigma$ statistical error on an \ha intensity in our map is $6.4\times10^{-18}$~erg~s$^{-1}$~cm$^{-2}$~arcsec$^{-2}$
, which translates to $3.2\times10^{37}$~erg~s$^{-1}$~kpc$^{-2}$. Given the $\approx 289$~pc$^{2}$ size of the average PSF in our MUSE data, this implies a $5\sigma$ point source sensitivity limit of $4.6\times10^{34}$~erg~s$^{-1}$. For a Case~B recombination, assumed temperature of $10^{4}$~K and no extinction, this translates to an ionizing photon production rate of $2.3\times10^{49}$~s$^{-1}$, which is about that of a main sequence O6 star \citep[][]{schaerer1997}.

\input{tab-galprops}

\subsection{HII Region Identification} 
\label{sec:regions}

As discussed in Section~\ref{sec:intro}, \hii\ regions physically trace ionization of gas by massive stars. Resolved information on the powering stellar populations is usually not available, however, and standard practice in the field has been to identify \hii\ regions as bright, compact regions of \ha\ emission \citep[e.g.,][]{thilker2000,wisnioski2012,fisher2017,limacosta2020}. We follow this approach and identify local maxima from an unsharp-masked version of the map (the ``difference'' map), then select the subset of significant peaks based on the contrast between the \ha\ peak and the local background. We then apply Baldwin-Phillips-Terlevich \citep[BPT;][]{baldwin1981} line-ratio cuts to identify likely \hii\ regions. Our final catalog contains \nregions \hii\ regions. We show their locations in the top panel of Figure~\ref{fig:peaks} and report their locations and properties in Appendix~\ref{ap:catalog}.

To identify local maxima, we first create a local background estimate by filtering the \ha map with a circular median filter that has a radius four times the PSF ($4''$ $\approx$ 65~pc). We subtract this from the original \ha image to create a difference map that highlights regions with intensities greater than their local background. Figure~\ref{fig:peaks} shows this difference map for one region of the galaxy. 

We identify local maxima in the difference map using a circular search kernel of radius $2''$ $\approx$ 32~pc. To classify a local maximum as a significant peak, we required that it show high contrast relative to the local background. Specifically, we require the value of the peak pixel in the native map to be $>1.65$ times its value in the median-filtered local background map. This eliminates extended, low-intensity features like ISM filaments but retains compact \hii\ regions. We also impose an intensity cut, requiring the difference map to have a value greater than $2.5\times 10^{38}$ erg~s$^{-1}$~kpc$^{-2}$ (corresponding to typical $S/N \approx 6.7$).  After selecting candidate maxima, we also exclude peaks within 8 pixels ($1.6''$) of the edge of the image, where the map becomes noisier; any peaks where \hb\ is not detected with at least 10$\sigma$ significance; and any peaks already identified as planetary nebulae in \citet{congiu2025}. Finally, we apply the BPT cuts described in Appendix~\ref{ap:bpt} to select regions likely to be photoionized by massive stars.

We chose the search kernel radius and the intensity and contrast thresholds by experimentation and visual inspection. Focusing on a rich field in the eastern disk (white box in top panel of Figure~\ref{fig:peaks}), several of the authors verified that the peaks selected corresponded well to the nebulae one would pick out by eye. In Appendix~\ref{ap:completeness}, we test the completeness and detection limits of this methodology by inserting synthetic sources into the \ha\ map. As expected, faint peaks with low contrast above the local background are the most difficult to detect, a significant challenge at small galactocentric radius where the detected regions tend to be brighter (Figure~\ref{fig:peaks}). Across the full disk, we miss about 10\% of injected sources with peak intensity $< 10^{40}$~erg~s$^{-1}$~kpc$^{-2}$. Larger regions  ($R_{\rm mom} >25$~pc) are also slightly more difficult to detect, especially those located in complex regions with many other nearby peaks. 

\subsection{H\texorpdfstring{$\alpha$}{alpha} Extinction Correction} 
\label{extinction}

After identifying peaks and applying BPT cuts, but before any subsequent analysis, we correct the H$\alpha$ map for the effects of extinction. Once the MUSE maps are smoothed to $1''$ resolution, the recovered signal in the \hb\ map was sufficient to perform a pixel-by-pixel extinction correction using the Balmer decrement assuming the \citet{odonnell1994} extinction curve with $R_{\text{V}}=3.1$ and an intrinsic \ha/\hb ratio of 2.86. Unless otherwise stated, the \ha fluxes and luminosities quoted throughout this paper are the extinction-corrected values. 

From the total extinction-corrected \ha luminosity of the galaxy and the conversion factor from \citet{kennicutt2012,murphy2011}, we calculate a star formation rate (SFR) of $2.78\pm0.56$~M$_\odot$~yr$^{-1}$, assuming 20\% uncertainty (Table~\ref{tab:ngc253_props}). Excluding the $\sim600\times600$~pc box around the galactic center (Figure~\ref{fig:peaks}), we calculate a disk SFR of $1.72\pm0.34$~M$_\odot$~yr$^{-1}$. The extreme values of extinction in the galactic center make \ha an unreliable tracer of star formation in this regime, but our calculations agree with \citet{bendo2015} ($1.73\pm0.12$~M$_\odot$~yr$^{-1}$) within $2\sigma$ uncertainty. Combining the disk SFR from MUSE and the central SFR from \citet{bendo2015}, we estimate the composite SFR to be 3.45~M$_\odot$~yr$^{-1}$.

\subsection{Radial Profiles} 
\label{sec:radialprofiles}

To characterize the structure of each \hii\ region, we build radial profiles around its \ha peak. To do this, we first calculate the distance from the peak to all nearby pixels. We assume zero inclination, i.e., that our $\approx 17$~pc resolution reaches a scale where the regions are approximately symmetric objects rather than foreshortened along the minor axis by the tilt of the galaxy. We verify in Appendix~\ref{ap:inclination} that the region extent appears nearly symmetric between the major and minor axes.

We bin pixels according to their distance from the peak into 6~pc ($\approx 0.35''$) wide annular bins out to 200~pc. We ensure that no pixel is included in more than one \hii\ region by creating masks that assign each pixel to at most one region (Appendix~\ref{ap:catalog}). When a pixel is contested between two regions, we assign it to the nearest peak. 

Within each bin, we calculate a set of statistics to describe the \ha intensity distribution, including mean, median, and various percentile ranges of intensity. We focus our analysis on the median profile because it reduces contamination from unmasked peaks in the bins at large distances. In Appendix~\ref{ap:meanvsmed}, we find good agreement between results calculated using the median and the mean profile. 

\subsubsection{Local Background and Aperture Definition}
\label{sec:localbackground}

\input{fig-cdf}

\hii\ regions often exist in regions with significant diffuse emission \citep[e.g.,][]{haffner2009,belfiore2022}. \citet{belfiore2022} showed that in PHANGS-MUSE, this diffuse emission appears as halos surrounding the \hii\ regions. This also appears to be the situation in NGC 253, where there are almost no lines of sight without \ha\ emission, and extended emission surrounds bright regions (Figures~\ref{fig:peaks}, \ref{fig:cdf}). Without a defined integration aperture, the luminosities and size estimates of many regions would diverge.

Based on visual inspection of both individual and stacked radial profiles, we judge that the binned profiles almost always fall to a stable background within 100~pc from the peak. Some regions drop off much faster. Other profiles show a slow, low-level decline in the background intensity level out to even larger scales, but this radius typically captures their bright inner emission. Therefore, we treat $R_{\rm lum}^{\rm max} = 100$~pc as an upper limit on the aperture definition.

Within this limit, we define a dynamically sized aperture and calculate and subtract a local background intensity for each region. For each region, we estimate the local background intensity as the 10th percentile of the binned median intensity profile within $R_{\rm lum}^{\rm max} = 100$~pc of the \ha peak. Based on visual inspection of a large number of individual and stacked profiles, this statistic does a good job of capturing the minimum intensity just outside the bright emission associated with the peak. We subtract this background value from the binned profiles and the peak intensity to yield our final set of measurements. Unless otherwise noted, we refer to these background-subtracted measurements below. 

After subtracting the background, we set the integration aperture for each region to the radius where the median intensity profile drops to $\leq 1\sigma$---the standard deviation estimated from the median absolute deviation. We call this radius \rlum\ and it represents the outer boundary for determining the region luminosity and sizes. 

Figure~\ref{fig:cdf} shows that on average this treatment leads to a flat enclosed flux over the range $\approx 50{-}150$~pc, indicative of a successful background subtraction. Though the presence of extended diffuse emission is widely acknowledged, the choice to subtract a local background is not standard in this field \citep[e.g., we often compare to][who do not subtract a background]{groves2023,barnes2025}. Appendix~\ref{ap:bgsub} examines the impact of this choice. We also highlight the impact at several points during the text and report several of our results with and without background subtraction to allow easier comparison with the literature. In these cases, we still use \rlum\ as the integration aperture.

We caution that some of our results are quite sensitive to how we define our integration apertures. Figure~\ref{fig:cdf} shows the cumulative flux profile as a function of radius with two different definitions of $R_{\rm lum}^{\rm max}$, the maximum background radius. The flux profile at large radius varies significantly between the two cases, and still diverges for $R_{\rm lum}^{\rm max} = 200$~pc, leading to larger region sizes. On the other hand, setting $R_{\rm lum}^{\rm max}$ to a lower value $< 100$~pc would lead to smaller regions.  
In our view, our approach is not any more arbitrary than using an image segmentation algorithm to define the edges (and the integration aperture) of the \hii\ regions, but it still represents a significant potential source of bias. In the specific case of the radius vs. luminosity relation (Section~\ref{sec:radius}), we verified that the choice of integration aperture (e.g., using one of our measured sizes in place of \rlum ) does not change the qualitative picture.

\subsubsection{Radius Measurements}
\label{sec:radius}

The literature uses several statistics to measure the sizes of \hii\ regions. These probe different aspects of region structure, and we measure several in our analysis: 
\begin{enumerate}
    \item \textbf{HWHM:} The half width at half maximum is the radius at which the azimuthally-averaged \ha intensity is half of the peak value. This is sometimes referred to as the ``core radius'' \citep[e.g.,][]{sandage1974,wisnioski2012}, and characterizes the inner structure of the emission. We measure this by interpolating the radial profile. For a Gaussian function, $2 \times \text{HWHM}=\sigma/2.355$.
    \item \textbf{\rmom:} The intensity weighted second moment size is commonly used to measure the sizes of ISM clouds \citep[e.g.,][]{rosolowsky2006} and has been increasingly applied to \hii\ regions \citep[e.g.,][]{giunchi2023,barnes2025}. \rmom is calculated as the intensity-weighted root mean square distance from the peak summing over the profile. 
    \item \textbf{\rf:} Sometimes referred to as the half-light or effective radius, \rf measures the radius enclosing 50\% of the total flux of the region. The approach to background subtraction and choice of \rlum\ significantly affects the resulting \rf. We find that \rf and \rmom are strongly correlated.
    \item \textbf{\rn:} Following the same procedure as \rf, we calculate \rn as the radius enclosing 90\% of the total flux of the region. The result is also sensitive to the selected \rlum.
    \item \textbf{\riso:} The isophotal radius measures the radius at which the surface brightness of the radial profile reaches a given value \citep{kennicutt1979}. This tends to be sensitive to the extended and diffuse emission at the outer edges of the regions, though this depends on the adopted isophote. The isophotal radius resembles the boundaries outlined by common image segmentation techniques such as CLUMPFIND and HIIPhot \citep{williams1994,thilker2000}. For our fiducial analysis, we choose a literature value for the background-subtracted surface brightness of $10^{38.8}$ erg~s$^{-1}$~kpc$^{-2}$, taken as the distinction between \hii\ regions and the surrounding DIG in \citet{belfiore2022}. We explore the effects of this choice in depth in Section~\ref{sec:isophotal}.
\end{enumerate}

Finally, we record the aperture size, \rlum, used to estimate the background and luminosity for each region. All radii are recorded in the region catalog (Appendix~\ref{ap:catalog}). In Appendix~\ref{ap:profileprops} we explore how the image resolution and the choices made in developing our methodology affect our conclusions.

\subsection{Peak Intensity Bins}
\label{sec:intbin}

We split our \nregions\ regions into 10 bins of $\approx250$ regions each based on their extinction-corrected, background-subtracted \ha peak intensity. Many key properties of our regions correlate with the \ha brightness of the peak, including luminosity, extinction, and optical line ratios. For example, Figure~\ref{fig:LvsI} shows a strong correlation between peak intensity and luminosity, but unlike the luminosity, the peak intensity is simple and independent of the adopted aperture size, \rlum.

\input{fig-LI}

In Figure~\ref{fig:peaks} we color the peaks by their intensity bin. The brightest regions are concentrated in the inner galaxy and spiral arms, while the faintest regions appear in the interarm and outer galaxy regions. Figure~\ref{fig:LvsI} and Table~\ref{tab:ngc253_props} report the distribution of background-subtracted peak intensities and luminosities. The distribution of peak intensity and luminosity within each bin is provided in Table~\ref{tab:binprops}. Bins 2-9, which capture typical \hii\ regions, span about two orders of magnitude in peak intensity and \ha\ luminosity.

\input{tab-binprops}

\subsection{Comparison to NGC 5068}
\label{sec:ngc5068}

We also apply our methodology to the HST narrowband H$\alpha$ observations of NGC\,5068 (Appendix~\ref{ap:ngc5068}) to demonstrate the generality of our results. At $D=5.2$~Mpc, NGC\,5068 is the nearest member of the H$\alpha$ survey by \citet{chandar2025} and \hii\ regions have been cataloged by \citet[][using MUSE]{groves2023} and \citet{barnes2025} (using these HST data). The HST data have $10$~pc resolution and worse surface brightness sensitivity than our NGC\,253 MUSE survey. NGC\,5068 is a low mass galaxy with moderate inclination and widely spaced \hii\ regions and so represents a useful counterpoint to NGC\,253. We reference the results from NGC\,5068 throughout Sections~\ref{sec:RL} and \ref{sec:profileshape}, and generally find our approach to be robust.

\section{Luminosity and Radius}
\label{sec:RL}

We identify \nregions \ha peaks likely to correspond to \hii\ regions (Figure~\ref{fig:peaks}). For comparison, the WISE IR-based catalog of \hii\ region candidates in the Milky Way contains about 8000 objects \citep{anderson2014} while the typical number of objects recovered in the more distant PHANGS-MUSE target galaxies with similar stellar mass to NGC\,253 is about 1500 \citep{groves2023}. We compare the luminosity distribution of our sample to PHANGS-MUSE in Section~\ref{sec:lumdist}.

In the rest of this section, we analyze the resolved structure of these regions. In Sections~\ref{sec:hwhm}, \ref{sec:effectiveradius}, and \ref{sec:isophotal} we examine different size measurements, focusing on their variation (or lack thereof) with region luminosity. As \hii\ regions increase in luminosity and the powering stellar population produces more ionizing photons, we expect the extent of the ionized gas to increase \citep[][]{stromgren1939}. This is expected to create a luminosity-size relation for \hii\ regions, which has been observed across redshift and galaxy types \citep[e.g.,][and references therein]{gutierrez2011,wisnioski2012,cosens2018}.

Beyond scaling relations, these profiles encode information on the structure of \hii\ regions, including the location and density of the ionized gas and the location where photoionizations occur. In Section~\ref{sec:profileshape}, we examine the resolved structure of our regions. Our results include measurements of a density-surface brightness relation that might be as fundamental as the luminosity-size relation, at least in NGC\,253 and the PHANGS-HST H$\alpha$ measurements from \citet{barnes2025}.

\subsection{Luminosity Distribution}
\label{sec:lumdist}

Figure~\ref{fig:LvsI} and Table~\ref{tab:ngc253_props} show the distributions of \ha luminosity and peak intensity for our regions. The cumulative distribution function (CDF) of extinction-corrected H$\alpha$ luminosity ($L_{\rm H\alpha}^{\rm corr}$, bottom right) is well-fit by a broken power law of slopes $-$0.92 and $-$1.86 with a luminosity break (L$_{\text{break}}$) at $10^{38.2}$~erg~s$^{-1}$. This agrees reasonably well with results from \citet{gutierrez2011} using high-resolution HST H$\alpha$ imaging of M51. They found that the \hii\ region luminosity CDF was well-described by a broken power law of slopes $-$0.84 and $-$1.26 with a break at $10^{38.5}$ erg~s$^{-1}$. 

For comparison, the CDF of $L_{\rm H\alpha}^{\rm corr}$ for \hii\ regions in the 19 PHANGS-MUSE galaxies \citep[][excluding galaxy centers]{santoro2022,groves2023} are plotted in gray in Figure~\ref{fig:LvsI}. Our NGC\,253 \hii\ regions span to lower luminosities and do not include analogs to the most luminous regions in PHANGS-MUSE. This partially reflects our use of background subtraction, which lowers the luminosity of faint regions. To see this, compare the black (with background subtraction) and gold (without) lines in Figure~\ref{fig:LvsI} and see Appendix~\ref{ap:bgsub}. The extension of our catalog to low luminosities reflects the closer distance of NGC\,253, which leads to improved luminosity sensitivity compared to \citet{groves2023}. At the high luminosity end, the higher physical resolution of our data compared to PHANGS-MUSE also lowers the typical region luminosity and size, because large complexes identified as one object at the $\approx 80$~pc MUSE resolution break into smaller \hii\ regions at $<20$~pc resolution. This is also seen in \citet{barnes2025}, where the largest PHANGS-MUSE regions separate into individual smaller regions. 

The brightest (bin 10) and faintest (bin 1) bins stand out, with their distributions quoted in Table~\ref{tab:binprops} and widespread profiles in Figure~\ref{fig:decile-profiles}. These bins each cover a range of $\approx 2$~dex in intensity and luminosity and capture the tails of these distributions. The faintest regions in bin 1 likely represent a mixture of mis-categorized DIG features and small, isolated \hii\ regions \citep[e.g., as in][]{barnes2022}. The brightest regions in bin 10 represent the most likely candidates for extranuclear starburst regions in NGC\,253. In particular, we flag the brightest subset of $\approx 17$ regions with luminosity $> 10^{38.7}$ erg~s$^{-1}$, which stand out in Figure~\ref{fig:LvsI} (bottom right). This break is consistent with the break associated with the transition from ionization-bounded to density-bounded \hii\ regions \citep[e.g.,][]{beckman2000}. Upon visual inspection, these regions appear to be bright knots in larger luminous complexes, perhaps analogous to regions like 30 Doradus in the Large Magellanic Cloud or NGC 604 in M33. Follow-up comparison to stellar sources and higher resolution recombination line imaging with HST or JWST will help clarify the nature of both populations.

Without background subtraction, our cataloged \hii\ regions account for $70\%$ of the total extinction-corrected \ha luminosity in the MUSE field (excluding the galactic center). Presumably, the remaining 30\% \ha luminosity arises from the DIG. Studying PHANGS-MUSE galaxies, \citet{belfiore2022} found $\sim40\%$ of the \ha luminosity to be associated with the DIG on average, with a range of $\sim 20{-}55\%$. The PHANGS-MUSE and NGC\,253 values agree with both the typical value and the wide spread in the DIG fraction found in the literature \citep[e.g., $f_{\text{DIG}}=7{-}69\%$,][]{thilker2000,oey2007,chevance2020}. When we apply a local background subtraction, we find the \hii\ regions account for $\sim 25\%$ of the \ha luminosity ($f_{\text{DIG}}=75\%$), again highlighting the large impact of the choice to background subtract.

Thus, we appear to recover a standard population of \hii\ regions, reaching lower luminosities and recovering larger numbers of objects than studies of more distant galaxies and including most of the \ha\ emission from the galaxy within our analyzed regions. In Appendix~\ref{ap:psf}, we show that the stack of all \hii\ regions is well resolved compared to the planetary nebulae. And because the \citet{congiu2025} mosaic covers these regions with spectroscopic mapping, this represents one of the largest spatially-resolved and spectroscopically recovered samples of extragalactic \hii\ regions \citep[along with][]{dellabruna2020,dellabruna2022}.

\input{fig-decileprofiles}

\subsection{HWHM Size vs. Luminosity}
\label{sec:hwhm}

In Figure~\ref{fig:decile-profiles}, the stacked profiles exhibit a consistent shape across all bins of \ha peak intensity. They appear strongly peaked in the center, with intensity dropping off quickly with increasing radius at first, then show a shallower slope outside the core until the profiles level off at $\sim$60~pc. At all \ha intensities, the median \hii\ region profiles are more extended than the MUSE PSF. Although the amplitudes of the profiles vary, regions show a nearly constant HWHM (pink, interpolated directly from the profile) of $\sim15$~pc (deconvolved HWHM $\approx12$~pc) across all bins. This can also be seen in Figure~\ref{fig:RL}, where we plot background-subtracted, extinction-corrected H$\alpha$ luminosity as a function of each radius measurement. The HWHM covers a relatively narrow range and shows only a weak correlation with luminosity ($\rho_{\text{HWHM}}=0.19$, Table~\ref{tab:RLcorrelations}). We also observe this nearly fixed HWHM about the H$\alpha$ peak in the HST H$\alpha$ observations of NGC\,5068 (Appendix~\ref{ap:ngc5068}). 

In NGC\,253, this behavior persists as we degrade the resolution of our data from $1''$ to $5''$ and remeasure profiles (Appendix~\ref{ap:res}). The lack of a clear core radius-luminosity relation also does not depend on our treatment of background subtraction (Appendix~\ref{ap:bgsub}). Thus, this appears to be a robust result. However, it contrasts with previous literature on the sizes of extragalactic \hii\ regions \citep[][and others]{wisnioski2012,cosens2018}, which have sometimes found strong correlations between the core radius and luminosity of large samples of \hii\ regions. We discuss this more in Section~\ref{sec:comparisons}.

Our best explanation for the stability of the HWHM is that the bright cores of our regions are not well-resolved even at our high physical resolution of 17~pc (FWHM, 8.5~pc HWHM). In Appendix~\ref{ap:res}, we show that because of the extended portion of the profile, our regions appear marginally resolved even if we blur the data by a factor of $\sim 5$, closer to the physical resolution of previous ground-based work targeting more distant galaxies. In this case, the HWHM captures the combined presence of a bright unresolved concentration of gas and a more extended component, and as a result, always appears at small multiples of the PSF size. Supporting this, in Section~\ref{sec:profileshape} we find that the surface brightness profile is well represented by a double Gaussian function with a compact inner component with deconvolved HWHM of $\sim 7$~pc (and extended component with HWHM of $\sim31$~pc), i.e., quite small. 

We emphasize that given the high physical resolution of our data, if the core radii in NGC\,253 remain only marginally resolved, then so do essentially all previous ground-based measurements of extragalactic \hii\ regions beyond the Local Group.

\subsection{Fractional Flux Sizes vs. Luminosity}
\label{sec:effectiveradius}

\input{fig-RLcombo}

Figures \ref{fig:decile-profiles} and \ref{fig:RL} and Table~\ref{tab:RLcorrelations} also report results for \rf and \rn , the radii enclosing fixed fractions of the flux. We also plot the closely related intensity weighted second moment \rmom . Compared to the HWHM, these size measurements are more sensitive to the full distribution of intensity within the region. The effective radii are larger than the HWHM and all show a stronger correlation with luminosity than the HWHM (Spearman correlation coefficient $\rho_{\text{\rf}} = 0.38$ compared to $\rho_{\text{HWHM}} = 0.19$), but still a much weaker correlation than the isophotal radius ($\rho_{\text{\riso}} = 0.89$, Table~\ref{tab:RLcorrelations}).

For both individual regions and the stacked profiles, \rmom and \rf show a strong correlation with one another ($\rho=0.99$, Table~\ref{tab:sizesize}). When binning the data by peak intensity or luminosity for these radii, we find almost no relationship between luminosity for either. We do observe a moderate correlation with luminosity in the individual data or when binning by radius. However, the relationship appears weaker than many of the luminosity-size relationships reported in the literature (Section~\ref{sec:comparisons}).

\rn\ shows a slightly stronger relationship to luminosity than \rf or \rmom ($\rho_{\text{\rn}}=0.43$, Table~\ref{tab:RLcorrelations}) and yields a wider overall range of sizes. Similar to \rf and \rmom , when binned by peak intensity or luminosity, the data show only a weak relation between \rn\ and luminosity. When binned by radius or fit to the data directly, however, the $L$-\rn\ appears similar to the $L$-\riso\ relationship but with more scatter. These trends appear consistent with the $L \propto R^2$ relationships often seen in the literature (compare Figures \ref{fig:RL} and \ref{fig:RLcomparisoniso}).

\input{tab-correlations}

Because these sizes depend on our estimate of the region luminosity, \rn\ and \rf\ depend on the adopted background treatment and aperture size, \rlum. In Appendix~\ref{ap:bgsub} we show that in the absence of background subtraction, \rn\ is almost completely set by the adopted aperture size, and it still correlates closely with \rlum\ after subtraction ($\rho=0.88$, Table~\ref{tab:sizesize}).  Because the second moment is sensitive to outliers, \rmom\ shows a similar sensitivity to the defined full extent of the region. In the cases where an image segmentation algorithm like HIIphot or selection based on dendrograms is employed, these size metrics will depend on the specifics of the image segmentation used to define the region footprint. That can be a strength if one trusts the image segmentation algorithm, e.g., because it has some physical motivation, but it can make these sizes unstable in the case where there is no strong constraint on the exact region footprint. 

Despite these caveats, ratios among these sizes, e.g., $R_{50} / R_{90}$ (often defined as the concentration index), can serve as a useful indicator of region morphology \citep[e.g.,][and Section~\ref{sec:density}]{hannon2022}. We also emphasize that despite the difficulty measuring \rf, this quantity has physical meaning as the two-dimensional radius within which half of the photoionizations, and so presumably much of the stellar feedback, occurs.

In Appendix~\ref{ap:ngc5068} we apply our methods to HST H$\alpha$ measurements of NGC\,5068. Compared to our results for NGC\,253, we find a much clearer relationship between luminosity and \rf, \rn, or \rmom. Broadly, we take these different results for the same method applied to different galaxies to reflect the instability in these size metrics. Specifically, we suspect that the cleaner background and more well-separated regions in NGC\,5068 contribute to these results, as does the use of the HIIphot-based \citet{groves2023} region definitions.

\input{tab_sizesize}

\subsection{Isophotal Size vs. Luminosity}
\label{sec:isophotal}

The most common measure of \hii\ region size is the isophotal radius. Most image segmentation algorithms at least approximately identify structures down to a fixed intensity contour or intensity gradient, so the area inside regions defined by HIIphot or CLUMPFIND will correspond well to \riso. For our measurements, we measure the radius at which the background-subtracted and extinction-corrected \ha\ profile matches a surface brightness of $10^{38.8}$~erg~s$^{-1}$~kpc$^{-2}$. In \citet{belfiore2022}, this represents approximately the intensity boundary between \hii\ regions and the surrounding DIG.

In Figures \ref{fig:decile-profiles} and \ref{fig:RL}, the isophotal radius shows a tight correlation with \ha luminosity across the full range of luminosity and intensity in our catalog. The trend is visible in both the distribution of individual data and the fits to the binned profiles, with all three binning methods (intensity, luminosity, and radius). In Figure~\ref{fig:RLcomparisoniso} we see that this luminosity-size relation measured using \riso\ matches the relation from previous extragalactic \hii\ region studies well. A power law that describes our data well is
\begin{equation}
\label{eq:lvsriso}
\log_{10} L\left({\rm H}\alpha^{\rm corr}\right)^{\rm bksub} = 
2.1\log_{10} R_{\rm iso}^{\rm 38.8} + 33.6,
\end{equation}
where $R_{\rm iso}^{38.8}$ is the isophotal radius of an \hii\ region at our fiducial $10^{38.8}$ erg~s$^{-1}$~kpc$^{-2}$ isointensity contour and $L\left({\rm H}\alpha^{\rm corr}\right)^{\rm bksub}$ is the extinction-corrected background-subtracted H$\alpha$ luminosity.

\input{fig-RLcomparisoniso}

The stronger luminosity-radius relationship for \riso\ contrasts with the absence of any strong relation between luminosity and HWHM and ambiguous results for the effective radius \rf. This reflects that the HWHM, \rf, and \rn\ measure the shape of the profile while \riso\ is also sensitive to its amplitude. In Figure~\ref{fig:decile-profiles}, \riso\ is the radius ($x$-axis value) at which the profile intercepts a fixed $y$-axis value. Although the profiles show a common shape, their amplitude varies dramatically. This leads to changes in \riso\ as a function of the peak \ha\ intensity. Because luminosity depends on peak intensity times area for a fixed shape profile, we then expect an isophotal size-luminosity relation even for a fixed region shape. For reference, we note the expectation for a Gaussian profile with width $\sigma$:
\begin{equation}
\label{eq:gaussriso}
R_{\rm iso} = \sigma \sqrt{2 \ln \frac{I_0}{I_{\rm iso}}},
\end{equation}
where $I_0$ is the central brightness and $I_{\rm iso}$ the isophotal threshold. In practice, the functional shape of the profile is more complex than a single Gaussian (see \S \ref{sec:profileshape}).

A corollary of the above is that tuning $I_{\rm iso}$ will change \riso, and our adopted isophotal contour is somewhat arbitrary. In \citet{belfiore2022}, there is galaxy-to-galaxy variation in this boundary and overlap between DIG and \hii\ region intensities. Moreover, \citet{belfiore2022} base their separation on the application of HIIphot to PHANGS-MUSE, so the sensitivity of the MUSE data and algorithmic biases from HIIphot inform this value. Our analysis here shows no sharp separation between the two, with the \hii\ regions steadily transitioning into the DIG (Figure~\ref{fig:cdf}, Appendix~\ref{ap:profileprops}). Other aspects of the \citet{belfiore2022} analysis support this view. Given this, in Figure~\ref{fig:RLcomparisoniso} we vary the isophotal intensity used to define \riso. In all cases that we study, we find a strong correlation between isophotal radius and luminosity. The specific sizes measured change, reflecting that a higher threshold samples the inner part of the profiles (recall Figure~\ref{fig:decile-profiles}).

By changing the isophote used in our analysis, we can match the trends in previous studies. For example, in Figure~\ref{fig:RLcomparisoniso} we plot the distribution of \hii\ region sizes from the PHANGS-MUSE catalog \citep[][brown contours]{groves2023} and the PHANGS-HST \ha catalog \citep[][gray contours]{barnes2025}, including NGC~5068 (see Appendix~\ref{ap:ngc5068}). With an isophote similar to our fiducial value, we match the high-sensitivity PHANGS-MUSE measurement well. A higher threshold produces a relation that matches the PHANGS-HST catalog, which has higher resolution but worse sensitivity ($1\sigma=6.3\times10^{-16}$~erg~s$^{-1}$~cm$^{-2}$~arcsec$^{-2} = 7.0\times10^{39}$~erg~s$^{-1}$~kpc$^{-2}$).

Note that this exercise breaks down for very high or very low isophotes. When the isophote is decreased below $\sim10^{38}$ erg~s$^{-1}$~kpc$^{-2}$ (e.g., purple lines in Figure~\ref{fig:RLcomparisoniso}), we reach a regime where the background-subtracted surface brightness of the high intensity \hii\ regions only reaches this threshold far out in the low-intensity, extended wing of the profile (Figure~\ref{fig:decile-profiles}, bottom panels). This leads the relation to steepen as high intensity regions yield sizes that approach the aperture definition \rlum . On the other hand, when the threshold is set too high (e.g., red lines in Figure~\ref{fig:RLcomparisoniso}), many \hii\ regions are unresolved because their peak intensities are fainter than the isophotal threshold (compare Figure~\ref{fig:LvsI}).

While \riso\ yields a clear relationship to luminosity, a major caveat is that \riso\ does not have any clear fixed physical meaning. This radius does not enclose a fixed fraction of the flux, nor does it represent a fixed fraction of the peak intensity. Nor does there appear to be good evidence that any particular value of \riso\ corresponds to a fixed physical boundary. This means that the relationship is also ``tunable'' via the choice of \riso.
 
\subsection{Comparisons to Previous Work}
\label{sec:comparisons}

In Figure~\ref{fig:RLcomparisoniso}, we plot our variation of isophotal radii in NGC\,253 alongside previous measurements for $z\approx0$ \hii\ regions. Most of the studies referenced here target more distant galaxies, and so have coarser physical resolution than our data. Few studies of the local Universe apply background subtraction in comparable ways to our approach \citep[see discussion of the topic in][]{kennicutt1979,kennicutt1989}. As we discussed in \S \ref{sec:effectiveradius} and Appendices~\ref{ap:bgsub} and \ref{ap:res}, both of these factors impact the measured sizes and luminosities. The literature studies also vary in their treatment of extinction.

In Figure~\ref{fig:RLcomparisoniso}, the density contour of star-forming regions from the PHANGS-MUSE \citep{groves2023} and PHANGS-HST \citep[][]{chandar2025,barnes2025} are shown with brown and gray contours, respectively. Black and gray lines show luminosity-size relations measured for local Universe extragalactic \hii\ regions by \citet[][W12, resolution $40-325$~pc]{wisnioski2012}, \citet[][C18, resolution $40-325$~pc]{cosens2018}, and \citet[][G23, resolution $\sim140$~pc]{giunchi2023}. Although most of the literature uses the measured size to calculate luminosity (where we use \rlum), this difference does not significantly change the relations presented in this section. The total luminosity of our regions decreases if $R<$\rlum but does not affect the overall trends. We discuss this more in Appendix~\ref{ap:bgsub}.

The literature luminosity-size relations show similar slopes to one another, with power law indices ranging from $\alpha = 1.48$ to 2.72, but a wide range of normalizations. Fitting the luminosity-isophotal radius relation to our catalog gives a similar result with index $\alpha=2.10$ when fit to all regions ($\alpha=3.45$ for binned values). The normalization of our fit using the fiducial intensity threshold lies intermediate between the low $\Sigma_{\rm SFR}$ and low $z$ samples from C18, the W12 sample, and the lower branch of the PHANGS-MUSE catalog.

Our fiducial results lie below the high $\Sigma_{\rm SFR}$ C18 data or G23 and the PHANGS-HST results, but tuning the threshold used to calculate the isophotal radius changes the normalization of the fit relation (Section~\ref{sec:isophotal}). These literature catalogs use higher thresholds to define their \hii\ regions, in good agreement with our results (e.g., see discussion of the worse surface brightness sensitivity of HST compared to MUSE in \citet{barnes2025}). If we adopted a higher threshold, our \riso\ results could match theirs (red line Figure~\ref{fig:RLcomparisoniso}).

As the physical resolution of the PHANGS-MUSE data degrades, the measured radii of \hii\ regions increase \citep{barnes2021,groves2023}. The data show a pile-up at radii approximately equal to the resolution for the low-luminosity regions, which can be seen in the contours (compared to the PHANGS-HST contours at fixed resolution). \citet{groves2023} note that the majority of the regions in the catalog are unresolved, resulting in unreliable sizes. 

The PHANGS radii are measured using a circularized area, which is analogous to our use of an isophotal radius, and not precisely analogous to our HWHM or \rf\ cases. At higher luminosity, where the PHANGS-MUSE regions appear extended, their regions show higher luminosity for the same size compared to ours.

Our HWHM, \rmom, and \rf\ measurements differ from the literature. Even as the luminosity changes by two orders of magnitude, these measurements remain nearly fixed. This notably differs from W12 and G23, who do use core radius measurements to define their results. Both of those studies also show good correlation between their core radius (roughly the HWHM) and \riso. Given that the relationship they find is almost 1-to-1 (Figure~3 G23, Figure~1 W12), our best explanation is that their segmentation algorithm leads their \riso\ to be defined as an intensity equal to some fraction of the peak and that their sample has a limited range in peak intensity. This would project the core and isophotal radius into similar terms, ensuring a correlation between the two measurements.

Following this point, we note that for many of these studies, the segmentation algorithm used to define the \hii\ regions plays a major role in defining the results \citep[as studied extensively in giant molecular cloud literature, e.g.,][]{pineda2009,hughes2013,leroy2016}. The studies in Figure~\ref{fig:RLcomparisoniso} variously use dendrograms (G23), HIIphot (PHANGS-MUSE), isophotal analysis \citep[W12,][]{barnes2025}, or a mixture of techniques (C18). The input parameters (e.g. minimum intensity thresholds or significance) in these algorithms reinforces the resulting correlations. 

NGC\,5068 shows HWHM and isophotal trends consistent with those presented for NGC\,253. As discussed above, \rf and \rmom show a different picture in NGC\,5068 than NGC\,253, with a good correlation visible between these fractional flux sizes and luminosity in NGC\,5068. A more complete discussion of the comparison tests performed can be found in Appendix~\ref{ap:ngc5068}. 

\section{Profile Shape and Physical Structure}
\label{sec:profileshape}

\subsection{Analytic Fits}
\label{sec:analytic}
\input{fig-profilefits}

We fit the stacked profiles of \ha\ intensity with analytic functions. Figure~\ref{fig:profile-fits} shows the median profile for all regions along with best fits and residuals for several functional forms. We report the best fit parameters in Table~\ref{tab:fitcoeffs} along with Bayesian Information Criteria (BICs), which we use to select which fit is statistically preferred. We assume the uncertainties in the data are random and uniform, and thus the log-likelihood is:
\begin{equation}
\label{eq:likelihood}
    \ln(\hat{L})=-\frac{1}{2\sigma_{\text{rms}}^2}\sum [y_{\rm data}-y_{\rm model}]^2,
\end{equation}
where $\sigma_{rms}$ is the error in the data, $y_\mathrm{data}$ is the \ha intensity data points, and $y_\mathrm{model}$ are the model values and the sum is carried out over the data. 

The double Gaussian function fits the profile the best, with a near-zero residual at all radii. A single Gaussian cannot fit the extended wings that dominate the profile beyond $\sim 20$~pc, while an exponential that fits the outer points fails to reproduce the core of the profile. Visually, a Moffat profile provides a similar quality fit to the double Gaussian, but we focus on the double Gaussian because it is easier to work with analytically and preferred by the BIC comparison. The BIC for the double Gaussian is lower than for all the other forms, and the difference is $>10$, indicating a statistical preference for this form. This agrees well with \citet{kennicutt1979}, who also found H$\alpha$ intensity profiles for \hii\ regions to be well represented by one or two Gaussian functions.

\input{tab-fitcoeffs}

For the best fitting double Gaussian, the profile consists of a core with HWHM $\approx 11$~pc. Deconvolving the PSF implies a size of HWHM $\approx 7$~pc for this central compact component. The larger, extended component has a deconvolved $\text{HWHM} = 31$~pc. The flux of an individual component $\propto A \sigma^2$, and most ($77\%$) of the total flux in the best fit double Gaussian lies in this extended component.

The best fitting double Gaussian for each of the peak intensity bins are provided in Table~\ref{tab:binprops}. We also fit each individual \hii\ region with a double Gaussian, simplifying to a single component when appropriate. Based on the same BIC criteria as above, we find that for approximately 50\% of all \hii\ regions the double Gaussian performs better than the single Gaussian. This multi-component structure of \hii\ regions is therefore not simply a manifestation of the stacking of an ensemble of regions with different sizes. The fraction of double Gaussian regions varies by \ha peak intensity, with the faintest regions preferring a single Gaussian more frequently than for the brightest regions (36\% in bin 2 vs. 63\% in bin 9). We include the best-fit parameters of the preferred function for each \hii\ region in the region catalog (Appendix~\ref{ap:catalog}).

Figure~\ref{fig:profile-fits} shows the native resolution data. In Appendix~\ref{ap:res} we show that the same functional form describes the median profile as we convolve the MUSE data to coarser resolution, but the distinction between a single and double Gaussian at low resolution is minimal. The deconvolved core sizes implied by these fits remain stable across resolution.

In our comparison analysis of NGC\,5068, we also find that a two-component Gaussian function well-approximates the average \hii\ region profile, with an inner HWHM of 6.2~pc (deconvolved HWHM$=3.6$~pc) and outer HWHM of 29.5~pc. Compared to the 5~pc HWHM of the PSF, the high-resolution HST \ha profile appears even more unresolved than in NGC\,253. We discuss this result further in Appendix~\ref{ap:ngc5068}.

The two-component structure of \hii\ regions could be interpreted as two fundamentally different physical components of the \hii\ regions, two regimes where the mechanisms producing the emission differ, but both are associated with the \hii\ region. This structure could also be interpreted as a dense core of gas producing emission on top of a broad background emission from the DIG. One of the leading theories for the origin of the DIG is leakage from nearby \hii\ regions \citep[e.g.,][]{ferguson1996,haffner2009,seon2009}. This could account for the shallow decline at the outer edges of the \hii\ regions, where the emission is escaping the compact region. The \ha emission is more diffuse outside the \hii\ region, steadily dropping but never reaching zero before encountering another nearby \hii\ region. 

\subsection{Implied Three-Dimensional Structure}
\label{sec:density}

We use our measured profiles (Table~\ref{tab:binprops}) to constrain the three-dimensional structure of \hii\ regions. First, we invert the measured intensity profiles to infer the three-dimensional density structure, then we test how well models that adopt this structure match our observations. Both approaches leverage the fact that H$\alpha$ emission traces the emission measure, $EM \propto \int n^2 dl$, and assume a radially varying density with no clumping. We compare to [\sii] line ratio measurements, which are sensitive to the physical density, in \S \ref{sec:sii}.

We analytically transform from the functional form of the luminosity surface density profile to a three-dimensional distribution of luminosity per volume. Then we convert from luminosity per volume (i.e., emissivity) to a density distribution assuming Case B recombination. To do this, we assume that in two dimensions the \ha luminosity surface density, $F_{\rm H\alpha}$, is distributed around the peak in an azimuthally symmetric distribution that follows the double Gaussian derived in Section~\ref{sec:profileshape}:

\begin{equation}
    F_{\text{\ha}}=A_1 e^{x^2/2\sigma_1^2}+A_2 e^{x^2/2\sigma_2^2}.
    \label{eq:haprofile}
\end{equation}

We invoke the inverse Abel transform of $F_{\text{\ha}}$, which converts the projected two-dimensional distribution into a three-dimensional distribution via the following integral:
\begin{equation}
    f_{\text{\ha,3D}}(r)\propto-\frac{1}{\pi}\int_r^\infty\frac{dF_{\text{\ha}}}{dx}\frac{dx}{\sqrt{x^2-r^2}},
\end{equation}
where $F_{\text{\ha}}$ is the functional form of the \ha profile (Equation \ref{eq:haprofile}), $x$ is a radial coordinate in two-dimensional space, and $r$ is the three-dimensional distance from the center of the \hii\ region. 

The \ha luminosity per volume, $f_{\text{\ha,3D}}$ at any distance $r$ from the center of the \hii\ region can then be written as 
\begin{equation}
    \begin{aligned}
    f_{\text{\ha,3D}}(r) =
    I_{\text{peak}}\Biggl( 
    \frac{A_{\text{f}}}{\sigma_1\sqrt{2\pi}}\text{exp}\left(-r^2/2\sigma_1^2\right)\\
    + \frac{1-A_{\text{f}}}{\sigma_2\sqrt{2\pi}}\text{exp}\left(-r^2/2\sigma_2^2\right)\Biggr).
    \end{aligned}
    \label{eq:3df}
\end{equation}
Here $A_{\text{f}}$ is the fraction of the total amplitude ($I_{\text{peak}}$) that the inner Gaussian accounts for, and $\sigma_1$ and $\sigma_2$ are the standard deviations of both Gaussians. 

We convert from $f_{\text{\ha,3D}}$ to density via
\begin{equation}
    n_{\text{e}}(r)=1.12\left(\frac{4\pi f_{\text{\ha,3D}}(r)}{h\nu_{\text{\ha}}\alpha_{\text{\ha}}^{\text{eff}}}\right)^{1/2},
    \label{eq:n}
\end{equation}
where $\alpha_{\text{\ha}}^{\text{eff}} = 1.17\times10^{-13}$~cm$^3$~s$^{-1}$ for Case~B recombination at T$=10^4$~K \citep[][]{drainebook}, with the factor of 1.12 to account for the contribution of electrons from ionized helium ($n_{\text{He}}/n_{\text{H}}=0.12$).

We perform this transformation for the best-fit deconvolved Gaussians describing the luminosity surface density profile in each \ha peak intensity bin and report the results in Table~\ref{tab:binprops}. The luminosity surface density profiles in different deciles appear similar to one another (Figure~\ref{fig:decile-profiles}) and therefore the density profiles also appear similar. The volume density profile is more extended than the luminosity surface density in both the core and extended components. This is expected because \ha scales as $n_{\rm e}^2$ and so preferentially arises from the denser inner regions. Although the shape of the density profiles shows no systematic variation with \ha peak intensity, the central values for the density vary significantly, from $n_e < 10$~cm$^{-3}$ in the faintest bin to $>100$~cm$^{-3}$ in the brightest bin. The median central density is $n_e \approx 25$~cm$^{-3}$.

We validate our calculations by forward modeling emission based on the fit volume density profiles and verifying that they reproduce the observed \ha profile. We construct a three-dimensional double Gaussian model for the volume density. The inner component has a deconvolved size of 8.4~pc and we vary the size of the outer component from 28-37~pc. We assume the same ratio of amplitudes as found for the one-dimensional \ha\ profile. 

Assuming this model for the volume density profile, we square the ionized gas density to correspond to the local contribution to the emission measure (i.e., the volumetric \ha\ emissivity) and then integrate along the line of sight to create a two-dimensional image that we expect to be proportional to the normalized \ha\ intensity image. We convolve this image with a Moffat function with a width of 17~pc and index 2.8 to simulate observing with the MUSE PSF. We then construct radial profiles and calculate the same size metrics that we derive for the observed profiles. To match the methodology used for the observations, we subtract a background equal to the value of the radial profile at 100~pc and assume 100\% of the luminosity is contained within this 100~pc radius.

\input{fig-models}

Figure~\ref{fig:model} compares the profiles and sizes predicted from this model to our observations. The far left panel shows that the \ha radial profile derived from the best-fitting density profile (red) matches the median observed \ha profile (black) well. In the two-dimensional image of the modeled and observed \ha emission (center left), we see that the contour values also agree with each other. 

In the right panels of Figure~\ref{fig:model}, we compare size measurements from the modeled profile to the observed distributions of \rf, \rn, and HWHM. This comparison tests whether the shape of the model profiles matches the real profiles (e.g., by comparing \rf/\rn, which is sometimes referred to as a concentration index). Our fiducial model profiles with outer width $\sigma_2=33$~pc reproduce radius measurements almost exactly equal to the median values of the observed distribution, reinforcing the similarity between the modeled profile and the observed one seen in Figure~\ref{fig:model}. This modeled value agrees well with the average $\sigma_2$ calculated from the \ha profile above. The fainter red points show the result if we vary the outer Gaussian width between $28-37$~pc. Doing so yields models that mirror some of the spread we see among real \hii\ regions.

\subsection{Density Based on the [SII] Ratio}
\label{sec:sii}

MUSE obtains full optical spectroscopy, which includes probes of the gas volume density that complement our structural analysis. The line ratio of the [\sii] doublet at $\lambda = 6716$ \AA\ and $\lambda = 6730$ \AA\ can be used to estimate the electron density because of the different critical densities of the two lines. This diagnostic is valid for densities between $\sim10$ and $10^4$~cm$^{-3}$.

\input{fig-siiratio}

Figure~\ref{fig:sii-ne} shows radial profiles of the [\sii] ratio and the calculated electron density for each peak intensity bin. To construct these, we calculate a line ratio profile for each region from the moment~0 maps. We do not apply any background subtraction to these, so that the profile should reflect all the emission along the line of sight and extend from the \hii\ region into the DIG. Then we calculate the median line ratio in each radial bin for each \ha peak intensity deciles. We use the PyNeb package \citep[][version 1.1.24]{luridiana2015} to convert the ratio to a density, assuming a constant temperature of 10000~K (choosing 6000 or 8000~K instead does not significantly affect the profile or implied densities).

The density profiles are shown in the bottom panel of Figure~\ref{fig:sii-ne}. The shaded region marks the low-density regime ([\sii] ratio$>1.44$), beyond which the line ratio asymptotes to a single value and we only recover an upper limit of $\sim10$~cm$^{-3}$ on the density. Because of this limit, the faintest few deciles have only upper limits on the density at all radii. We see that the brightest regions in \ha are also the densest, with enhancements in the [\sii] line ratios visible in the inner regions of deciles 8-10. Away from the peaks, all regions fall to the same low-density limit that also describes the DIG.

\input{tab-siidensity}

We fit the density profiles of the deciles that show clear density enhancements (8-10) with Gaussian functions and report the results in Table~\ref{tab:density-fits}. We quote the central density and the standard deviation of a single Gaussian that describes the spatial profile ($\sigma$) for each decile, along with the measured [\sii] ratio at the peak. The mean $\sigma$ for three deciles is $\approx 15$~pc (HWHM $\approx 18$~pc, FWHM $\approx 36$~pc), notably larger than the $17$ pc FWHM PSF and comparable to the HWHM values of the modeled 3D density profiles for these deciles. In this small sample, the width of the Gaussian fit increases with \ha peak intensity, indicating that the brightest regions have higher density in the center and take longer to drop to the low density regime compared to the fainter regions. 

This density measurement is likely an underestimate of the true density due to DIG contamination. The low density of the DIG blending with the higher density of the \hii\ decreases the averaged density along the line of sight as traced by the [\sii] line ratio. 

\subsection{Density, Luminosity, and Surface Brightness}
\label{sec:denssb}

\input{fig-densityluminosity}

Both structural analysis of \ha\ emission and [\sii] line ratios show that regions with higher peak intensity and higher luminosity show higher ionized gas volume density. In other words, our results imply a density-luminosity relation for \hii\ regions in NGC\,253. The left panel of Figure~\ref{fig:nL} shows the central densities derived from the \ha profile of all \hii\ regions (gray points) and the median [\sii] ratio profile for the brightest three deciles (colored squares). For reference, we compare these calculations to the density-luminosity relation expected for a \stromgren sphere with fixed radius. This line assumes Case~B recombination, $T=10^4$~K, and radius 8.5~pc, i.e., the HWHM of the PSF. 

Figure~\ref{fig:nL} shows good overlap between the structurally inferred densities, the [\sii] estimates, and the \stromgren\ toy model in NGC\,253. This is consistent with both the structural and [\sii] estimates being sensitive to the density of a central core of denser ionized gas different density estimates suggests that the inner component may be relatively filled by ionized gas, because clumping within the region would lead the [\sii] (which accesses the luminosity-weighted local physical density) to yield a higher density compared the structural estimate (which assumes a smooth gas distribution). 

In green contours, Figure~\ref{fig:nL} also shows the [\sii]-derived density vs. \ha luminosity relation for the PHANGS HST+MUSE catalog of \citet{barnes2025}. These show an offset normalization with much higher luminosity at a given density and a poor correlation between [\sii]-based density and luminosity. We also inspected individual galaxies in the \citet{barnes2025} catalog and found significant scatter in the $n$-$L$ relation between galaxies, with none appearing as strongly correlated as NGC\,253. Along with Figure~\ref{fig:nL}, this suggests that there is not a universal peak density-luminosity relation in local galaxies \citep[indeed sometimes an anti-correlation is claimed, e.g.,][]{hunt2009}.

However, in both our NGC\,253 data and the PHANGS HST+MUSE data we do observe a strong correlation between peak ionized volume density and the region-averaged \ha surface brightness inside \rmom ,
\begin{equation}
\label{eq:sb}
\overline{\text{SB}}_{\text{\ha}}=L_{\text{\ha}}^{\text{corr}}/{2\pi R_{\text{mom}}^2}~.
\end{equation}
The observed relation, shown in the right panel of Figure~\ref{fig:nL}, appears consistent across the full PHANGS-HST+MUSE sample and both the [\sii]\ and structural density estimates in NGC\,253,
\begin{equation}
\label{eq:sbdensreln}
\frac{n_{e,\text{ SII}}}{\rm cm^{-3}} = 10^{-23.2} \times \left(\frac{\overline{\text{SB}}_{\text{\ha}}}{\rm erg~s^{-1}~kpc^{-2}} \right)^{0.61} 
\end{equation}
Individual relationships are also visible for individual galaxies within the \citet{barnes2025} catalog. We use the region average SB within \rmom here, but we see similar agreement for a variety of radius measurements. 

Here we show the peak of the [\sii]-derived density for NGC\,253, whereas the PHANGS-HST densities come from the [\sii] ratio of the full region in MUSE. We compare $\overline{\text{SB}}_{\text{\ha}}$ from HST and a density from MUSE, which has lower resolution and therefore includes more area in the [\sii] detections. 

A density-surface brightness relation makes physical sense, since the surface brightness captures the concentration of H$\alpha$ emission $\propto \int n_e^2 dl$ per area. Given that H$\alpha$ surface brightness is significantly more accessible than direct density diagnostics, the surface brightness-density relation may offer a useful way to predict ionized gas volume densities in addition to yielding a basic constraint on the structure of \hii\ regions. If this is indeed the more physical relation, then we observe a density-luminosity relation in NGC~253 because of the tight correlation between intensity and luminosity in this galaxy (Figure~\ref{fig:LvsI}), which does not appear to hold in such a universal way for other galaxies.

\subsection{Implications}
\label{sec:synthesis}

A double Gaussian structure is preferred in all of our stacked profiles, most individual profiles, and also in the NGC\,5068 comparison data. In most regions, we find evidence for compact cores with radius $< 10$~pc that are still unresolved or marginally resolved at our $17$~pc (FWHM) resolution. It will be interesting to see whether these peaks are only knots of ionized gas or also correspond to the locations of the powering stellar sources, and perhaps even shell-like structure \citep[as in][]{pedrini2024}. From preliminary inspection of multiwavelength emission, in many cases, they do seem to also be peaks in the infrared and optical.

The presence of significant flux in the extended component means that most ($\sim 80\%$ on average) ionizing photons escape from the core to ionize the surrounding halo structure. This is before accounting for overall escape from \hii\ regions to the DIG.

From our three-dimensional estimates of luminosity density, we inferred volume density assuming a smooth distribution. In Figure~\ref{fig:mass}, we show the normalized radial profiles of enclosed mass for each decile of \ha peak intensity in color and the enclosed number of photoionizations (or recombinations) per second in gray (the number of ionizations traces the luminosity per volume and scales as n$_{\rm e}^2$). The vertical lines mark the radius within which 50\% of the total mass is enclosed (M$_{50}$, dark blue) and where 50\% of the photoionizations occur (gray). The shapes of the profiles are extremely similar across peak intensity, resulting from the extremely similar \ha profiles. 

We see that 50\% of the total ionizations occur within $\sim 6$~pc of the \ha peak. This is smaller than the HWHM of the MUSE PSF for these observations, thus we are not able to resolve half of all ionization events that take place inside these \hii\ regions. The location of emission is strongly concentrated at the center of the regions. One might assume that half of the photoionization heating and perhaps a similar fraction of radiation pressure might be exerted within this volume, which might have important implications for the dominant mode of stellar feedback \citep[e.g., see][]{pathak2025}.

\input{fig-massion}

The mass profile is significantly more extended, though we caution that the assumption of a smooth density distribution makes the plotted mass profiles an outer limit. The median value of M$_{50}$ is $\approx 50$~pc. Thus M$_{50} \approx 1.1$\rn $\approx 2.2$\rf $\approx 3.3$HWHM, after taking the median of the radius values across the bins of \ha peak intensity. For a smooth density distribution, 50\% of the total mass of an \hii\ region is found within the same region as approximately 90\% of the \ha flux. Note that if the ionized gas in this extended distribution is clumpy, i.e., concentrated into locally denser regions, the mass of the extended component will drop and become more centrally concentrated. It will be interesting to test this further, e.g., by assessing the patchiness of the emission.

Finally, the common density-surface brightness relation observed between PHANGS MUSE+HST and our NGC\,253 analysis suggests some common structural behavior for \hii\ regions. This may offer a useful way to coarsely estimate the peak density in a region, and will be interesting to compare to high-redshift observations where density diagnostics are increasingly available from JWST spectroscopy \cite[e.g.,][]{topping2025}.

\section{Conclusions} 
\label{sec:conclusions}

Using the largest MUSE spectral map of a nearby, massive star-forming galaxy, we study \nregions \hii\ regions identified as \ha peaks and confirmed through line ratio diagnostics in the nearby spiral NGC\,253. We release the catalog of \hii\ regions and their properties with this paper (Appendix~\ref{ap:catalog}). Our false-source tests suggest that our catalog is $>90\%$ complete down to I$_{\text{\ha}}=10^{40.3}$~erg~s$^{-1}$~kpc$^{-2}$, corresponding to $L_{\text{\ha}}=10^{37}$~erg~s$^{-1}$. The luminosity function is well represented by a broken power law with indices $\alpha_1=-0.92$ and $\alpha_2=-1.86$. Before subtracting any local background, 70\% of the total \ha luminosity of NGC~253 is contained within our \hii\ regions, in good agreement with the median compact \ha fraction of 63\% found in \citet{belfiore2022} and other literature studies. Thus, these \hii\ regions represent a typical sample of \hii\ regions comparable to those recovered in the PHANGS nebular catalogs \citep[][]{barnes2025,groves2023} but observed at high physical resolution of $\approx 17$~pc (FWHM). 

We use extinction-corrected, background-subtracted radial profiles of individual and stacked \hii\ regions to study the structure of resolved \hii\ regions. 

\begin{enumerate}
\item Profiles of \ha emission are well-fit by a double Gaussian function, with a compact core surrounded by an extended component. The deconvolved HWHM size of the cores is $\sim$7~pc, and the surrounding broad Gaussian has a deconvolved HWHM of $\sim$30~pc. The extended component contains $\approx 80\%$ of the total flux in the profile (Sections~\ref{sec:analytic}, \ref{sec:synthesis}).

\item This double Gaussian profile is preferred over a single Gaussian for all stacked profiles and $\sim50\%$ of individual \hii\ regions in NGC 253, with some dependence on \ha peak intensity. We recover a similar best-fit functional form for \hii\ regions in NGC\,5068, with the inner Gaussian appearing even more concentrated than in NGC\,253 (Sections~\ref{sec:analytic}, \ref{sec:synthesis}, Appendix~\ref{ap:ngc5068}).
\end{enumerate}

An important next step will be to test how these results change if we instead construct profiles around concentrations of massive stars rather than H$\alpha$ peaks \citep[e.g.,][]{hannon2022}. With upcoming HST \ha\ coverage of NGC\,253 (GO 18086, PI McClain), we also expect to test how often the marginally resolved \ha\ cores seen here resolve into the shell-like structures seen on small scales in Galactic and HST observations \citep[e.g.,][]{anderson2014,hannon2022}.

We use our profiles to measure the relationship between the radius and luminosity of \hii\ regions. This measurement is sensitive to methodological choices, which we discuss in detail in the appendices (\ref{ap:res},\ref{ap:bgsub}). For our fiducial background-subtracted, native resolution measurements, we find:

\begin{enumerate}
\setcounter{enumi}{2}

\item The HWHM or core radius appears approximately fixed as a function of \ha peak intensity and luminosity, indicating that \hii\ regions are strongly cored and this core is unresolved. This result also appears to hold in 10~pc resolution HST H$\alpha$ observations of NGC\,5068, suggesting that it represents a general conclusion (Section~\ref{sec:hwhm}).

\item We recover the expected luminosity-size relationship for isophotal radius measurements. The normalization of the relationship is tunable via the choice of isophotal threshold, and we show that our data can match the HST H$\alpha$ results of \citet{barnes2025} or the MUSE results of \citet{groves2023} as well as several literature studies (Section~\ref{sec:isophotal}, \ref{sec:comparisons}). However, we caution that the isophotal radius has an ambiguous physical meaning, capturing a variable fraction of the flux and peak intensity across our sample.
    
\item Three measures of effective radius (\rmom, \rf, \rn) show a weak correlation with luminosity, and are strongly sensitive to the treatment of the background (Section~\ref{sec:effectiveradius}, \ref{sec:comparisons}). However, we recover stronger relations in complementary analysis of NGC\,5068 (Appendix~\ref{ap:ngc5068}). Some of this difference can be attributed to target-specific traits, the sensitivity of the data, or differences in \hii\ region identification and segmentation algorithms. 
\end{enumerate}

Analyzing the radial profiles in \ha and [\sii], we estimate physical parameters of the \hii\ regions in our sample. 
\begin{enumerate}
\setcounter{enumi}{5}
\item We convert the best-fit \ha profiles to three dimensional electron density profiles, which imply central densities of $\approx 10{-}110$~cm$^{-3}$ (Section~\ref{sec:density}, Table~\ref{tab:binprops}). Assuming a smooth distribution, the three dimensional density profile is a double Gaussian with core width $\approx 10$~pc and extended Gaussian with HWHM $\approx 40$~pc.

\item The density-sensitive [\sii] ratio yields a value above the low density limit for the binned profiles of the brightest three deciles of \hii\ regions. The brighter \hii\ regions have broader density distributions than the fainter regions (Section~\ref{sec:sii}, Table~\ref{tab:density-fits}).

\item We find a good correlation between the [\sii]-derived densities and the region-averaged surface brightness (measured within \rmom) with index $\approx 0.61$ (Section~\ref{sec:denssb}, Figure~\ref{fig:nL}). Approximately the same correlation holds for densities inferred from H$\alpha$ profiles in NGC\,253, [\sii]\ in NGC\,253, and using [\sii]\ in the HST+MUSE sample of \citet{barnes2025}. In NGC\,253 but not the broader HST+MUSE sample, this relationship manifests as a strong density-luminosity relation.
\end{enumerate} 

For this analysis, we choose to center \hii\ regions at the location of brightest \ha emission. However, an obvious improvement to this assumption is to center the radial profiles on the underlying powering stellar clusters and associations \citep[e.g.,][]{whitmore2011}. Incoming HST and future JWST imaging will provide sufficient wavelength coverage and resolution to locate and characterize the stellar populations powering the identified \hii\ regions. 

With sufficient data, we can determine the ages and masses associated with the identified \hii\ regions, revealing the evolution of \hii\ region structure across the full galaxy. Expanding this analysis to other nearby galaxies with high physical resolution imaging (e.g., full PHANGS-HST sample) will provide key insight into the generality of the results presented in this paper and the role of small-scale stellar feedback in shaping galaxies.

\section{Acknowledgments} 
\label{sec:acknowledgements}

\input{acknowledgements}

\bibliography{ref}{}
\bibliographystyle{aasjournal}

\appendix

\section{Region catalog}
\label{ap:catalog}

We provide our region catalog and a corresponding spatial mask as a data release accompanying this paper. Table~\ref{tab:catalogcolumns} describes the columns included in the catalog. The entries in the catalog are explained throughout the main text. We include a flag (Flag$_{\text{bg}}$) to indicate isolated regions with a clean background, after all other regions in the catalog are masked. This is determined based on the median absolute deviation of the unbinned profile between \rlum and 200~pc. These regions tend to be at the edges of spiral arms or in the interarm region. 

We also provide a spatial mask on the astrometric grid of the MUSE survey. For each region, this mask identifies the pixels that lie within \rlum\ and so contribute to the luminosity and size measurements. When a pixel lies within \rlum\ for multiple regions, we assign it to the nearest peak. The rest of this appendix expands on details related to catalog construction and validation.

\input{tab-catalog-columns}

\subsection{BPT Cuts}
\label{ap:bpt}

\input{fig-bpt}

As discussed in Section~\ref{sec:regions}, we perform spectroscopic line ratio cuts to limit the sample of \ha peaks to those most likely to be \hii\ regions. At the location of brightest pixel in each region, we measure the strengths of the following emission lines from the moment-0 maps (before any correction for extinction): \hb, [\oiii]$\lambda5007$ (\oiii), [\nii]$\lambda6583$ (\nii), and [\ion{S}{2}]$\lambda6716$ + [\ion{S}{2}]$\lambda6730$ (\sii). Using the relative strengths of these lines and \ha with the diagnostics presented in \citet{kewley2001} and \citet{kauffmann2003}, we separate \hii\ regions powered by young stars from regions likely to be powered by other ionizing sources. 

The N2-BPT and S2-BPT diagrams for all \ha peaks are shown in Figure~\ref{fig:bpt}. We consider the regions that fall below the \citet{kauffmann2003} line in the N2-BPT diagram and below the \citet{kewley2001} line in the S2-BPT diagram likely to be \hii\ regions. Those that fall between the two lines in the N2-BPT diagram are considered composite objects \citep{kewley2006}. We remove all peaks above either line from our sample, retaining only likely \hii\ regions. These cuts reject a total of 554 peaks, or 18.2\% of our catalog, removing supernova remnants, planetary nebulae, composite objects, and objects highly contaminated by the DIG. 

\citet{groves2023} invokes an S/N requirement greater than 5 for the emission lines used for BPT cuts. We note here that only 370 regions in the final sample have [\oiii] S/N $<5$, all of which belong in the faintest two bins of \ha peak intensity. All other lines are significantly detected at the location of the \ha peaks. We choose to leave these regions in our sample, but note that there may be additional contamination affecting the results of the faintest two bins.

There is a noticeable separation of [\sii]/\ha (S2) across \ha intensity bins in the S2-BPT diagram, which does not appear for [\nii]/\ha (N2). This is due to the fact that S2 is more sensitive to \ha intensity than N2, changing more steeply with increasing \ha surface brightness, as seen in the PHANGS-MUSE galaxies in \citet{belfiore2022}. S2 is also more strongly affected by DIG contamination than the metallicity-sensitive N2 ratio, and the faintest regions show less contrast with the DIG than the brightest regions. 

\subsection{Completeness}
\label{ap:completeness}

We test the completeness of our \hii\ region catalog by injecting false sources into the \ha extinction-corrected map. We assume the sources are symmetric two-dimensional Gaussians with peak intensity and size randomly chosen to lie in the linear range  $10^{39.1}<I_{\text{peak}}<10^{40.9}$~erg~s$^{-1}$~kpc$^{-2}$ and $10<$~HWHM~$<32$~pc, which correspond to the 16-84th percentile of the observed \hii\ region properties. 

\input{fig-completenessscatter}

We add 100 sources to the map at a time, and repeat this exercise 100 times. Each time, we re-identify peaks and record whether the false sources have been recovered. Figure~\ref{fig:completeness} shows the results of this test. Overall, we recover 98\% the 10,000 false sources. Fainter sources are more challenging to recover, with only $\approx 90\%$ completeness for sources with peak intensity $< 10^{40.3}$ erg s$^{-1}$ kpc$^{-2}$, following the fit in Figure~\ref{fig:LvsI} this corresponds to better than 90\% completeness for regions with luminosity $> 10^{36.8}$ erg s$^{-1}$. We are also 90\% complete for regions with a contrast $>75$. 

There is also a weak trend that smaller regions are recovered better. The coloring of the points and location of undetected regions in Figure~\ref{fig:completeness} show that the detectability of the false sources is dependent most strongly on contrast with the local background. This is, in general, worse for low-intensity regions, but also depends on galactic position because faint sources are difficult to detect close to bright sources. We expect that there is thus some bias in our catalog such that we miss faint regions in the spiral arms and inner galaxy.

Once the false sources have been detected, we build radial profiles around them and find a strong 1:1 correlation between the measured HWHM and \rf, consistent with the predictions for a Gaussian source. We also find a strong 1:1 correlation between the HWHM from the profile and the input source size. We are able to recover the input parameters for a sample of false sources injected into the map in isolated regions.

\subsection{PSF Characterization}
\label{ap:psf}

\input{fig-pneresolutions}

\begin{figure}
    \begin{center}
    \includegraphics[width=0.5\linewidth]{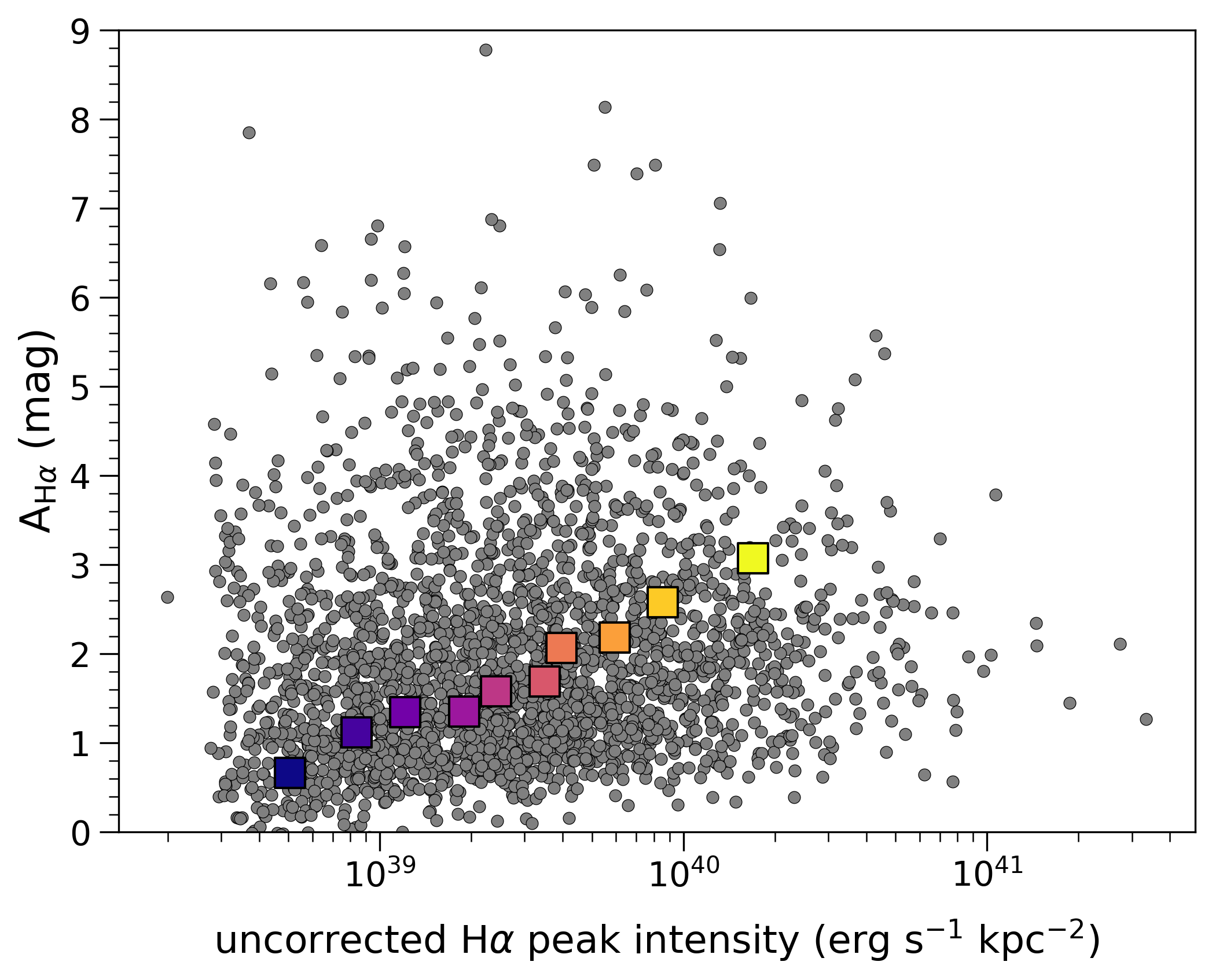}
    \caption{A$_{\text{\ha}}$ at the location of the \ha peak, as a function of background-subtracted \ha peak intensity before extinction-correction.}
    \label{fig:AVpeakint}
    \end{center}
\end{figure}

Many of our regions show compact cores, and their extent relative to the PSF is of interest. We expect the PSF to vary across the map, because the data were collected over the course of more than one year, and observing conditions varied across the many observing sessions. We use 571 planetary nebulae visually identified in \citet{congiu2025} to estimate the PSF of each pointing. Because the PNe are physically very small \citep[$<$1~pc,][]{osterbrock1964}, they are expected to appear as unresolved point sources at the distance of NGC\,253.

The functional form of the MUSE PSF is a Moffat function, given as 
\begin{equation}
\label{eq:moffat}
    f(x) = A\left(1+\left(\frac{x-x_0}{\alpha}\right)^2 \right)^{-\beta},
\end{equation}
where $A$ is the amplitude, $x_0$ is the center of the distribution (0, in this case), $\alpha$ is the width of the distribution, and $\beta$ is the index controlling the sharpness of the distribution. In \citet{congiu2025}, each PN was fit with a Moffat profile with index $\beta=2.8$, except in the two central pointings, which used an index $\beta=2.3$. The central pointings were observed with the adaptive optics mode (AO), and the rest were observed with the natural seeing mode (NOAO). 

Assuming that all PNe are point sources in the MUSE map, we take the \citet{congiu2025} fits to reflect the local PSF. For those pointings with more than one PN, we use the mean width of all of the relevant PNe to estimate the PSF. For the three pointings with no detected PNe (pointings 13, 41, and 50 at the outer edge of the galaxy), we take the average width of the full catalog as the estimated PSF.  

The mean FWHM of the entire PN catalog is 0.80$^{\prime\prime}$ and the mean of the pointing PSFs is 0.78$^{\prime\prime}$. To calculate the effective PSF, we add the \citet{congiu2025} measurements in quadrature with the $0.6''$ smoothing kernel used in our analysis (\S \ref{sec:musedata}). We report the median and range of PSF widths in Table~\ref{tab:ngc253_props}, and the local PSF for each region is provided in our catalog. Based on this, our typical working PSF is a Moffat function with FWHM $\approx 1''$ and index $\beta=2.8$ (Figure~\ref{fig:pn-profile}).

We test our radial profile calculations and check the accuracy of the PSF estimates by applying our profile methods to the identified PNe. In Figure~\ref{fig:pn-profile}, we show the median stacked profile combining all PNe. The profile appears well described by a Moffat profile with width $1''$. For comparison, we show the median profile of all \hii\ regions in blue. The \hii\ regions appear extended compared to the PNe, supporting their treatment as extended sources in our analysis. This conclusion also holds if we degrade the resolution of our images (Appendix~\ref{ap:res}), indicating that the emission around \hii\ regions shows extended structure on a range of scales. 

In Figure~\ref{fig:LvsI}, we plot the expected $L_{\text{\ha}}$ vs. $I_{\text{peak}}$ for a point source with a profile that matches the MUSE Moffat PSF profile. We account for the variation in the PSF by shading the region spanned by a PSF of 0.86$''$ (14.6~pc FWHM) and 1.15$''$ (19.5~pc FWHM), the 16th-84th percentile range of the PN FWHMs discussed in the Appendix~\ref{ap:psf}. A large majority of the regions are resolved and more extended than the PSF profile. However, 205 of our regions have a luminosity less than the 1$''$ PSF would predict for the corresponding peak intensity, and 121 have luminosities less than the 0.86$''$ PSF. Some of these regions have over-subtracted backgrounds due to, for example, unmasked peaks nearby or imperfect extinction-correction. These regions are scattered evenly throughout the disk, but are generally surrounded by regions where the \hb S/N is low.

\section{Impact of Methodology, Data Quality, and Galactic Properties}
\label{ap:profileprops}

\subsection{Median vs. Mean}
\label{ap:meanvsmed}
We use the median statistic to construct individual radial profiles and to stack regions to determine aggregate behaviors. For individual regions, this minimizes the effect of filamentary ISM structure or nearby unmasked regions on the backgrounds. In the stacks, it down-weights unusually compact or extended regions. We verified the use of the median to construct the profiles in individual regions via inspection of cutouts and profiles for many regions using different approaches. We show that our results are robust to stacking methodology in Figure~\ref{fig:meanvsmed-bins}.

\input{fig-meanvsmed}

Figure~\ref{fig:meanvsmed-bins} compares the mean and median-based stacked profiles for each peak intensity bin, without background subtraction. For bins 2 through 9 (our normal \hii\ regions), the two approaches match each other closely. In the central-most regions of the profiles, the two approaches show little difference. The background level does vary slightly between the two methods, with the mean picking up more surrounding emission and thus yielding a higher background. Upon visual inspection, much of the emission picked up in the mean profile is unassociated with the \hii\ region and thus, the median stack is preferred. 

There is only a strong difference in the results for the brightest and the faintest \ha intensity bin. The faintest bin has the lowest contrast with the background. The brightest bin contains some extremely bright and extremely cored regions as outliers, which are visible in the traces shown in Figure~\ref{fig:decile-profiles}. This causes a difference between the mean and the median at the very centers of the regions.

\subsection{Effect of Background Subtraction and \texorpdfstring{\rlum}{Rlum}}
\label{ap:bgsub}

\input{fig-RLbackground}

Applying a local background subtraction lowers the measured luminosity of our \hii\ regions by a factor of $\approx 3$, which affects the luminosity function and luminosity-size relations. Here, we explore the impact of the background subtraction on our size measurements. We use the same methodology adopted for our fiducial results, but set the background intensity to be zero and so build non-background-subtracted profiles. We maintain the same aperture radius, $R_{\text{lum}}^{\text{max}}$, used for the background-subtracted calculations.

Figure~\ref{fig:RLbackground} shows the luminosity-size relations with and without background subtraction. Qualitatively, the trends remain the same with (solid colors) and without (faded colors) background subtraction. Subtracting a local background makes the regions appear less luminous, moving all relations down in luminosity-size space. This impacts the faintest regions the most, where the background intensity is more comparable to the peak intensity of the \hii\ region. The background subtraction has the effect of lowering the measured \rmom, \rf, and \rn\ significantly, because it removes low-intensity extended flux from the profile. For bright regions, the HWHM is largely unaffected, because the background represents a modest fraction of the peak. For fainter regions, the HWHM radius decreases with the background subtraction, which enhances the contrast between the peak and the surrounding profile. The isophotal radius decreases significantly with background subtraction in the faintest bins, reflecting that our adopted intensity threshold is low, in line with its goal of capturing the \hii\ region-DIG divide. 

Despite shifts in the measured quantities, the relationships between luminosity and size are not strongly affected. After binning by intensity or luminosity, \rmom, \rf\ and the HWHM do not show a strong positive correlation with luminosity regardless of background treatment, and \riso\ does show a strong relationship to the luminosity regardless of our treatment of the background (Table~\ref{tab:RLcorrelations}). The best fit relations shift somewhat in slope and significantly in normalization depending on this treatment, as recorded in Table~\ref{tab:RLcorrelations}.

In Figure~\ref{fig:RLbackground}, we adopted the same radius to calculate luminosity in the cases with and without background. If, instead, we choose a fixed aperture size for all regions, then we see significant pile-up in fractional flux sizes \rf\ and \rn\ for sources without background subtraction, with little spread in the distribution. Without background subtraction, the aperture used to make the size and luminosity measurements plays a critical role. In the case where a segmentation algorithm is used, this often sets the effective luminosity radius.

When calculating the luminosity and radius of the \hii\ regions, we use \rlum as the outer boundary. Comparing this luminosity, $L(\text{\rlum})$ to the luminosity within our radius measurements (e.g., $L$(HWHM) or $L$(\rmom)) there is only a slight increase in the strength of the $R-L$ relationship. For example, the Spearman rank correlation coefficient $\rho$ for all regions changes from 0.19 to 0.39 for $L$(HWHM) and from 0.39 to 0.43 for $L$(\rmom). 

\subsection{Resolution}
\label{ap:res}

The HWHM, \rmom, \rf, and \rn show a strong imprint of the resolution, manifesting at $\sim$2, 3, and 4 times the PSF size. As a more robust check on the potential effects of resolution on our results, we convolve our data to a coarser resolution and repeat our analysis. We use a Gaussian kernel to convolve the extinction-corrected \ha\ map to final FWHM resolution of 1.5\arcsec, 2\arcsec, 2.5\arcsec, 3\arcsec, 3.5\arcsec, 4\arcsec, 4.5\arcsec, and 5\arcsec\ corresponding to 26, 34, 42, 51, 59, 68, 76, and 85~pc. These are comparable to the typical physical resolution of seeing-limited studies targeting galaxies at $\sim 5{-}30$~Mpc, including PHANGS-MUSE and SINGS \citep{groves2023,kennicutt2003,wisnioski2012}.

We plot the median profiles of all \hii\ regions for each resolution in Figure~\ref{fig:convolved-profiles}. We use the same \ha peaks and the same region masks as our native resolution catalog. Because the resolution of the convolved data becomes larger than the size of the initial search kernel, some of the \hii\ regions are blended together and thus potentially double-counted. To construct the profiles, we consider circular areas with a radius of 200~pc centered on each peak. The aperture and background radius \rlum\ is still measured as the 10th percentile of the profile. The maximum allowed \rlum\ increases from 100~pc by the convolved PSF value. This allows the aperture to be larger for lower resolution profiles and therefore increases the total luminosity of each region.

\input{fig-profilesres}

\input{tab-respowers}

In Figure~\ref{fig:convolved-profiles} at all resolutions, the radial profiles of \ha\ intensity always show a barely resolved core, only slightly more extended than the current PSF (dashed line in each panel) that transitions into an extended, shallowly declining component at larger radii. We present the best-fit double Gaussian function for resolutions $1''-5''$ in Table~\ref{tab:resfits}. Once convolved, the outer Gaussian becomes larger compared to the native resolution profile, with the deconvolved sizes increasing until FWHM resolution $3.5''$. We estimate the flux contributed from each component $\propto A\sigma^2$. We report the fraction of the flux associated with the inner Gaussian and the outer Gaussian components as f$_{\text{P}}$ in Table~\ref{tab:resfits}. 

Table~\ref{tab:resfits} also includes the best-fit single Gaussian at each resolution and $\Delta$BIC comparing the double and single Gaussian. Here $\Delta$BIC$>10$ prefers a double Gaussian over a single Gaussian and $\Delta$BIC$<-10$ prefers a single Gaussian over a double Gaussian. From $1''-4''$, the double Gaussian is significantly preferred, but at the lowest resolutions, the two functions are statistically indistinguishable. This could reflect that the kernel becomes comparable to the \hii\ region sizes at the lowest resolutions. For example, the deconvolved HWHM of the best-fitting single Gaussian at $5''=42.42$~pc resolution is 45.65~pc, greater than the size of the extended component fit at the native resolution. This trend towards increasing simple Gaussian forms may relate to the results in \citet{wisnioski2012}, which finds most of the luminosity in the core of their \hii\ regions at $1''-3''$ resolution (physical resolution of $40-325$~pc). Assuming the trends presented here, at this physical resolution, the regions would likely be represented well by a single Gaussian, which would contain the majority of the flux in the core.  

Figure~\ref{fig:convolved-profiles} shows how the sizes determined for the binned profiles change as a function of resolution. As the resolution of the data degrades, the actual measured sizes increase, but the overall luminosity-radius trends for each metric remain similar to what we measure at the native resolution. The HWHM, \rmom, and \rf show no clear trend with radius when binned by peak intensity. They remain approximately constant at multiples of the (increasing) PSF. Meanwhile, the isophotal radius shows a strong correlation with luminosity, with the slope of $L$ as a function of $R$ becoming shallower as the resolution becomes coarser (for the binned values). This reflects that the convolution increases the isophotal size for most bright regions but can drop the intensity in some faint regions enough to decrease their size at the isophotal intensity threshold. 

The bottom panel of Figure~\ref{fig:convolved-profiles} shows the full distribution of regions at 5" resolution with the same binning methods as Figure~\ref{fig:RL}. The trends at this coarse resolution do not vary strongly from the native resolution, although the $I_{\text{peak}}$ and $L_{\text{\ha}}$ bins track each other better at low resolution than at high resolution.  
We quote the Spearman rank correlation coefficient ($\rho$) and the power law index ($\alpha$) and normalization ($A$) for the full sample of \hii\ regions across resolution in Table~\ref{tab:resfits}.

\subsection{Galactic Inclination}
\label{ap:inclination}

NGC\,253 is highly inclined, with $i=76\degree$ \citep{mccormick2013}. Radial profiles of extended objects in the galactic disk could be affected by this high inclination. The viewing angle will also lead the minor axis profile to reflect a convolution of the vertical structure of regions, not only their in-plane extent. However, \hii\ regions should mostly be a local phenomenon, e.g., Milky Way regions showing sizes much smaller than the $\approx 100$~pc scale height of the cold gas \citep{kalberla2009,heyer2015,anderson2019}. Therefore, we expect that their shapes may not reflect the large-scale geometry of the system and instead be more symmetric.

\input{fig-2dcutoutscombo}

We test this expectation by examining the structure of the radial profiles. The mean and median two-dimensional stacks of all \hii\ regions are shown in Figure~\ref{fig:2dcutouts}. The $x$-axis is aligned to the in-plane direction (i.e., the major axis) and the $y$-axis aligned to the vertical direction (i.e., the minor axis), with the local background subtracted before stacking. Profiles below both stacks show cuts through the axes. Both stacks appear circular, suggesting that on average, the inclination of the galaxy does not affect the two-dimensional structure of the \hii\ regions on these scales. We also examined two-dimensional cutouts at a range of galactocentric radii and position angles. We did not observe any systematic distortions in the shapes of stacked sets of regions as a function of these parameters.

\section{Comparison to HST H\texorpdfstring{$\alpha$}{alpha} Measurements of NGC 5068}
\label{ap:ngc5068}

With the recent publication of the PHANGS-HST \ha narrowband observations from \citet[][]{chandar2025}, we have access to a complementary sample of \ha images for \hii\ regions. NGC\,5068 ($D$=5.2~Mpc) is the nearest galaxy in the \citet{chandar2025} sample, and is less massive and less inclined than NGC\,253. Because the physical resolution of these HST data resembles that of our NGC\,253 MUSE map, we replicate our measurements in NGC\,5068 to test the stability and generality of the results presented in the main text. 

We use the catalog of 126 \hii\ regions and \ha map presented in \citet{barnes2025}, which has a resolution of 10~pc. We correct the \ha flux for extinction based on a single value of $E(B-V)$ for each region calculated from MUSE \citep[][]{groves2023} and center the profiles on the \ha peak. We increase the isophote used for \riso to match the effective isophote presented in \citet{barnes2025}, which reflects the higher noise level in the HST H$\alpha$ maps compared to MUSE. We extract source properties at $R_{\text{lum}}^{\text{max}} = 100~\mathrm{pc}$. Due to the lower sensitivity of HST, the sample of regions identified in NGC\,5068 is most comparable to the high luminosity distribution of regions seen in NGC\,253.

The PHANGS-HST catalog was built around the PHANGS-MUSE \citet{groves2023} \hii\ region catalog, and thus inherits some of the biases associated with their image segmentation. Thus, the approach here is not an exact comparison to the analysis presented for NGC\,253, but we believe the effects to be small. 

\subsection{Size-Luminosity Relations}
\input{fig-HSTRL}

In Figure~\ref{fig:hstrl}, we find that the HWHM radius remains stable, showing no strong relation with \ha luminosity. This agrees with the results we find for NGC\,253 (Section~\ref{sec:hwhm}). When centered around \ha peaks, there is always some unresolved cored distribution of emission, which is most closely probed by the HWHM. This relation shows no dependence on the aperture chosen or the background intensity.

We also see consistency between NGC\,5068 and NGC\,253 in the \riso trend, which shows little variation with aperture changes. The slope of the relation is similar to that found for NGC\,253, but the normalization is shifted to a higher value due to the increased isophote intensity threshold that we use in NGC\,5068 (based on \citet{barnes2025} due to the higher noise in the HST data). 

When we investigate the trends in \rmom and \rf, however, a clear, strong size-luminosity relation emerges in NGC\,5068. This contrasts with NGC\,253, where these size measurements show only a scattered, weak relationship to luminosity. In every case, the radius trends in NGC\,5068 are consistent with the best-fit \riso-$L$. The Spearman rank correlation coefficients for each radius are higher in NGC\,5068 than in NGC\,253 (Table~\ref{tab:sizesize}).

\subsection{Functional Form}

\input{fig-ngc5068profile}

As in NGC\,253, we fit the average NGC\,5068 \ha profile with a double Gaussian and a Moffat. The Moffat profile is visually and statistically better than a double Gaussian, but we discuss the results of the double Gaussian here for simpler comparison to NGC\,253 results. We find that this function fits the profile well with $\Delta\text{BIC}=3147$ compared to a single Gaussian. With an inner Gaussian HWHM of 6.2~pc (compared to a PSF FWHM of 10~pc), the core of the \hii\ regions in NGC\,5068 appear to be less resolved than those in NGC\,253. Bin 9 in NGC\,253 has a deconvolved inner Gaussian HWHM of $\approx 5.2$~pc, while the deconvolved HWHM in NGC\,5068 is $\approx 3.6$~pc. Like NGC\,253, the stacked profile in NGC\,5068 shows an extended component, with width $\sigma_2 \approx 25$~pc, similar to the $\sigma_2 \approx 30$~pc extended component seen in NGC~253.

We perform the same fitting on the individual \hii\ region profiles and find that a double Gaussian function is a better fit to 83\% of all \hii\ regions in NGC\,5068, based on the BIC criteria outlined in Section~\ref{sec:profileshape}. This is comparable to 63\% and 79\% of regions fit best by a double Gaussian in bins 9 and 10 in NGC\,253. We conclude that, in general, bright \hii\ regions identified as \ha peaks are well-fit by a double Gaussian function. 

\end{document}

%% file: affils.tex
\newcommand{\ANU}{Research School of Astronomy and Astrophysics, Australian National University, Canberra, ACT 2611, Australia}   

\newcommand{\OSU}{Department of Astronomy, The Ohio State University, 140 West 18th Avenue, Columbus, OH 43210, USA}

\newcommand{\CCAPP}{Center for Cosmology and Astroparticle Physics (CCAPP), 191 West Woodruff Avenue, Columbus, OH 43210, USA}

\newcommand{\nrao}{National Radio Astronomy Observatory, 520 Edgemont Road, Charlottesville, VA 22903, USA}
\newcommand{\UCT}{Department of Astronomy, University of Cape Town, Rondebosch 7701, South Africa}

\newcommand{\Virginia}{Department of Astronomy, University of Virginia, Charlottesville, VA 22904, USA}

\newcommand{\JHU}{Department of Physics and Astronomy, The Johns Hopkins University, Baltimore, MD 21218 USA}

\newcommand{\UConn}{Department of Physics, University of Connecticut, 196A Auditorium Road, Storrs, CT 06269, USA}

\newcommand{\UWyoming}{Department of Physics and Astronomy, University of Wyoming, Laramie, WY 82071, USA}

\newcommand{\Oxford}{Sub-department of Astrophysics, Department of Physics, University of Oxford, Keble Road, Oxford OX1 3RH, UK}

\newcommand{\eso}{European Southern Observatory, Karl-Schwarzschild-Stra{\ss}e 2, 85748 Garching, Germany}

\newcommand{\esochile}{European Southern Observatory (ESO), Alonso de C\'ordova 3107, Casilla 19, Santiago 19001, Chile}

\newcommand{\ULyon}{\label{ULyon} Univ Lyon, Univ Lyon 1, ENS de Lyon, CNRS, Centre de Recherche Astrophysique de Lyon UMR5574,\\ F-69230 Saint-Genis-Laval, France}

\newcommand{\UOA}
{Department of Physics, University of Arkansas, 226 Physics Building, 825 West Dickson Street, Fayetteville, AR 72701, USA}

\newcommand{\CfA}
{Center for Astrophysics $\mid$ Harvard \& Smithsonian, 60 Garden St., 02138 Cambridge, MA, USA}

\newcommand{\nraoABQ}
{National Radio Astronomy Observatory, 800 Bradbury SE, Suite 235, Albuquerque, NM 87106 USA}

\newcommand{\UniCA}{Université Côte d'Azur, Observatoire de la Côte d'Azur, CNRS, Laboratoire Lagrange, 06000, Nice, France}

\newcommand{\UNAM}{Instituto de Astronom\'{\i}a, Universidad Nacional Aut\'onoma de M\'exico, Ap. 70-264, 04510 CDMX, Mexico}

\newcommand{\ITA}{Universit\"{a}t Heidelberg, Zentrum f\"{u}r Astronomie, Institut f\"{u}r Theoretische Astrophysik, Albert-Ueberle-Str.\ 2, 69120 Heidelberg, Germany}

\newcommand{\Arcetri}{INAF -- Osservatorio Astrofisico di Arcetri, Largo E. Fermi 5, I-50157, Firenze, Italy}

\newcommand{\ARI}{Astronomisches Rechen-Institut, Zentrum f\"{u}r Astronomie der Universit\"{a}t Heidelberg, M\"{o}nchhofstra\ss e 12-14, D-69120 Heidelberg, Germany}

\newcommand{\UAlberta}{Dept.\ of Physics, University of Alberta, 4-183 CCIS, Edmonton, AB, T6G 2E1, Canada}

\newcommand{\ucsd}{Department of Astronomy \& Astrophysics, University of California, San Diego, 9500 Gilman Dr., La Jolla, CA 92093, USA}

\newcommand{\ghent}{Sterrenkundig Observatorium, Ghent University, Krijgslaan 281-S9, B-9000 Ghent, Belgium}

\newcommand{\STScI}{Space Telescope Science Institute, 3700 San Martin Drive, Baltimore, MD 21218, USA}

\newcommand{\stanford}{Kavli Institute for Particle Astrophysics and Cosmology, Stanford University, 452 Lomita Mall, Stanford, CA\,94305, USA}

%% file: authors.tex
\author[0000-0002-6187-4866]{Rebecca L. McClain}
\affiliation{\OSU}
\affiliation{\CCAPP}
\email{mcclain.378@osu.edu}

\author[0000-0002-2545-1700]{Adam K. Leroy}
\affiliation{\OSU}
\affiliation{\CCAPP}
\email{leroy.42@osu.edu}

\author{Enrico Congiu}
\affiliation{\esochile}
\email[]{econgiu@eso.org}

\author[0000-0003-0410-4504]{Ashley.~T.~Barnes}
\affiliation{\eso}
\email{ashley.barnes@eso.org}

\author{Francesco Belfiore}
\affiliation{\Arcetri}
\email[]{francesco.belfiore@inaf.it}

\author[0000-0002-4755-118X]{Oleg Egorov}
\affiliation{\ARI}
\email[]{oleg.egorov@uni-heidelberg.de}

\author{Eric Emsellem}
\affiliation{\eso}
\affiliation{\ULyon}
\email{eric.emsellem@eso.org}

\author[0000-0002-5204-2259]{Erik Rosolowsky}
\affiliation{\UAlberta}
\email{rosolowsky@ualberta.ca}

\author[0000-0002-8553-1964]{Amirnezam Amiri}
\affiliation{\UOA}
\email{amirnezamamiri@gmail.com}

\author[0000-0003-0946-6176]{Médéric Boquien}
\affiliation{\UniCA}
\email[]{mederic.boquien@oca.eu}

\author[]{J\'er\'emy Chastenet}
\affiliation{\ghent}
\email{jeremy.chastenet@ugent.be}

\author[0000-0001-8241-7704]{Ryan~Chown}
\affiliation{\OSU}
\email[]{chown.5@osu.edu}

\author[0000-0002-5782-9093]{Daniel~A.~Dale}
\affiliation{\UWyoming}
\email{ddale@uwyo.edu}

\author[0000-0002-9069-7061]{Sanskriti Das}
\altaffiliation{Hubble Fellow}
\affiliation{\stanford}
\email{}

\author[0000-0001-6708-1317]{Simon C.~O.\ Glover}
\affiliation{\ITA}
\email{glover@uni-heidelberg.de}

\author[0000-0002-3247-5321]{Kathryn~Grasha}
\altaffiliation{ARC DECRA Fellow}
\affiliation{\ANU}   
\email{kathryn.grasha@anu.edu.au}

\author[0000-0002-4663-6827]{R\'emy Indebetouw}
\affiliation{\Virginia}
\affiliation{\nrao}
\email{indebetouw@gmail.com}

\author[0000-0001-9605-780X]{Eric W. Koch}
\affiliation{\nraoABQ}
\affiliation{\CfA}
\email{eric.koch@cfa.harvard.edu}

\author{Smita Mathur}
\affiliation{\OSU}
\affiliation{\CCAPP}
\email{mathur.17@osu.edu}

\author[0000-0002-6972-6411]{J. Eduardo M\'endez-Delgado}
\affiliation{\UNAM}
\email[]{jemd@uni-heidelberg.de}

\author[0000-0002-0119-1115]{Elias~K.~Oakes}
\affiliation{\UConn}
\email[]{elias.oakes@uconn.edu}

\author[0000-0002-1370-6964]{Hsi-An Pan}
\affiliation{Department of Physics, Tamkang University, No.151, Yingzhuan Road, Tamsui District, New Taipei City 251301, Taiwan} 
\email{hapan@gms.tku.edu.tw}

\author{Karin Sandstrom}
\affiliation{\ucsd}
\email[]{kmsandstrom@ucsd.edu}

\author[0000-0002-6313-4597]{Sumit K. Sarbadhicary}
\affiliation{\JHU}
\email{ssarbad1@jh.edu}

\author[0000-0002-3784-7032]{Bradley~C.~Whitmore}
\affiliation{\STScI}
\email{snskriti@stanford.edu}

\author[0000-0002-0012-2142]{Thomas~G.~Williams}
\affiliation{\Oxford}
\email[]{thomas.williams@physics.ox.ac.uk}

%% file: fig-peakmapcutouts.tex
\begin{figure*}
    \centering
    \includegraphics[width=\linewidth]{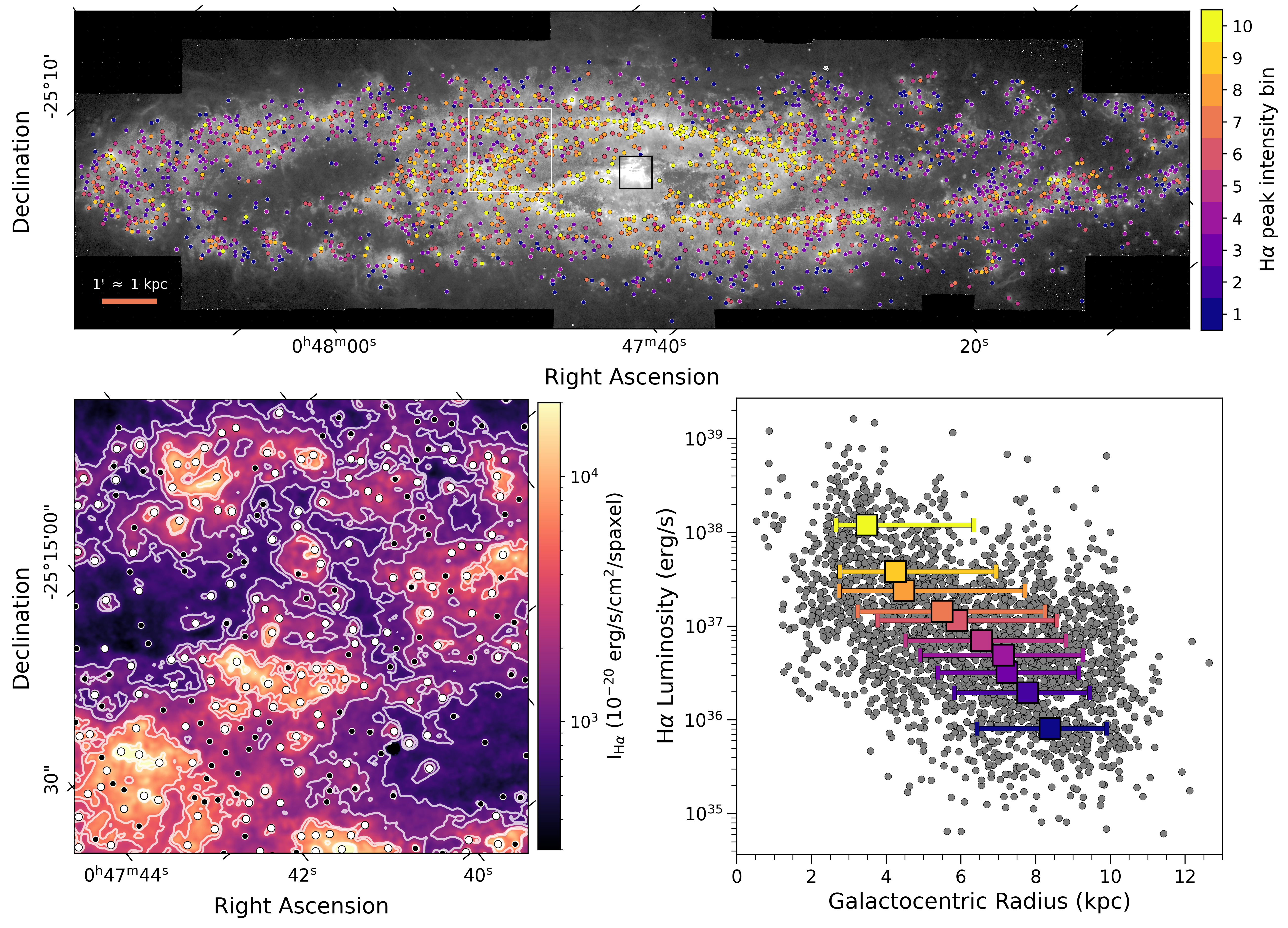}
    \caption{Top: MUSE extinction-corrected \ha map of NGC\,253 in grayscale \citep{congiu2025}. Points show identified \ha peaks colored by decile of background-subtracted \ha peak intensity (Section~\ref{sec:intbin}, Figure~\ref{fig:LvsI}). The brighter \hii\ regions tend to be located in the inner spiral arms while the fainter regions tend to be located in the outer spiral arms or in the interarm region. The black box shows the galactic center excluded from our analysis. Bottom Left: Cutout of the \ha intensity map before extinction-correction to illustrate peak identification for one complex region
    (white box in top panel). White contours show the 25, 50, 84, 90, 99, and 99.5\% intensity levels. White points show the peaks included in the final catalog and the black points show peaks removed by intensity, background contrast, and line ratio diagnostic cuts (Appendix \ref{ap:bpt}). Bottom Right: Luminosity of \hii\ regions (gray points) as a function of galactocentric radius. Colored points represent the median and 16-84\% range for regions binned by \ha peak intensity. The brightest regions tend to lie in the inner galaxy, reflecting the difficulty of detecting faint \hii\ regions in complex galactic environments.}
    \label{fig:peaks}
\end{figure*}

%% file: tab-galprops.tex
\begin{table}
    \begin{center}
    \caption{NGC 253 galaxy and \hii\ region catalog}
    \begin{tabular}{cc}
    \toprule
    \multicolumn{2}{c}{Properties of NGC\,253} \\
    \hline\hline
    Distance & 3.50 $\pm$ 0.22 Mpc (\textit{a, b}) \\
    Position Angle & 52.5$\degree$ (\textit{c}) \\
    Inclination & 76$\degree$ (\textit{d})  \\
    $R_{25}$ & 13.5 arcmin  (\textit{c}) \\
    Stellar Mass & $4.37^{+1.13}_{-0.90}$ $\times 10^{10}$ M$_{\odot}$ (\textit{e}) \\
    SFR & $2.78 \pm0.56$ M$_{\odot}$ yr$^{-1}$ \\
    Disk SFR & $1.72 \pm 0.34$~M$_{\odot}$~yr$^{-1}$\\
    \hline\hline
    \multicolumn{2}{c}{Properties of the \hii\ region catalog} \\
    \hline\hline
    Sample size & \nregions\\
    Total \ha flux$^*$ & $4.67\times10^{45}$ erg~s$^{-1}$~kpc$^{-2}$ \\
    Total \ha$_{\text{corr}}$ flux & $1.94\times10^{46}$ erg~s$^{-1}$~kpc$^{-2}$ \\
    $L_{\text{\ha}}$ function$^{\dagger}$ & $\alpha_1 = -0.92,~\alpha_2 = -1.86$ \\
    $I_{\text{\ha peak}}$ distribution$^{\dagger}$ & $\alpha_1 = -1.21,~\alpha_2 = -2.45$ \\
    Completeness limit & 10$^{37.3}$ erg/s, 10$^{41}$ erg/s/kpc$^{2}$ \\
    $L_{\text{break}}$ & 10$^{38.2}$ erg~s$^{-1}$\\
    $I_{\text{break}}$ & 10$^{41.9}$ erg~s$^{-1}$~kpc$^{-2}$\\
    Native PSF & 0.73$''$ [0.62$''$, 0.95$''$]$^{\ddagger}$\\
    Convolved PSF & 0.95$''$ [0.86$''$, 1.13$''$]$^{\ddagger}$\\
    Typical $A_{\text{\ha}}$ & 1.73 [0.86, 3.09]$^{\ddagger}$\\ 
    \bottomrule
    \label{tab:ngc253_props}
    \end{tabular}
    \end{center}
    \tablecomments{References for galaxy properties: (a) \citet{okamoto2024}, (b) \citet{congiu2025}, (c) \citet{makarov2014}, (d) \citet{mccormick2013}, (e) \citet{leroy2021}.
    $^*$The total \ha flux for \hii\ regions, corrected and uncorrected for the effects of extinction, is calculated without background subtraction and accounts for 70\% and 59\% of total detected \ha flux in the map. $^{\dagger}$The fits are performed on the cumulative distribution of luminosity and intensity (Figure \ref{fig:LvsI}) $^{\ddagger}$Values represent the median [16th, 84th percentile] of the full distribution. We calculate the SFR and disk SFR from the extinction-corrected \ha luminosity of the galaxy, including and excluding the center respectively, using the conversion factor from \citet{murphy2011,kennicutt2012}.
    }    
\end{table}

%% file: fig-cdf.tex
\begin{figure}
    \centering
    \includegraphics[width=\linewidth]{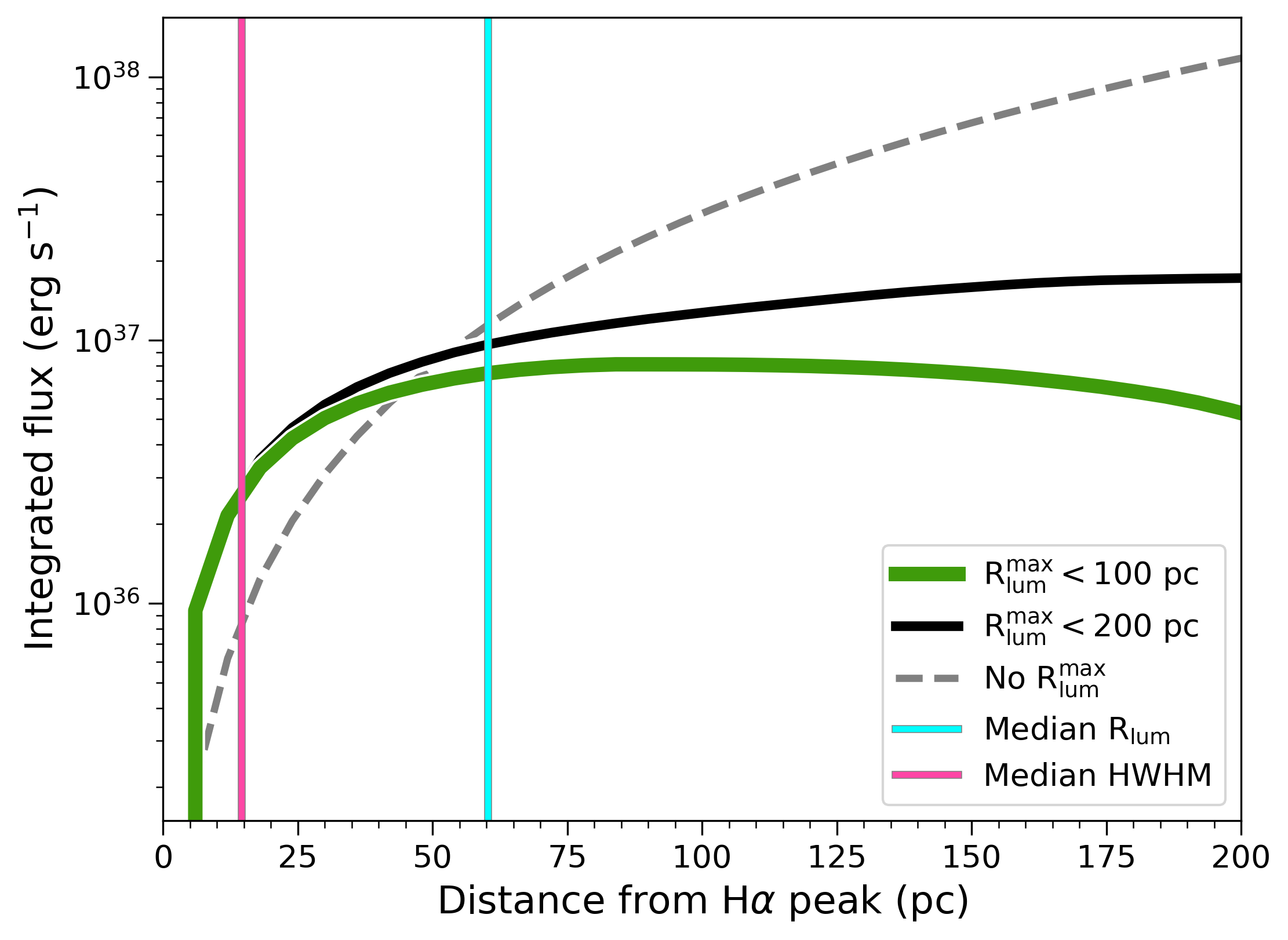}
    \caption{Median integrated flux profile, i.e., flux enclosed within the radius shown on the $x$-axis, over the full sample of \hii\ regions for several treatments of the background. Green shows our fiducial treatment, with $R_{\text{lum}}^{\text{max}} = 100$~pc, black shows $R_{\text{lum}}^{\text{max}} = 200$~pc, and the gray dashed line shows the integrated flux profile with no background subtraction.
    Here $R_{\text{lum}}^{\text{max}}$ is the maximum radius allowed when determining the aperture used to integrate the flux, calculate the background, and measure sizes for each \hii\ region. The gray dashed line shows that with no background subtraction the flux (and so also \rf , \rn , and \rmom) rises monotonically with distance from the peak. Our fiducial background subtraction leads to a flat enclosed flux, on average, between $\sim 50{-}150$~pc. Using a larger maximum aperture of $R_{\text{lum}}^{\text{max}} = 200$~pc (black line) yields a profile that rises out to $200$~pc, though more slowly than the unsubtracted case. Given this behavior, both our inferred luminosity and sizes based on fractions of enclosed flux, e.g., \rf and \rn, are sensitive to the definition of background. The light blue line shows the median aperture radius for the whole sample using our fiducial treatment.
    }
    \label{fig:cdf}
\end{figure}

%% file: fig-LI.tex
\begin{figure*}
    \begin{center}
    \begin{tabularx}{\linewidth}{cc}
    \includegraphics[width=0.475\linewidth]{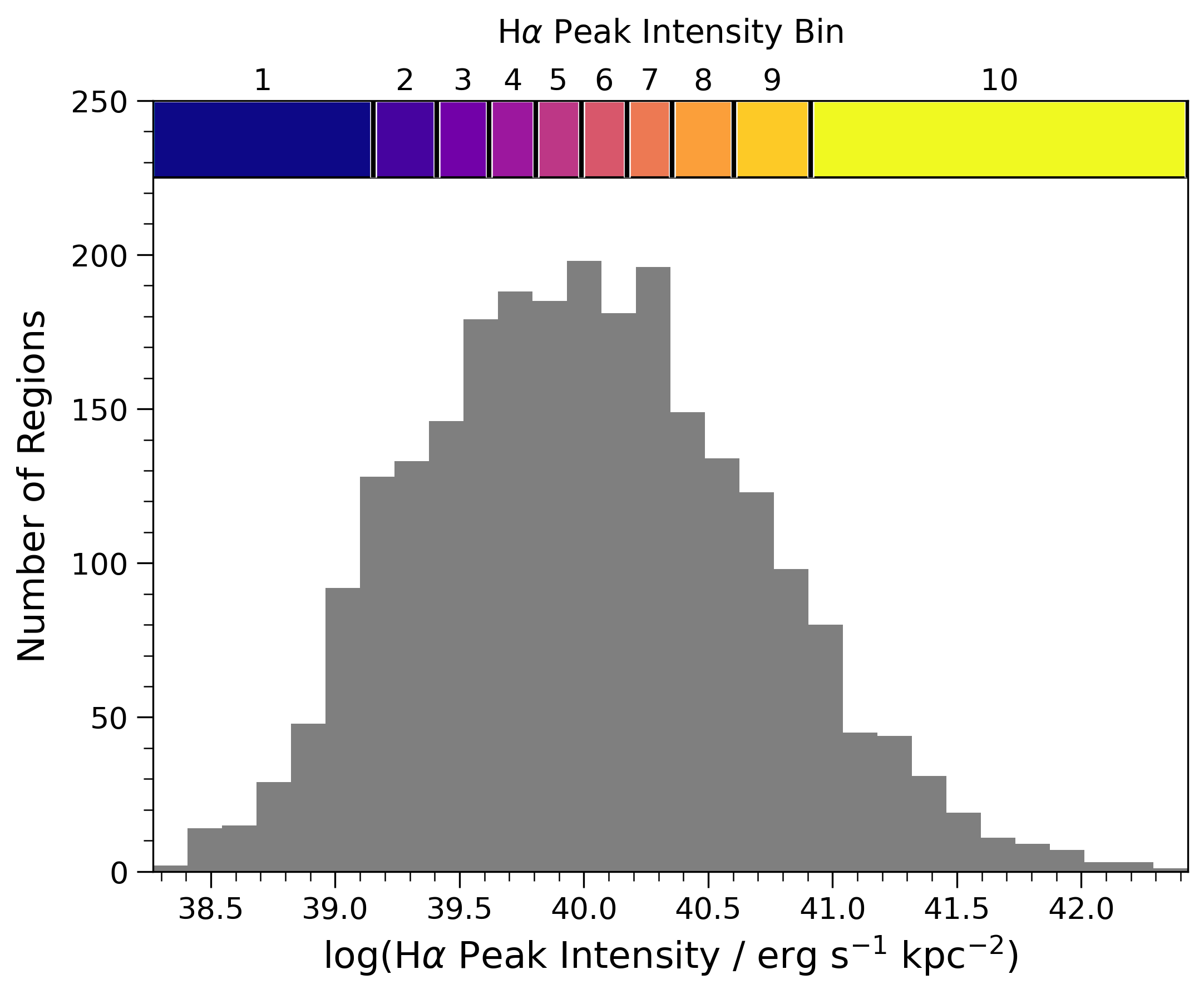} &
    \includegraphics[width=0.475\linewidth]{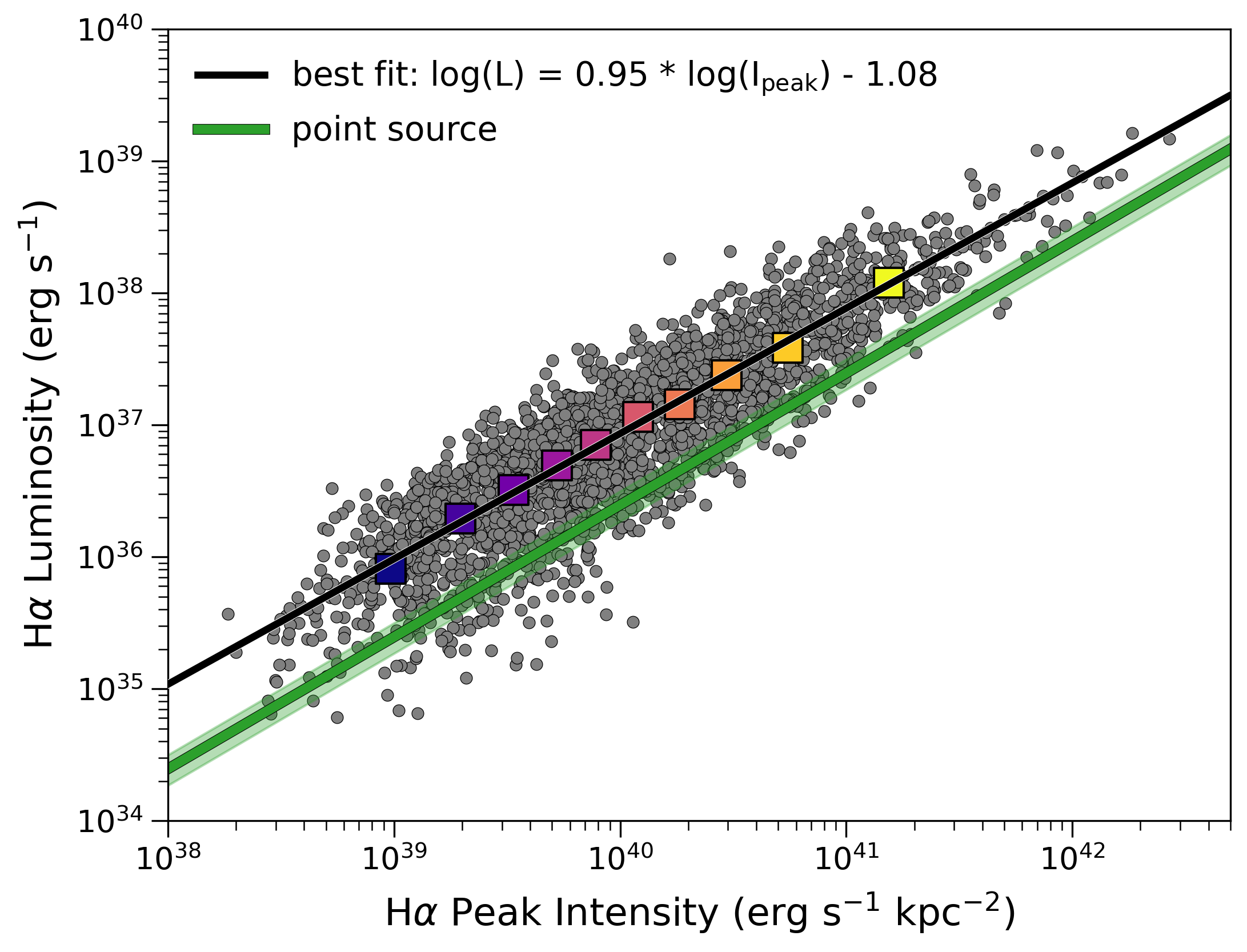} \\
    \includegraphics[width=0.475\linewidth]{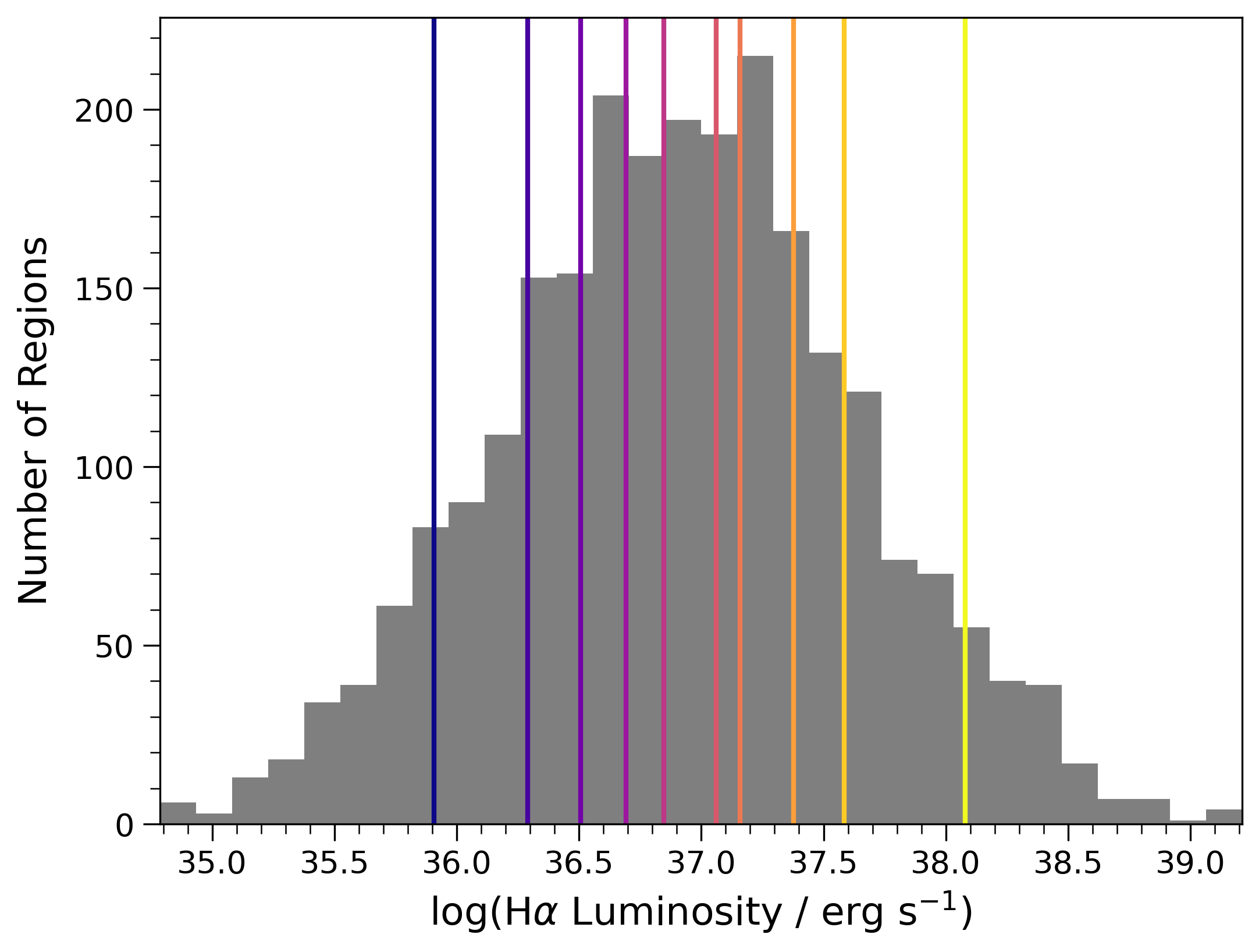} & \includegraphics[width=0.475\linewidth]{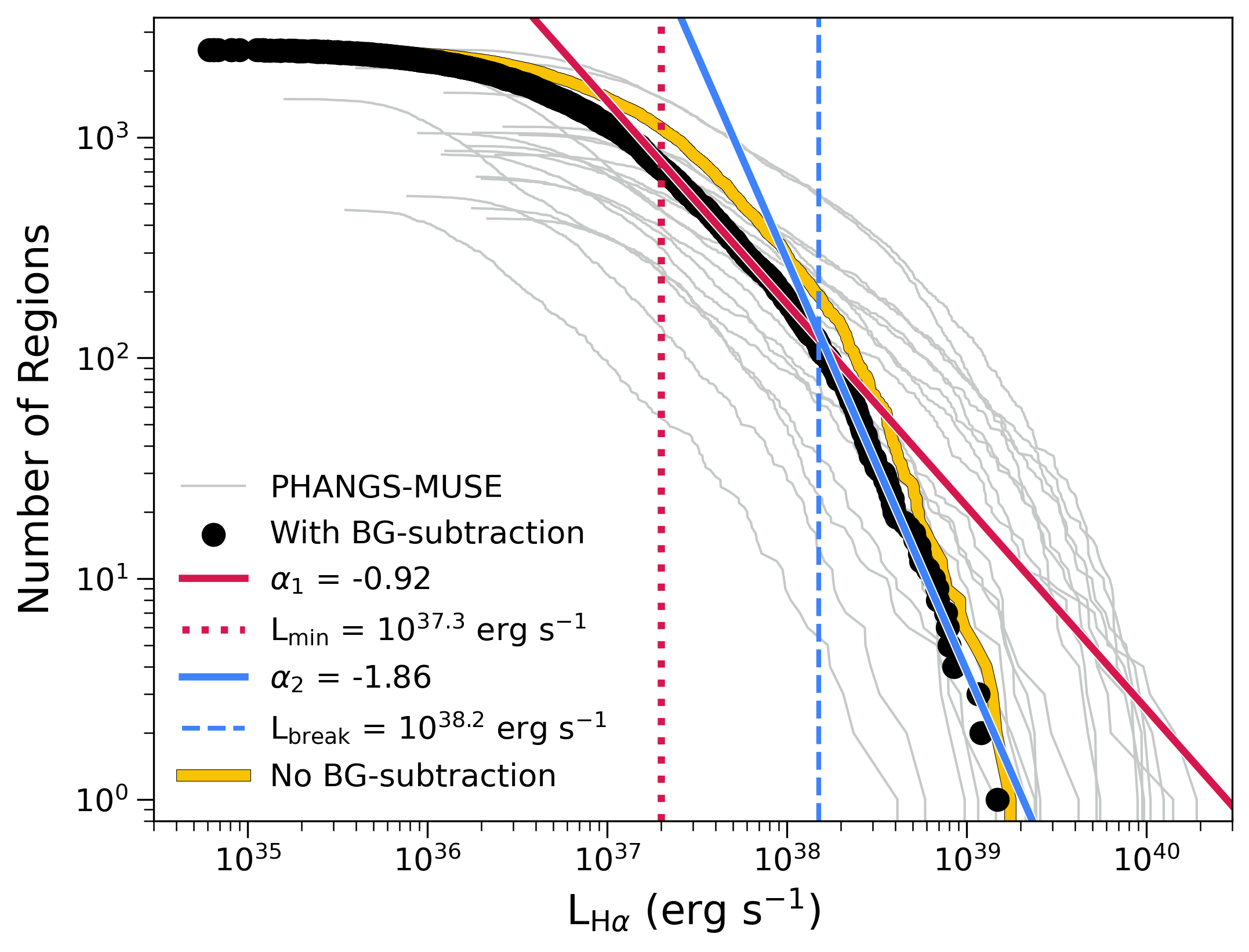}
    \end{tabularx}
    \caption{Luminosity and intensity distributions in the \hii\ region catalog. Top left: Distribution of extinction-corrected H$\alpha$ intensities at the peaks that define our regions. The color bar indicates the ranges spanned by the peak intensity bins (deciles) used in our analysis. Top right: Relationship between extinction-corrected, background-subtracted L$_{\rm H\alpha}$ and peak intensity for individual regions (gray points) and median values within intensity bins (squares). In NGC\,253, peak intensity and L$_{\rm H\alpha}$ are tightly correlated, and the black line shows the best fit to the bins. The green line shows expectations for an unresolved point source with individual traces for each field. Bottom left: Distribution of extinction-corrected, background-subtracted \ha luminosities (L$_{\rm H\alpha}$) for our regions. Colored lines indicate the median luminosity within each peak intensity bin. Bottom right: The \ha luminosity function for our regions with (black) and without (gold) background subtraction. For comparison, gray lines show the luminosity functions for each PHANGS-MUSE galaxy \citep[][]{groves2023}. Blue and red lines show power laws that match the upper and lower luminosity ranges for NGC~253. \label{fig:LvsI}}
    \end{center}
\end{figure*}

%% file: tab-binprops.tex
\begin{table*}
    \begin{center}
    \caption{Properties of peak intensity bins}

    \begin{tabular}{c|cccc|cccc|cccc}
    \toprule
    Bin & \multicolumn{4}{c|}{\ha Distribution} & \multicolumn{4}{c|}{\ha Profile} & \multicolumn{4}{c}{Density Profile} \\
    \hline
    \hline
     & Range $I_{\text{peak}}$ & Med $I_{\text{peak}}$ & Range $L_{\text{\ha}}$ & Med $L_{\text{\ha}}$ & $A$ & $A_{\text{f}}$ & $\sigma_1$ & $\sigma_2$ & $A$ & $A_{\text{f}}$ & $\sigma_1$ & $\sigma_2$ \\
     & $\log\left[\frac{\text{erg}}{\text{s}~\text{kpc}^{2}}\right]$ & $\log\left[\frac{\text{erg}}{\text{s}~\text{kpc}^{2}}\right]$ & $\log\left[\frac{\text{erg}}{\text{s}}\right]$ & $\log\left[\frac{\text{erg}}{\text{s}}\right]$ & 
     $\log\left[\frac{\text{erg}}{\text{s}~\text{kpc}^{2}}\right]$ & & pc & pc & cm$^{-3}$ & & pc & pc \\
    \midrule
    1 & [38.27,39.15] & 38.99 & [34.79,36.64] & 35.91 & 38.98 & 0.65 & 10.49 & 24.28 & 7.68 & 0.65 & 10.76 & 32.79 \\
    2 & [39.15,39.41] & 39.29 & [35.08,37.07] & 36.29 & 39.29 & 0.44 & 7.38 & 23.87 & 19.22 & 0.44 & 2.18 & 32.18 \\
    3 & [39.41,39.62] & 39.53 & [35.18,37.14] & 36.51 & 39.52 & 0.62 & 9.28 & 27.94 & 15.76 & 0.62 & 8.24 & 38.18 \\
    4 & [39.62,39.81] & 39.72 & [35.19,37.49] & 36.69 & 39.72 & 0.62 & 9.70 & 25.79 & 18.82 & 0.62 & 9.17 & 35.02 \\
    5 & [39.81,39.99] & 39.89 & [35.56,37.58] & 36.85 & 39.89 & 0.66 & 10.06 & 26.20 & 22.59 & 0.66 & 9.90 & 35.62 \\
    6 & [39.99,40.17] & 40.08 & [35.51,37.72] & 37.06 & 40.08 & 0.67 & 9.52 & 27.25 & 29.62 & 0.67 & 8.79 & 37.16 \\
    7 & [40.17,40.35] & 40.26 & [36.26,38.26] & 37.16 & 40.26 & 0.67 & 9.26 & 24.34 & 37.98 & 0.67 & 8.21 & 32.87 \\
    8 & [40.35,40.60] & 40.47 & [36.40,38.31] & 37.38 & 40.47 & 0.64 & 8.74 & 23.09 & 51.37 & 0.64 & 6.98 & 31.02 \\
    9 & [40.60,40.91] & 40.74 & [36.79,38.39] & 37.58 & 40.73 & 0.70 & 8.47 & 23.19 & 75.48 & 0.70 & 6.27 & 31.17 \\
    10 & [40.91,42.43] & 41.19 & [37.18,39.21] & 38.08 & 41.19 & 0.69 & 8.55 & 24.02 & 124.03 & 0.69 & 6.48 & 32.40 \\
    \bottomrule
    \label{tab:binprops}
    \end{tabular}  
    \end{center}
    \tablecomments{Properties of the \ha peak intensity bins used throughout the analysis. In the first four columns, we quote the range of \ha peak intensity ($I_{\text{peak}}$) and \ha luminosity ($L_{\text{\ha}}$) as [min,max] and the median of each bin. In the middle four columns, we provide the parameters of the best-fit double Gaussian to the median one-dimensional \ha profile. The final four columns are the parameters of the three-dimensional density profile derived from the deconvolved one-dimensional \ha profile (Equations~\ref{eq:3df} and \ref{eq:n}). For both the \ha and the density profile, $A$ is the peak density at the center of the profile, $A_{\text{f}}$ is the fraction of amplitude corresponding to the inner Gaussian, $1-A_{\text{f}}$ to the outer Gaussian of the density distribution. $\sigma_1$ and $\sigma_2$ are the HWHM of the inner and outer Gaussian.}
\end{table*}

%% file: fig-decileprofiles.tex
\begin{figure*}
    \centering
    \includegraphics[height=0.845\textheight]{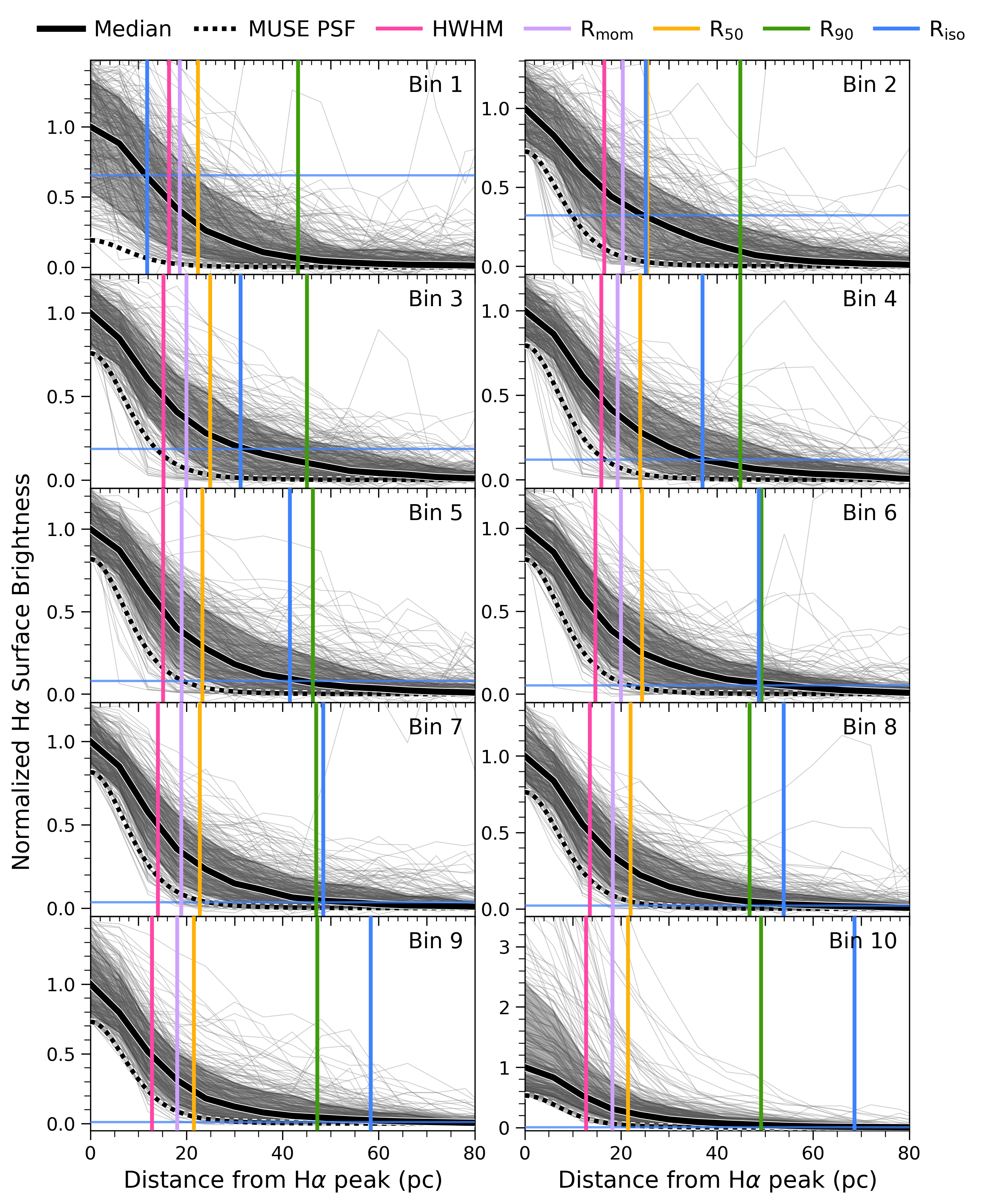}
    \caption{Background-subtracted radial profiles of extinction-corrected \ha intensity for \hii\ regions grouped by their peak \ha intensity. The individual profiles are normalized by the median peak intensity within the bin. The profiles of individual regions are plotted in gray and the median (solid black line) profile for each bin is drawn on top with the 16th to 84th percentiles shaded. The average MUSE PSF is shown with a black dotted line for comparison, normalized to the faintest region in the bin. Five radius measurements--HWHM (pink), \rmom (purple), \rf (yellow), \rn (green), and \riso (blue)--are represented with vertical lines. A horizontal line is plotted at the fiducial isophote used to measure \riso , $10^{38.8}$ erg~s$^{-1}$~kpc$^{-2}$. The profile shapes appear consistent across peak intensity bins. 23 \hii\ regions in bin 10 have profile peaks that extend off the plot.}
    \label{fig:decile-profiles}
\end{figure*}

%% file: fig-RLcombo.tex
\begin{figure*}
    \centering
    \includegraphics[width=\linewidth]{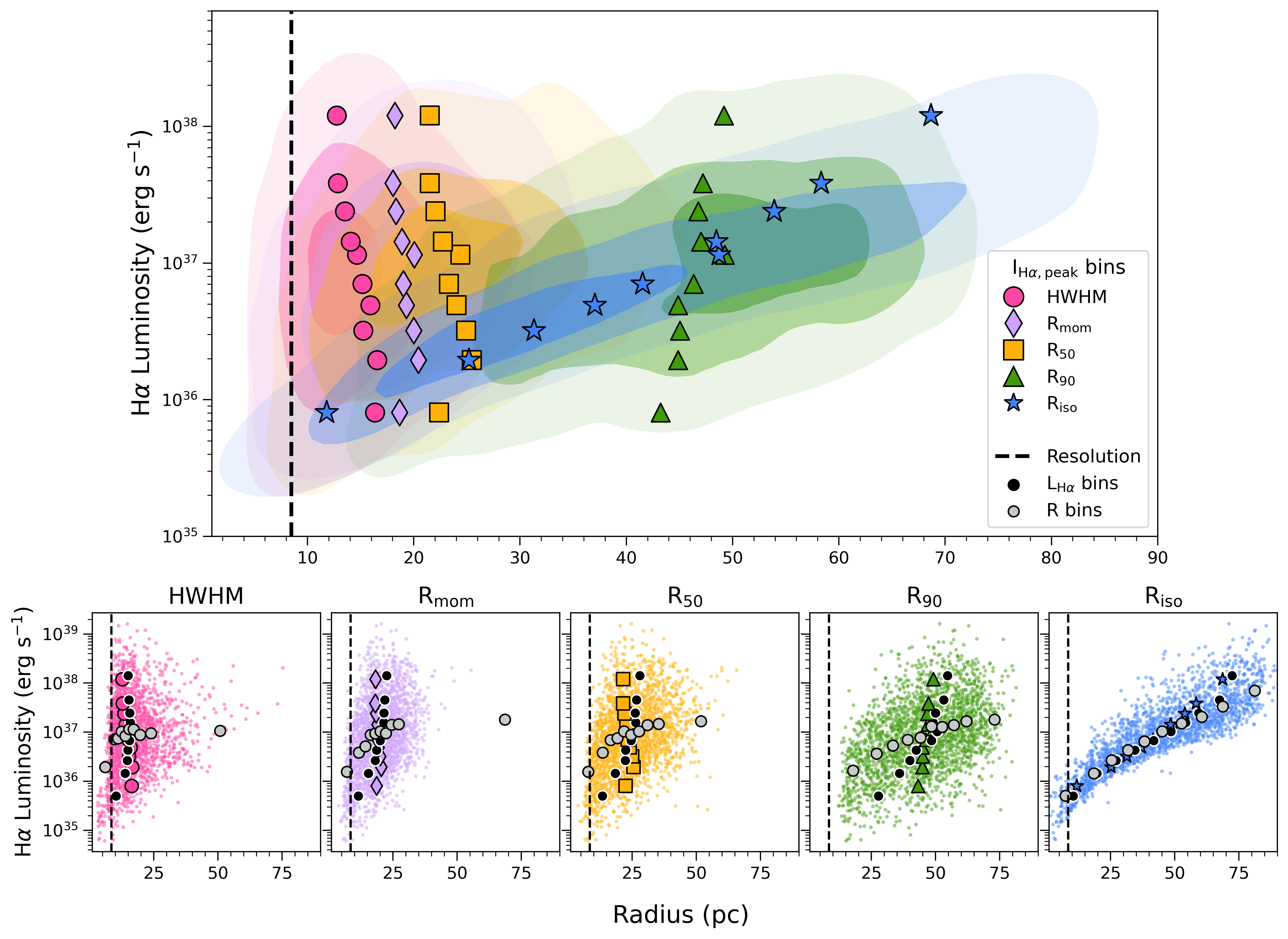}
    \caption{Luminosity and size for \hii\ regions in NGC 253. Top: 16th, 50th, and 84th percentile density contours of extinction-corrected, background-subtracted \ha luminosity as a function of several radius measurements (Section \ref{sec:radius}). The large points show the median values of regions binned by peak \ha intensity (Section \ref{sec:intbin}). There is no strong trend between HWHM (or core) radius and luminosity (Section~\ref{sec:hwhm}), while the isophotal radius (blue) shows a strong correlation with luminosity (Section~\ref{sec:isophotal}). The second moment radius, \rmom, and effective radius, \rf , also show large scatter and little clear correlation with radius, while \rn shows more ambiguous results that reflect its sensitivity to the adopted integration aperture (Section~\ref{sec:effectiveradius}) Bottom: Luminosity-radius relations for each metric showing individual data points and results for different binning approaches: peak intensity bins (large colored markers), bins of \ha luminosity (black circles), and bins of size (gray circles). 
    }
    \label{fig:RL}
\end{figure*}

%% file: tab-correlations.tex
\begin{table*}
    \begin{center}
    \caption{Radius-Luminosity correlations.}
    \begin{tabular}{cccccccccc}
    \toprule
    \multicolumn{10}{c}{With Background Subtraction}\\
    \hline
    \hline
     & \multicolumn{3}{c}{All regions} & \multicolumn{2}{c}{I$_{\text{peak}}$ Bins} & \multicolumn{2}{c}{L$_{\text{\ha}}$ Bins} & \multicolumn{2}{c}{R Bins}\\
    Radius & $\rho$ & $\alpha$ & $\log(A)$ & $\alpha$ & $\log(A)$ & $\alpha$ & $\log(A)$ &$\alpha$ & $\log(A)$ \\
    \midrule
    HWHM & 0.19 & 0.93 & 35.82 & - & - & - & - & - & - \\
    \rmom & 0.39 & 1.82 & 34.62 & - & - & - & - & 1.10 & 35.48 \\
    \rf  & 0.38 & 1.56 & 34.81 & - & - & - & - & 1.29 & 35.17 \\
    \rn & 0.43 & 1.95 & 33.75 & - & - & - & - & 1.79 & 33.99 \\
    \riso & 0.89 & 2.10 & 33.64 & 3.45 & 31.36 & 2.70 & 32.55 & 2.04 & 33.69 \\
    \midrule
    \multicolumn{10}{c}{Without Background Subtraction}\\
    \hline
    \hline
     & \multicolumn{3}{c}{All regions} & \multicolumn{2}{c}{I$_{\text{peak}}$ Bins} & \multicolumn{2}{c}{L$_{\text{\ha}}$ Bins} & \multicolumn{2}{c}{R Bins}\\
    Radius & $\rho$ & $\alpha$ & $\log(A)$ & $\alpha$ & $\log(A)$ & $\alpha$ & $\log(A)$ &$\alpha$ & $\log(A)$ \\
    \midrule
    HWHM & 0.13 & 0.54 & 36.50 & - & - & - & - & - & - \\
    \rmom & 0.38 & 1.71 & 34.88 & - & - & - & - & 1.50 & 35.15\\
    \rf  & 0.36 & 1.48 & 35.03 & - & - & - & - & 1.32 & 35.25 \\
    \rn & 0.42 & 1.96 & 33.88 & - & - & - & - & 1.80 & 34.13 \\
    \riso & 0.77 & 1.99 & 33.92 & 4.92 & 28.77 & 3.22 & 31.67 & 1.88 & 34.05\\
    \midrule
    \multicolumn{10}{c}{Literature Fits}\\
    \hline
    \hline
    Radius &  & $\alpha$ & $\log(A)$ & & \multicolumn{2}{c}{Label} &  \multicolumn{3}{c}{Reference} \\
    \midrule
    Core/Isophotal & & 2.72 & 31.99 &  & \multicolumn{2}{c}{W12} & \multicolumn{3}{c}{\citet{wisnioski2012}} \\
     Core/Isophotal & & 2.45 & 33.30 &  & \multicolumn{2}{c}{All $z\approx0$ C18} & \multicolumn{3}{c}{\citet{cosens2018}} \\
    Core/Isophotal & & 2.66 & 32.92 &  & \multicolumn{2}{c}{Low $\Sigma_{\text{SFR}}$ C18} & \multicolumn{3}{c}{\citet{cosens2018}} \\
     Core/Isophotal & & 1.48 & 37.53 &  & \multicolumn{2}{c}{High $\Sigma_{\text{SFR}}$ C18} & \multicolumn{3}{c}{\citet{cosens2018}} \\
    Core & & 2.04 & 34.69 &  & \multicolumn{2}{c}{G23} & \multicolumn{3}{c}{\citet{giunchi2023}} \\
    \bottomrule
    \label{tab:RLcorrelations}
    \end{tabular}
    \end{center}
    \tablecomments{Spearman rank correlation coefficient $\rho$ and best fit power law index $\alpha$ for each radius-luminosity relation in Figure~\ref{fig:RL}. The correlation coefficients are calculated on both the full catalog and the binned data points, excluding the faintest (1) and brightest (10) bins. We record the values binned by \ha peak intensity (I$_{\text{\ha}}$), \ha luminosity (L$_{\text{\ha}}$), and radius (R). In the cases where $\alpha$ is physically unreasonable (e.g., negative or $>10$), we do not quote a best fit powerlaw index. The power law fits for a comparative sample of the literature are also provided.}
\end{table*}

%% file: tab_sizesize.tex
\begin{table}
    \begin{center}
    \caption{Radius-Radius correlations.}
    \begin{tabular}{ccccccc}
    \toprule 
        \multicolumn{7}{c}{Rank correlation between quantities}\\
    \midrule
 & $R_{\rm mom}$ & $R_{50}$ & $R_{90}$ & $R_{\rm iso}$ & $R_{\rm lum}$ & $L_{\rm H\alpha}$ \\
HWHM & 0.60 & 0.54 & 0.36 & 0.31 & 0.29 & 0.19 \\
$R_{\rm mom}$ &  & 0.99 & 0.90 & 0.60 & 0.73 & 0.39 \\
$R_{50}$ &  &  & 0.87 & 0.58 & 0.68 & 0.38 \\
$R_{90}$ &  &  &  & 0.62 & 0.88 & 0.43 \\
$R_{\rm iso}$ &  &  &  &  & 0.60 & 0.89 \\
$R_{\rm lum}$ &  &  &  &  &  & 0.45 \\
\bottomrule
    \label{tab:sizesize}
    \end{tabular}
    \end{center}
    \tablecomments{Spearman rank correlation coefficient $\rho$ relating different size metrics (Section \ref{sec:radius}) and luminosity.}
\end{table}

%% file: fig-RLcomparisoniso.tex
\begin{figure*}
    \centering
    \includegraphics[width=\linewidth]{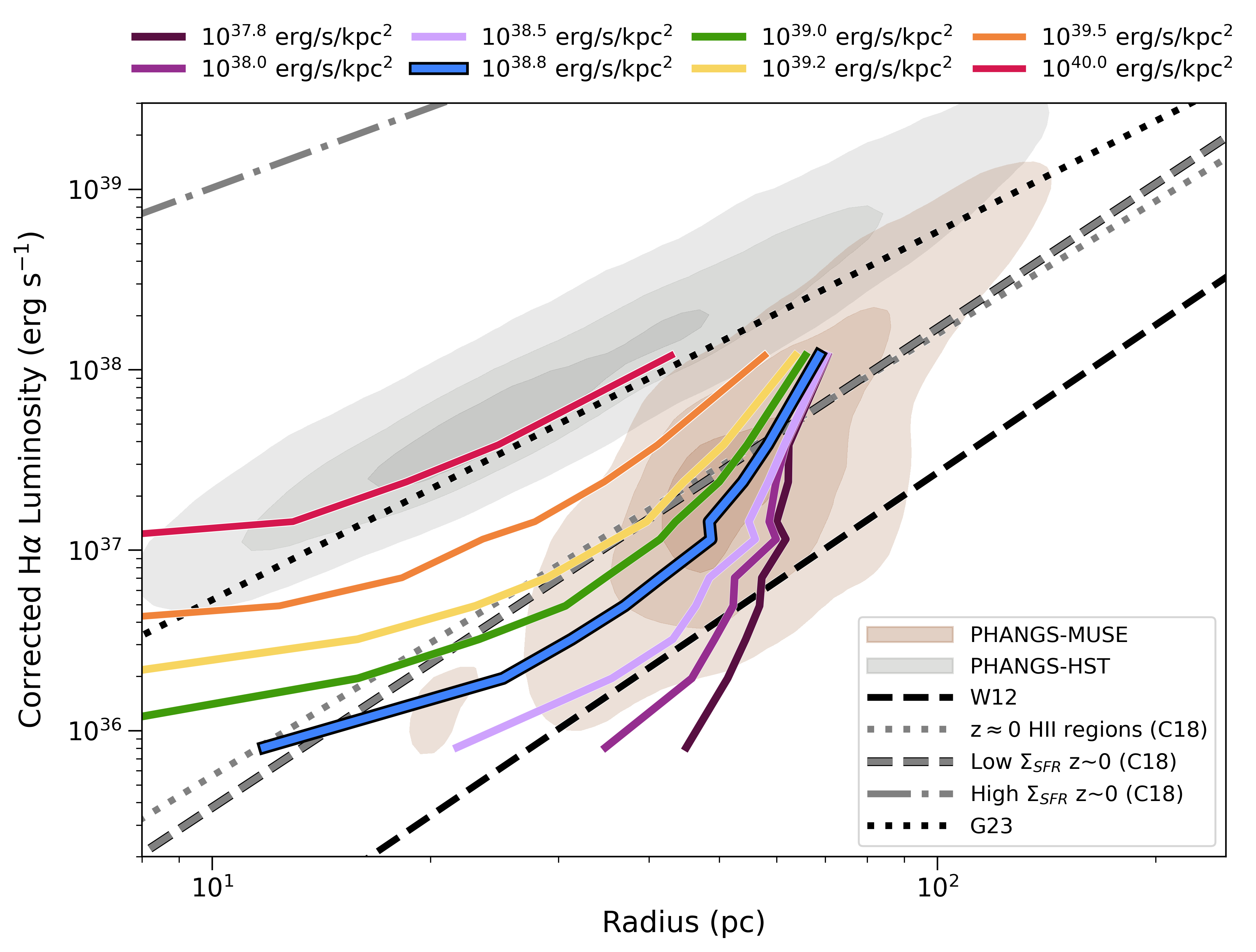}
    \caption{\hii\ region luminosity-radius relation for our isophotal radius and literature measurements. Contours show data density for the PHANGS-MUSE \citep[brown,][]{groves2023} and PHANGS-HST (gray, A. Barnes et al. submitted) catalogs. Solid colored lines represent the $L$-\riso relation that we measure in NGC\,253 for varying isophotes with the thick blue line showing our fiducial value of $10^{38.8}$ erg~s$^{-1}$~kpc$^{-2}$. Lines show luminosity-radius relations measured for local $z\approx0$ \citep[][(C18, G23)]{cosens2018,giunchi2023} and $z=0-2$ \hii\ regions \citep[][(W12)]{wisnioski2012}. Our fiducial results match the approximate slope of the literature relations. By changing the threshold isophote used to calculate the sizes, we can adjust the NGC\,253 relation to match most of the literature studies (NGC\,253 does not reach high enough intensities to match the \citet{cosens2018} high $\Sigma_{\text{SFR}}$ relation). The same impact of changing threshold can depend on resolution, which can be seen contrasting PHANGS-HST (sharper resolution, high isophotal threshold) and PHANGS-MUSE (coarser resolution, lower intensity region boundaries).   
    }
    \label{fig:RLcomparisoniso}
\end{figure*}

%% file: fig-profilefits.tex
\begin{figure}
    \centering
    \includegraphics[width=\linewidth]{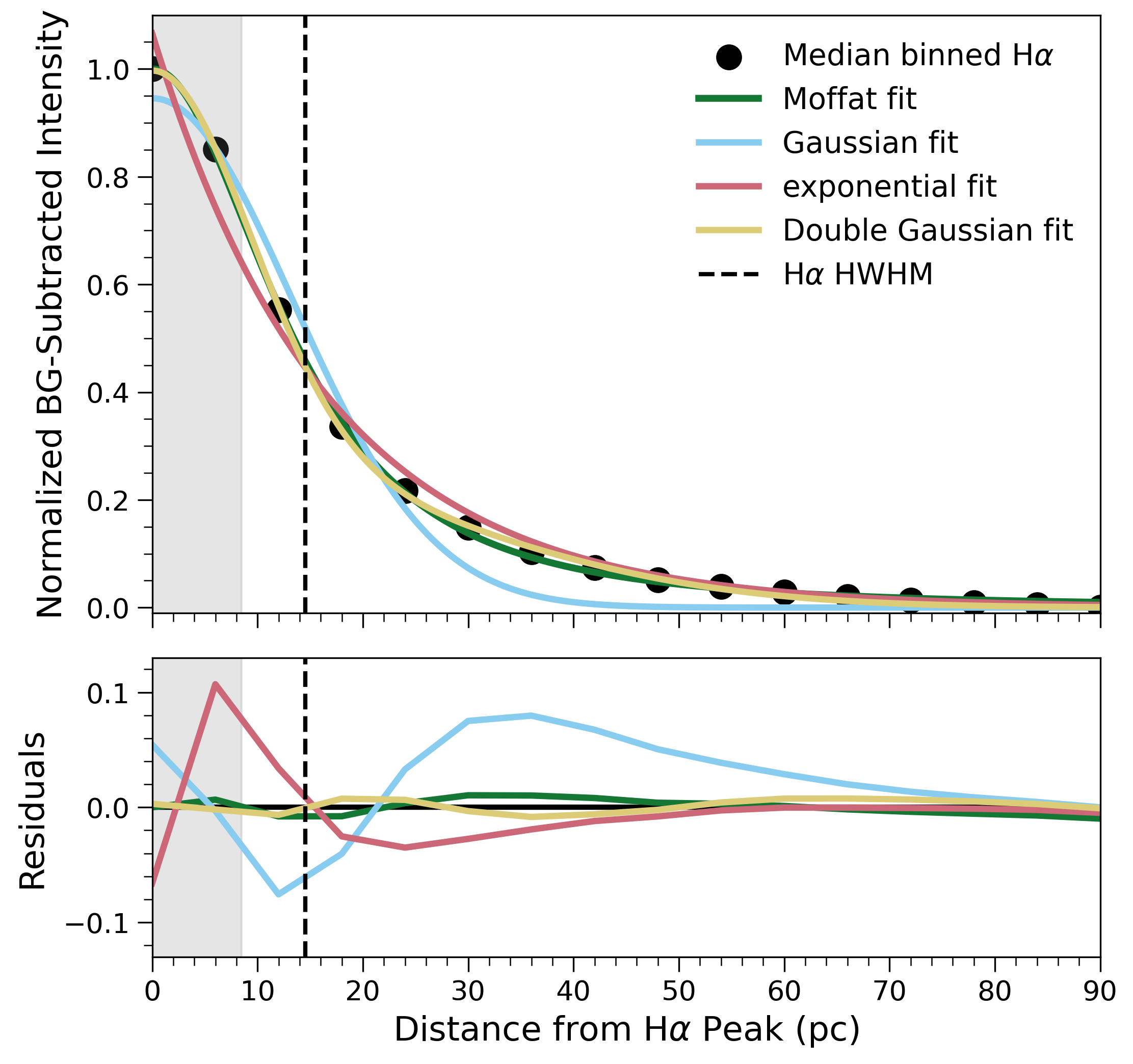}
    \caption{Top: Median \ha profile for all \hii\ regions fit with several functional forms (Table~\ref{tab:fitcoeffs}). The HWHM of the PSF is shaded in gray and the median HWHM of all \hii\ regions is marked as a vertical line. Bottom: Residuals for each model. The double Gaussian function fits the data best at all radii.}
    \label{fig:profile-fits}
\end{figure}

%% file: tab-fitcoeffs.tex
\begin{table}
\begin{center}
    \caption{Functional fits to sample-wide stacked \ha profile.}
    \begin{tabular}{c||cccc}
    \toprule
     & Moffat & Gauss. & Exp. & Double Gauss. \\
     \hline\hline
    HWHM$_1$ & & 15.63 & - & 10.81\\
    $A_1$ & 1.00 & 1.00 & 1.07 & 0.71\\
    HWHM$_2$ & - & - & - & 31.03\\
    $A_2$ & - & - & - & 0.28\\
    $\alpha$ & 16.34 & - & 16.66 & - \\
    $\beta$ & 1.34 & - & - & - \\
    \bottomrule
    \label{tab:fitcoeffs}
    \end{tabular}
    \end{center}
    \tablecomments{Coefficients for functional fits to the normalized median \ha radial profile averaged over the whole sample (shown in Figure~\ref{fig:profile-fits}). Formulae --- Gaussian: $A_1 \exp(-x^2/2\sigma_1^2)$; Exponential: $A_1 \exp (-x/\alpha)$; Double Gaussian: $A_1 \exp(-x^2/2\sigma_1^2)+A_2 \exp(-x^2/2\sigma_2^2)$; Moffat as Equation \ref{eq:moffat} with $x_0 = 0$. $\sigma=2\times\text{HWHM}/2.355$, and all scale widths are in units of parsecs.} 
\end{table}

%% file: fig-models.tex
\begin{figure*}
    \centering
    \begin{tabularx}{\linewidth}{cc}
    \includegraphics[width=0.51\linewidth]{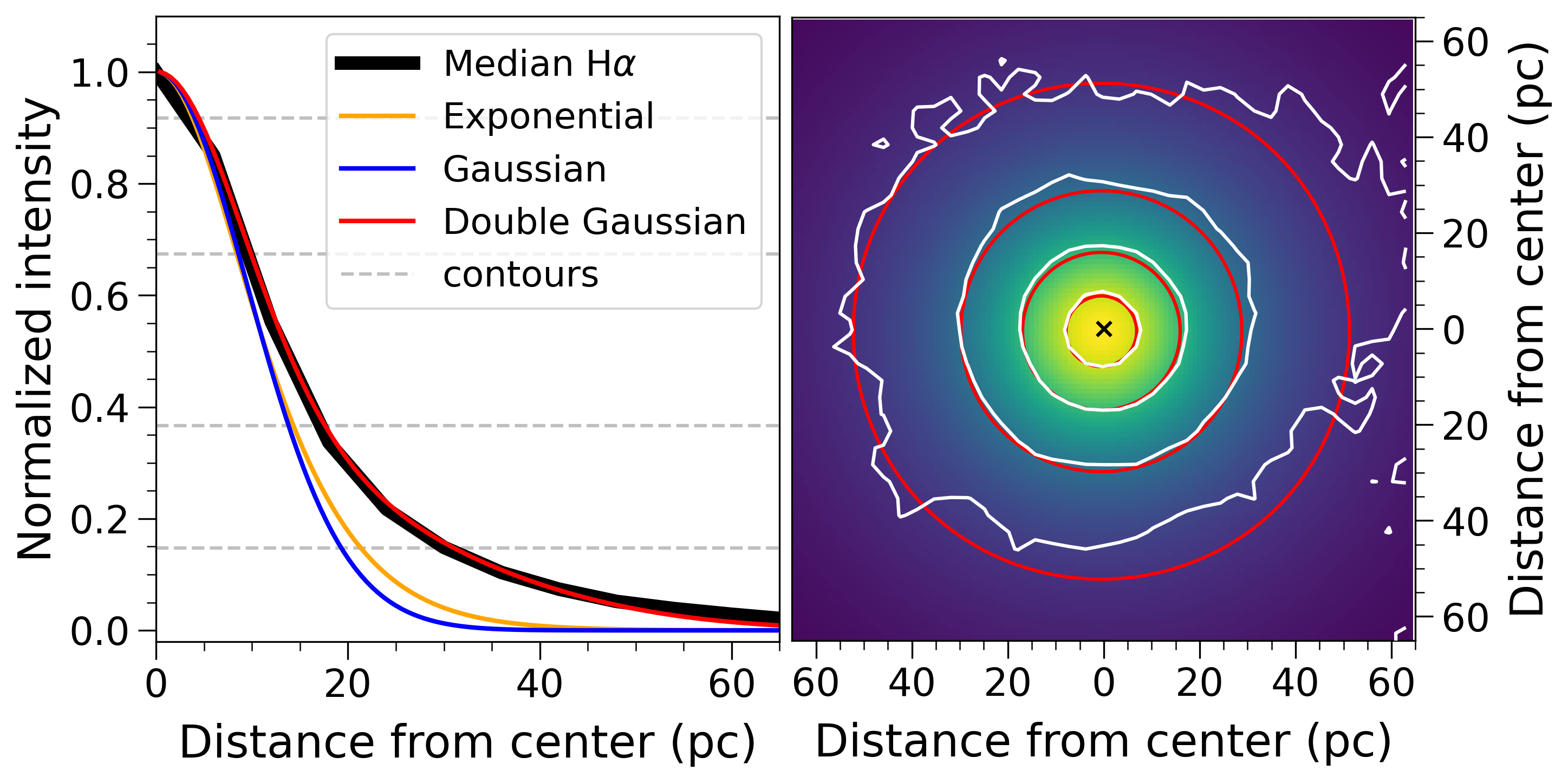} &
    \includegraphics[width=0.45\linewidth]{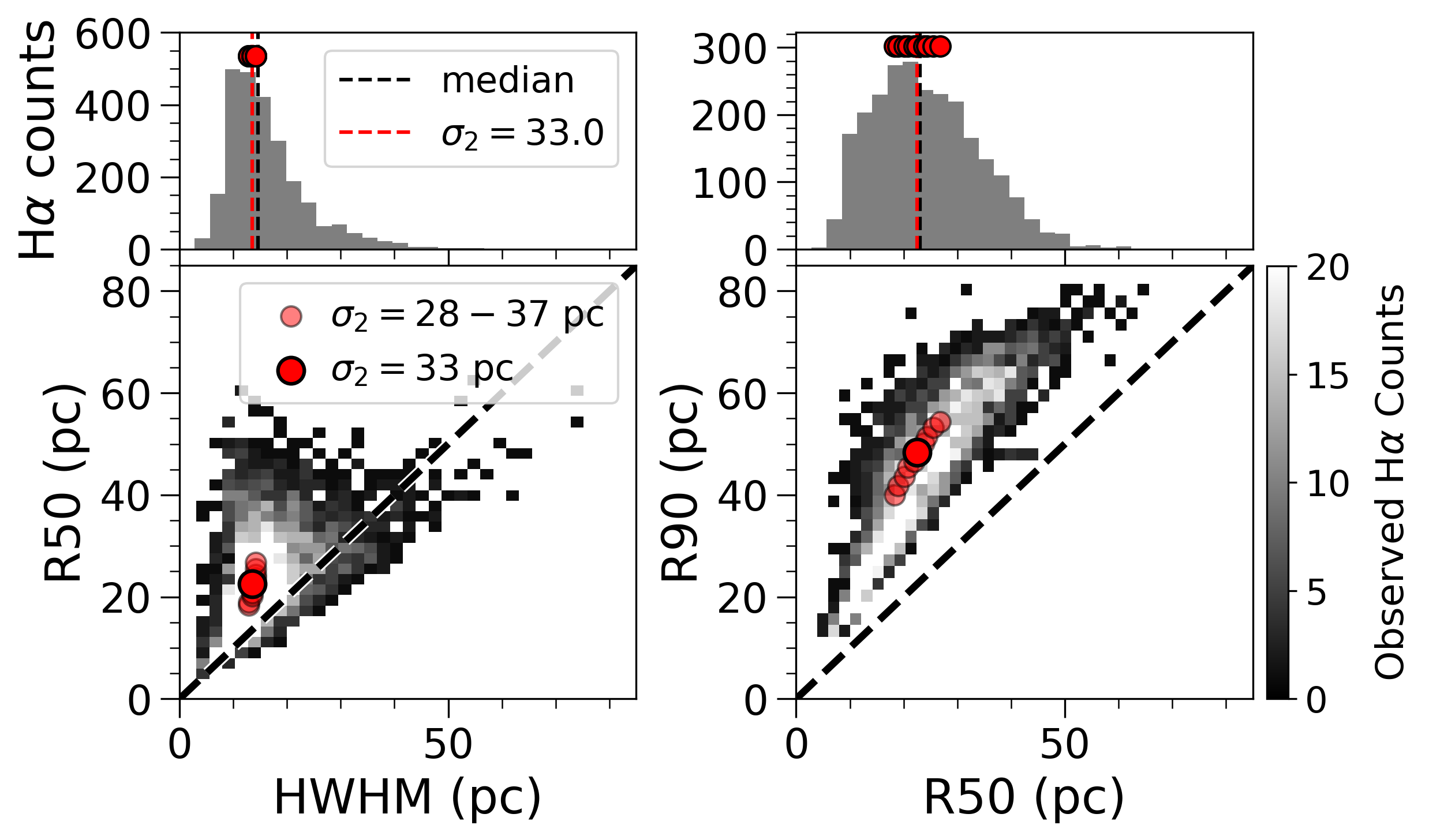}\\
    \end{tabularx}
    \caption{Predicted emission based on model density profiles. Left: The \ha profile implied from the modeled double Gaussian density profile (red) is a close match to the observed median \ha profile (black, left panel) and median stacked \ha image. White and red contours represent the 99th, 95th, 84th, and 50th percentile. The best fit exponential and single Gaussian functions are also shown in orange and blue respectively and fit the observations less well.
    Right: Plots comparing different radius metrics, i.e., effectively showing concentration indices for our measured regions (gray data density plots) and model \ha profiles assuming a double Gaussian density distribution with a compact inner core and an outer Gaussian width varying between 28 and 37~pc (red points) with the best-fit model, which has $\sigma_2=33$~pc, marked. The models match the distributions of the observed data well.}
    \label{fig:model}
\end{figure*}

%% file: fig-siiratio.tex
\begin{figure}
    \centering
    \includegraphics[width=\linewidth]{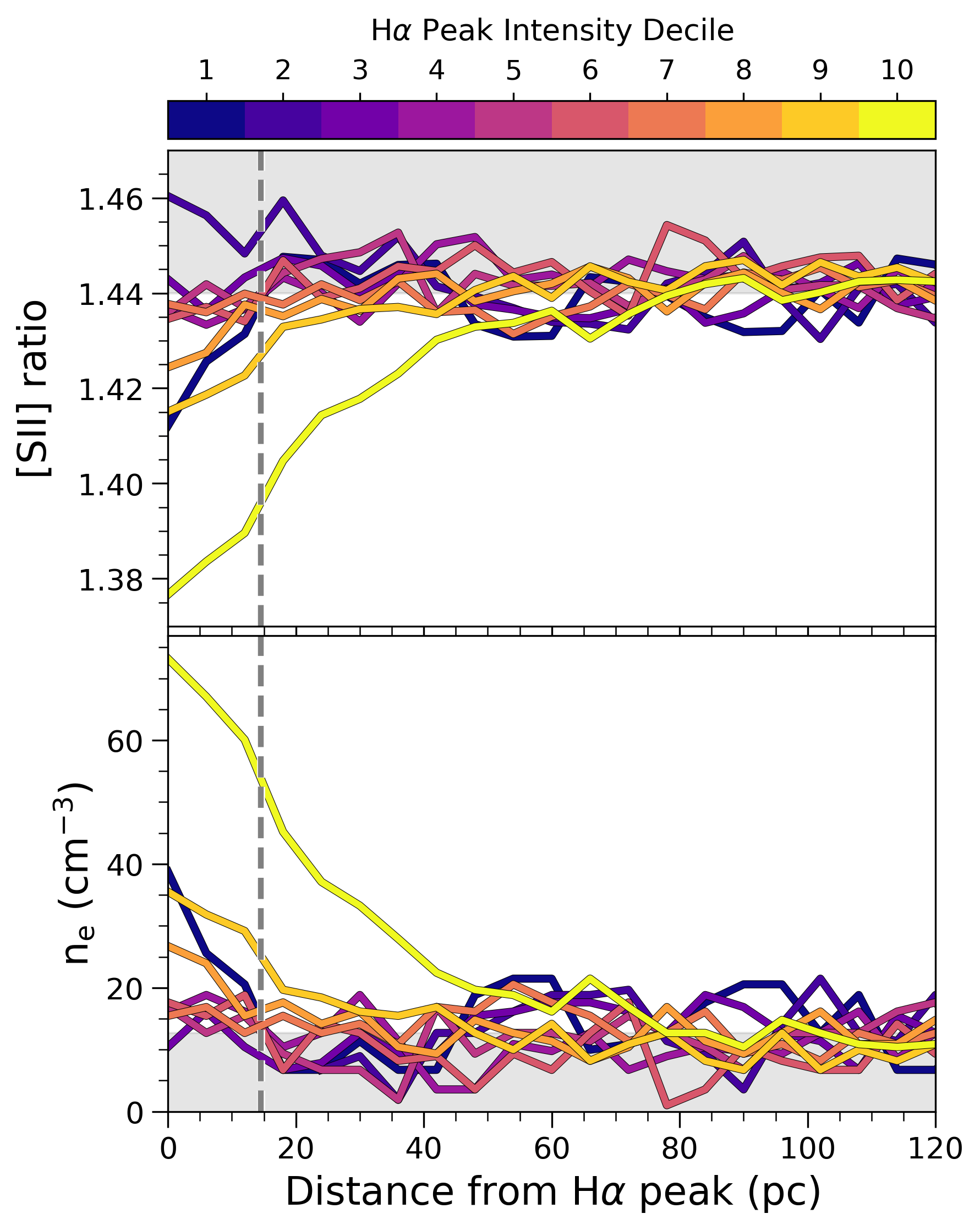}
    \caption{Radial profiles of [\sii] flux ratio (top) and electron density (bottom) for each decile of \ha peak intensity. The vertical black line marks the median HWHM of all \hii\ regions and the shaded gray regions marks the low density regime where the [\sii] ratio only provides an upper limit on the density.}
    \label{fig:sii-ne}
\end{figure}

%% file: tab-siidensity.tex
\begin{table}
    \begin{center}
    \caption{Best-fit [\sii]-derived density profiles.}
    \begin{tabular}{cccc}
        \toprule
        Bin & \shortstack{Peak [\sii]\\ ratio} & \shortstack{Peak density\\ (cm$^{-3}$)} & \shortstack{$\sigma$\\ (pc)}  \\
        \midrule
        8 & 1.4244 & $26.21\pm2.28$ & $9.17\pm2.09$\\
        9 & 1.4158 & $34.54\pm2.05$ & $14.30\pm1.72$\\
        10 & 1.3814 & $68.58\pm2.39$ & $20.99\pm1.12$\\
        \bottomrule
    \label{tab:density-fits}
    \end{tabular}
    \end{center}
    \tablecomments{Best-fit parameters for Gaussian fits to the density profiles derived from the [\sii] ratio (T$_{\text{e}}=10^4$ K) of the brightest \textbf{three} bins of \hii\ regions. The fits to the fainter regions are not well constrained.}
\end{table}

%% file: fig-densityluminosity.tex
\begin{figure*}
    \centering
    \includegraphics[width=\linewidth]{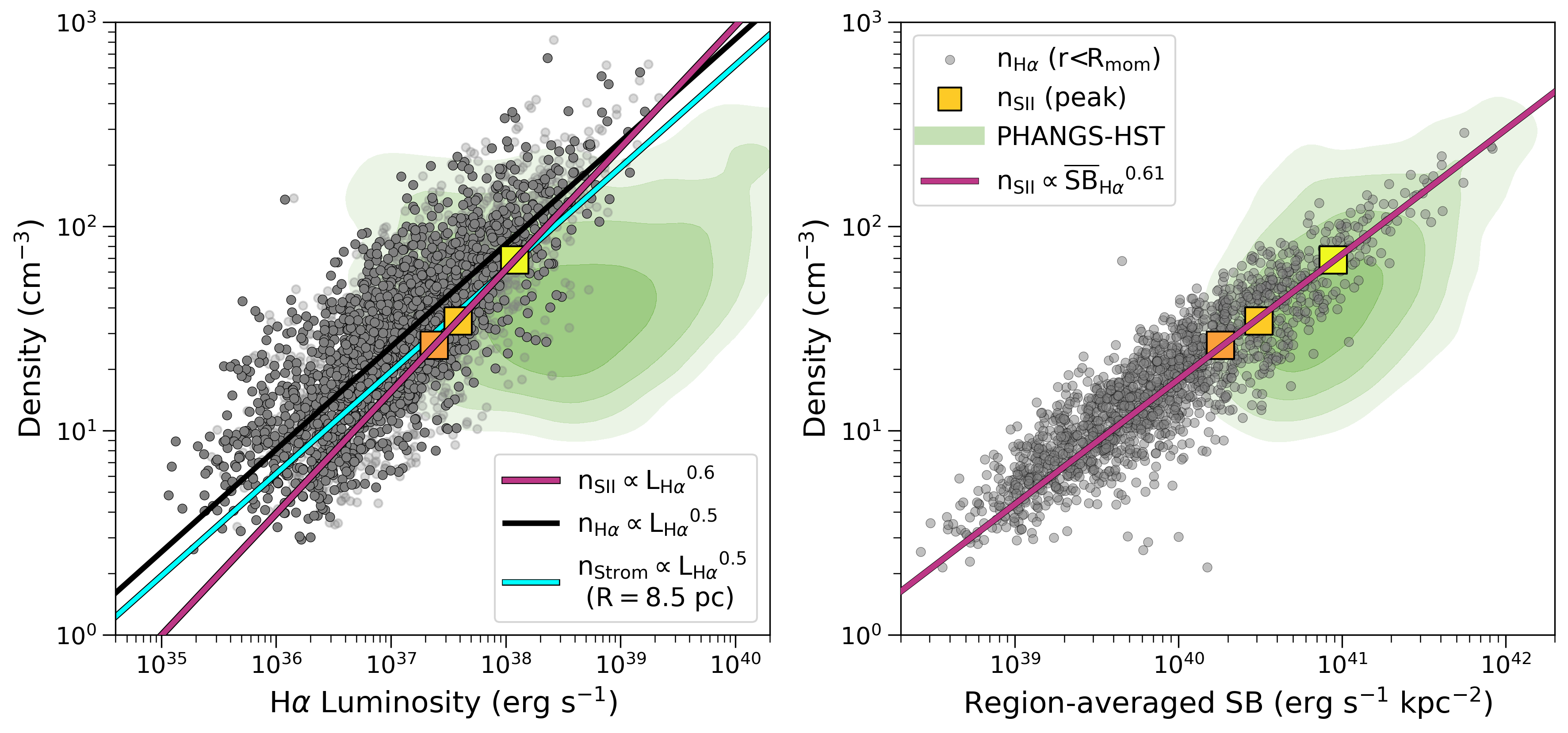}
    \caption{Left: Density-luminosity relation for the two methods of density determination. The large colored points represent the densities calculated from the [\sii] ratio for the brightest four deciles (n$_{\text{SII}}$). The solid gray points represent the \ha peak intensity converted to density through emission measure of \ha for each \hii\ region (n$_{\text{\ha}}$). A powerlaw is fit to both sets (only for points with L$_{\text{\ha}}>10^{36}$~erg~s$^{-1}$), showing that the [\sii] ratio yields a steeper density-luminosity relation than \ha. The predicted relation for a classical \stromgren sphere with radius R$=8.5$~pc (cyan line, PSF HWHM) well-matches the observations. Dark gray points include background subtraction, while the lighter, faded gray points in the background represent the \ha intensity-to-density conversion without background subtraction. Background subtraction has a negligible effect on the overall density-luminosity distribution. Right: We see no similar n-L relation in the PHANGS-HST sample. Instead we see strong agreement in the relation between density and the region-averaged surface brightness ($\overline{\text{SB}}_{\text{\ha}}$) as determined by the region luminosity and \rmom. Both the structural (gray points, average density within \rmom) and spectroscopic density measurements for NGC\,253 (colored squares, peak density) agree with the full sample of PHANGS galaxies without background subtraction in all cases.}
    \label{fig:nL}
\end{figure*}

%% file: fig-massion.tex
\begin{figure}
    \centering
    \includegraphics[width=\linewidth]{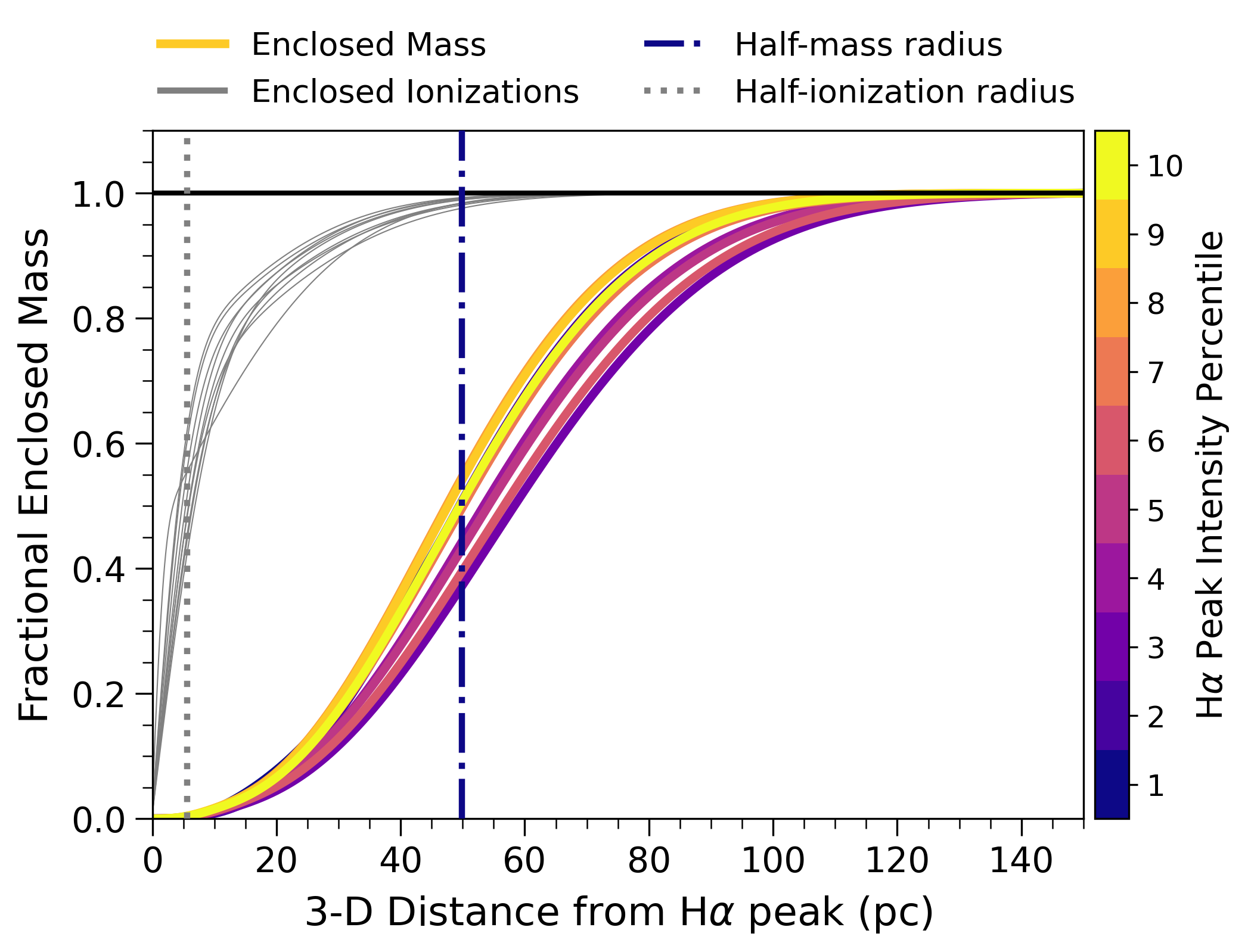}
    \caption{Color: Enclosed ionized gas mass profile of \hii\ region deciles assuming the double Gaussian density profile in Table~\ref{tab:binprops} and a smooth density distribution. Gray: Cumulative enclosed ionizations per second assuming the number of ionizations trace the luminosity density of \ha . The profiles are normalized at 150~pc where they level off. Vertical lines mark the median radius at which half of the mass (dark blue dash-dotted line) and ionizations (gray dotted line) are enclosed. 50\% of the ionizations occur within one resolution element of MUSE, therefore we cannot resolve the majority of the flux production.}
    \label{fig:mass}
\end{figure}

%% file: acknowledgements.tex
Based on observations collected at the European Organisation for Astronomical Research in the Southern Hemisphere under ESO programmes 108.2289 (PI: Congiu) and 0102.B-0078(A) (PI: Zschaechner).

This work was carried out as part of the PHANGS collaboration.

RLM is supported by the NSF GRFP. 
RLM, AKL, and RC gratefully acknowledge support from NSF AST AWD 2205628, JWST-GO-02987.001-A, and JWST-GO-03707.001-A. 
AKL also gratefully acknowledges support by a Humbolt Research Award. 
FB acknowledges support from the INAF Fundamental Astrophysics program 2022. 
ER acknowledges the support of the Natural Sciences and Engineering Research Council of Canada (NSERC), funding reference number RGPIN-2022-03499.
MB acknowledges support from the ANID BASAL project FB210003. This work was supported by the French government through the France 2030 investment plan managed by the National Research Agency (ANR), as part of the Initiative of Excellence of Université Côte d’Azur under reference number ANR-15-IDEX-01.
JEMD acknowledges support from project UNAM DGAPA-PAPIIT IG 101025, Mexico.
HAP acknowledges support from the National Science and Technology Council of Taiwan under grant 113-2112-M-032-014-MY3.
OE acknowledges funding from the Deutsche Forschungsgemeinschaft (DFG, German Research Foundation) -- project-ID 541068876.
KG is supported by the Australian Research Council through the Discovery Early Career Researcher Award (DECRA) Fellowship (project number DE220100766) funded by the Australian Government.
SD is grateful for the support provided by NASA through the NASA Hubble Fellowship grant HST-HF2-51551.001-A awarded by the Space Telescope Science Institute, which is operated by the Association of Universities for Research in Astronomy, Incorporated, under NASA contract NAS5-26555.

This research has made use of the Astrophysics Data System, funded by NASA under Cooperative Agreement 80NSSC21M00561.
This research made use of Astropy, a community-developed core Python package for Astronomy \citep{Astropy13, Astropy18}, Matplotlib \citep{Hunter07}, NumPy \citep{Harris20}, Pyneb \citep{luridiana2015}, SciPy \citep{Virtanen20}.

%% file: tab-catalog-columns.tex
\begin{table}
    \begin{center}
    \caption{Region catalog columns}
    \begin{tabular}{ccl}
    \toprule
        Column Name & Units & Description \\
    \hline\hline
        Region ID & - & Region identifier \\
        $I_{\text{peak}}$ Bin & - & Peak intensity decile for this region \\
        RA & degrees & Right ascension of \ha\ peak \\
        Dec & degrees & Declination of \ha\ peak \\
        $R_{\rm gal}$ & kpc & Galactocentric radius in the plane of the galaxy \\
        PSF & arcseconds & Estimated PSF of the working \ha\ map at this location \\
        $A_{\rm H\alpha}$ & mag & Extinction at the H$\alpha$ peak \\
        $I_{\text{peak}}$ & erg~s$^{-1}$~kpc${-2}$ & Extinction corrected background subtracted peak intensity \\
        $I_{\text{bg}}$ & erg~s$^{-1}$~kpc${-2}$ & Extinction corrected intensity of subtracted local background \\
        HWHM & pc & Half-width half maximum size \\
        \rmom & pc & Intensity-weighted root mean square size \\
        \rf & pc & Effective radius encompassing 50\% of the luminosity \\
        \rn & pc & Effective radius encompassing 90\% of the luminosity \\
        \riso & pc & Isophotal radius for fiducial isophote \\
        \rlum & pc & Radius used to define background aperture and luminosity integral \\
        $L_{\text{H}\alpha}$ & erg~s$^{-1}$ & H$\alpha$ luminosity not corrected for extinction, background subtracted  \\
        $L_{\text{H}\alpha\text{,corr}}$ & erg~s$^{-1}$ & H$\alpha$ luminosity corrected for extinction, background subtracted \\
        $L_{\text{H}\alpha +\text{BG}}$ & erg~s$^{-1}$ & H$\alpha$ luminosity not corrected for extinction, without background subtraction  \\
        $L_{\text{H}\alpha\text{,corr}+\text{BG}}$ & erg~s$^{-1}$ & H$\alpha$ luminosity corrected for extinction, without background subtraction \\
        HWHM$_{1,\rm fit}$ & pc & HWHM of the best-fit inner Gaussian\\
        HWHM$_{2,\rm fit}$ & pc & HWHM of the best-fit outer Gaussian\\
        $A_{\text{ratio}}$ & - & Amplitude ratio of the inner and outer Gaussian\\
        Flag$_{\rm bg}$ & - & True when \hii\ region is isolated with clean background after masking\\
    \bottomrule
    \label{tab:catalogcolumns}
    \end{tabular}  
    \end{center}
\end{table}

%% file: fig-bpt.tex
\begin{figure*}
    \centering
    \includegraphics[width=\linewidth]{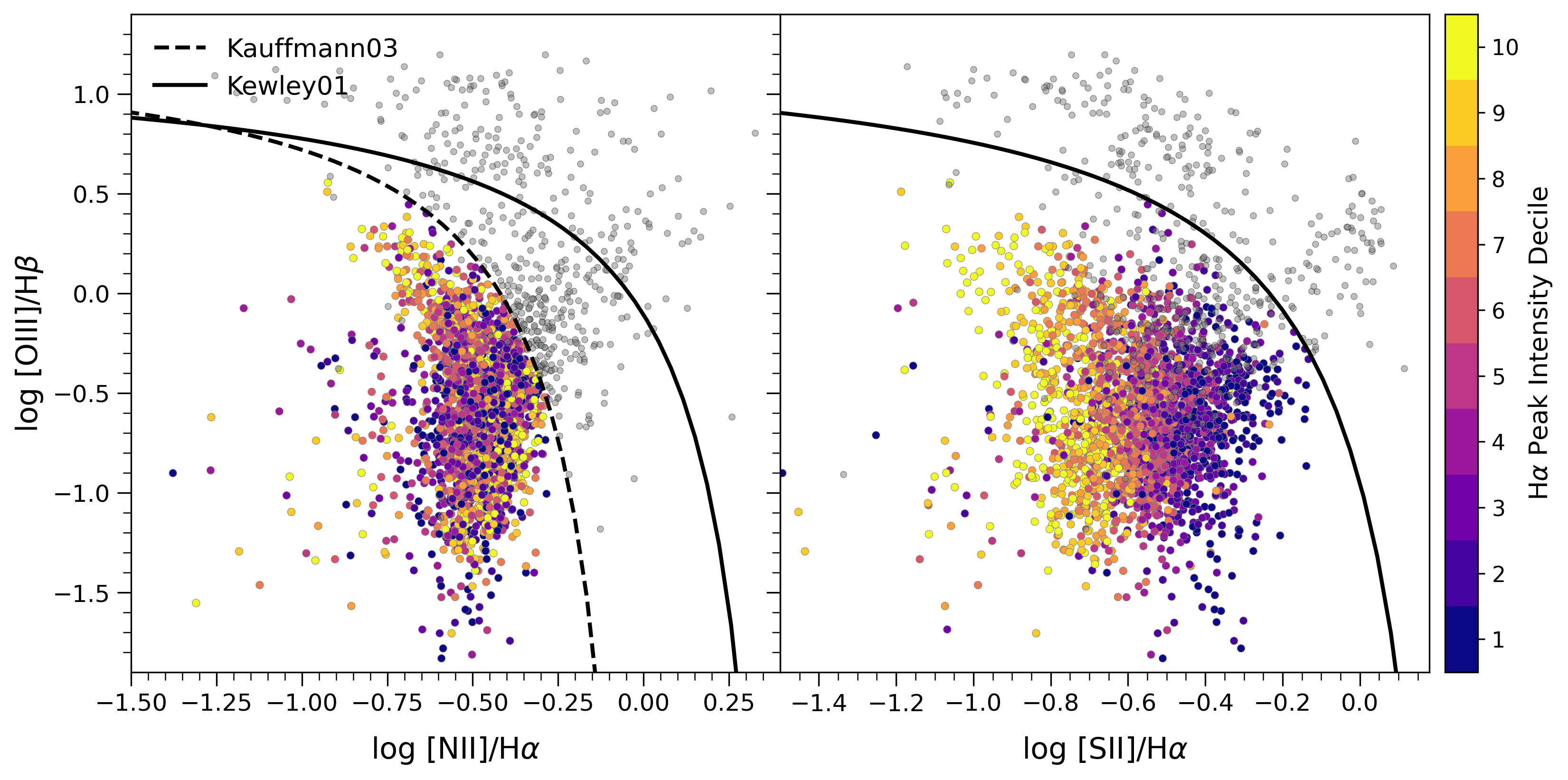}
    \caption{N2-BPT (left) and S2-BPT (right) diagrams for \ha peaks. Colored points show our sample of \hii\ regions, colored by their extinction-corrected background-subtracted \ha\ peak intensity bin (\S \ref{sec:intbin}). Gray points show regions identified as \ha peaks, but excluded from our catalog because they fall above the \citet{kauffmann2003} (dashed black, left panel) and/or \citet{kewley2001} lines (solid black, both panels).}
    \label{fig:bpt}
\end{figure*}

%% file: fig-completenessscatter.tex
\begin{figure*}
    \centering
    \includegraphics[width=0.95\linewidth]{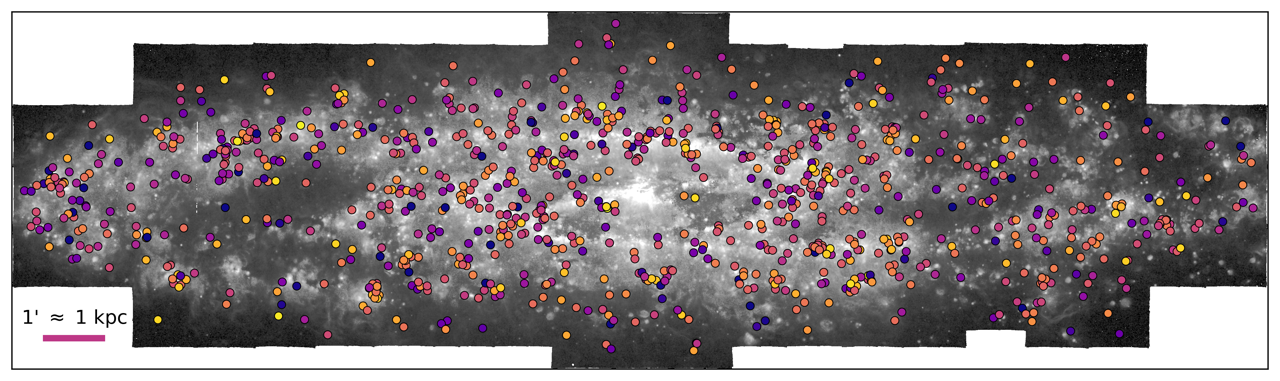}
    \includegraphics[width=0.55\linewidth]{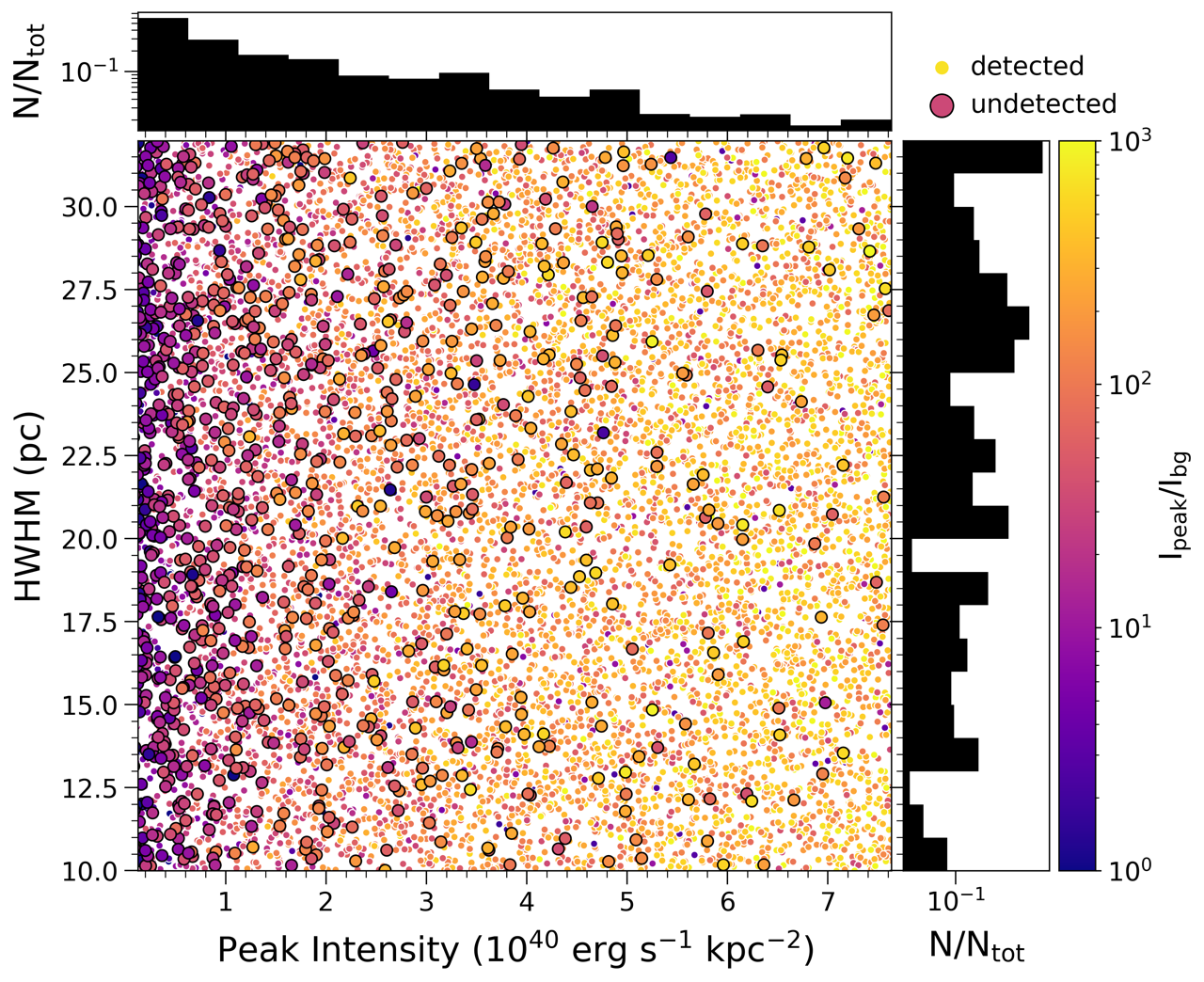}
    \caption{Top: The locations of undetected false sources, colored by their contrast above the background. The regions that we fail to detect tend to have low contrast in bright areas of the galaxy. Bottom: Parameters of detected (small colored points) and undetected (large points with black outline) false sources colored by their contrast above the local background. Histograms show the fraction of undetected sources as a function of region size ($y$-axis) and peak intensity ($x$-axis). Faint regions are the most difficult to detect, and we miss $\sim10$\% of regions with peak intensity below $10^{40.3}$ erg~s$^{-1}$~kpc$^{-2}$. Large regions are also somewhat harder to detect than smaller regions. Contrast with the local background plays an important role, with faint regions in regions with high background the most difficult to detect. }
    \label{fig:completeness}
\end{figure*}

%% file: fig-pneresolutions.tex
\begin{figure}
    \centering
    \includegraphics[width=0.5\linewidth]{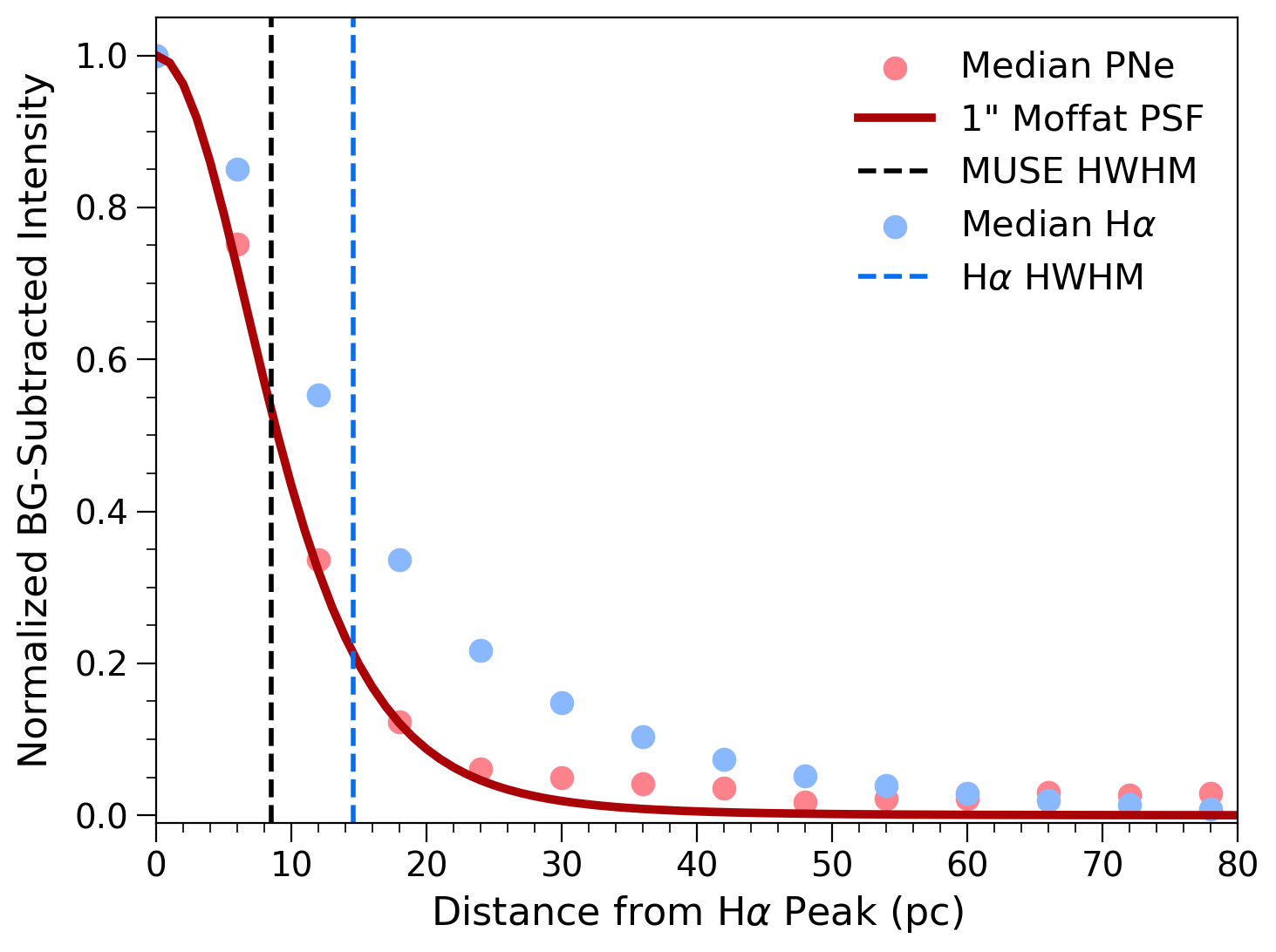}
    \caption{Background-subtracted and normalized median radial profile of all planetary nebulae (red) and \hii\ regions (blue) across the disk of NGC\,253. The Moffat function describing the typical PSF measured by \citet{congiu2025} is shown with a solid red line. This matches the median PN profile, validating our profiling method and the PSF estimate. The median HWHM of the \ha profiles is larger than the PSF validating our treatment of the \hii\ regions as resolved sources.}
    \label{fig:pn-profile}
\end{figure}

%% file: fig-meanvsmed.tex
\begin{figure*}
    \centering
    \includegraphics[width=\linewidth]{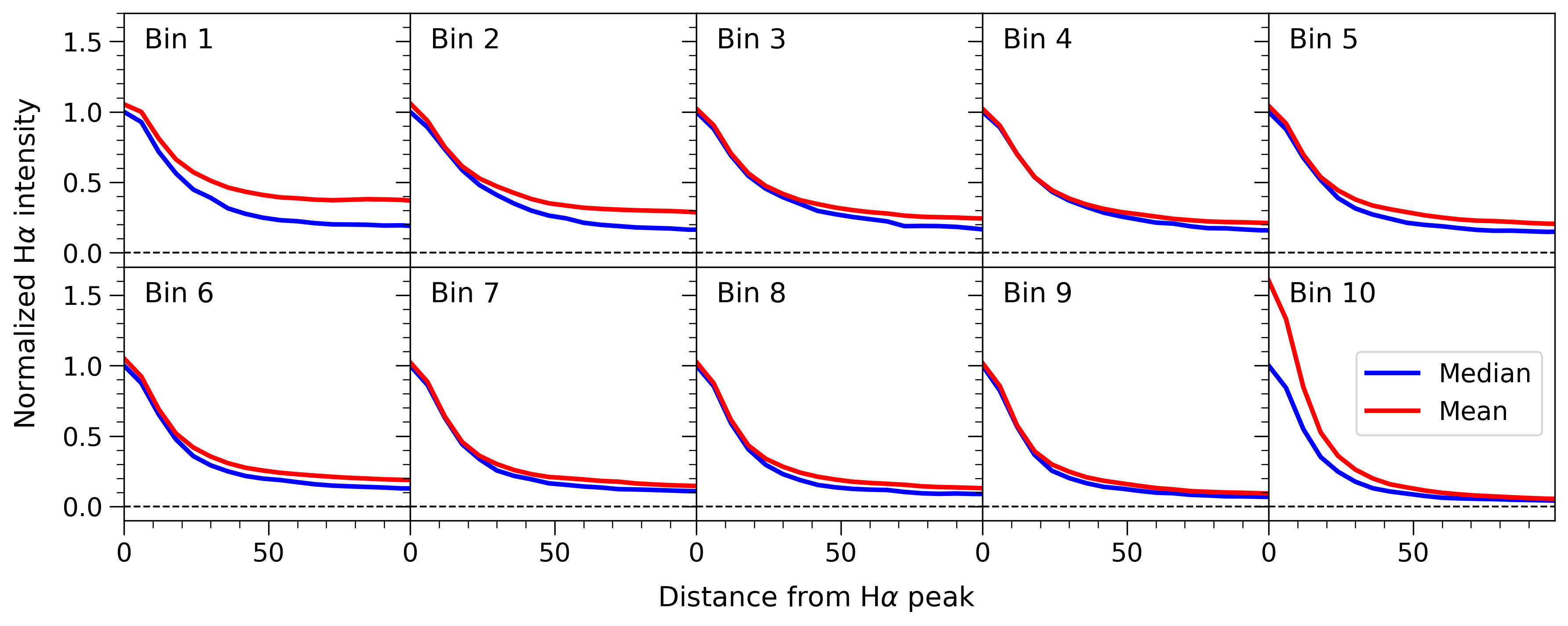}
    \caption{Median (blue) and mean (red) profiles of all \hii\ regions in each bin of \ha peak intensity. We use the median of the radial bins for the individual profile, and take the corresponding statistic of the stack of profiles in each bin. Then both profiles are normalized to the peak value of the median profile.}
    \label{fig:meanvsmed-bins}
\end{figure*}

%% file: fig-RLbackground.tex
\begin{figure}
    \centering
    \includegraphics[width=\linewidth]{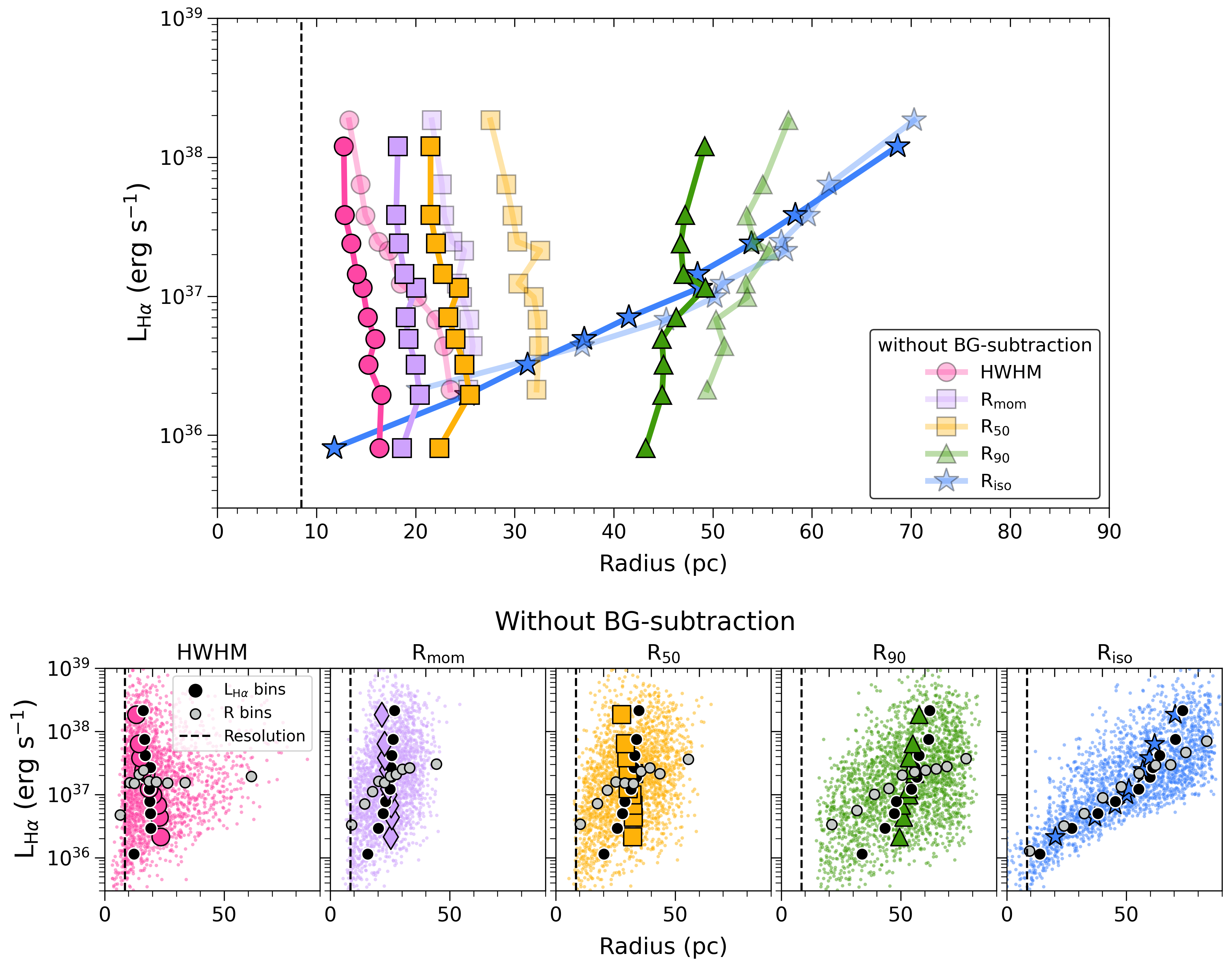}
    \caption{Top: Luminosity-radius relation for the median profiles binned by peak intensity with (solid) and without (faded) background subtraction. Bottom: Full distribution of all regions without background subtraction. The binning methods are the same as Figure~\ref{fig:RL}. Without background subtraction, the luminosity of the faintest bins increases and the regions get larger, but the overall trends remain the same. In both methods, the radius is measured within the same \rlum and the maximum radius allowed is 100~pc. See Table~\ref{tab:RLcorrelations} for a comparison of fits with and without background subtraction.}
    \label{fig:RLbackground}
\end{figure}

%% file: fig-profilesres.tex
\begin{figure*}
    \centering
    \includegraphics[width=\linewidth]{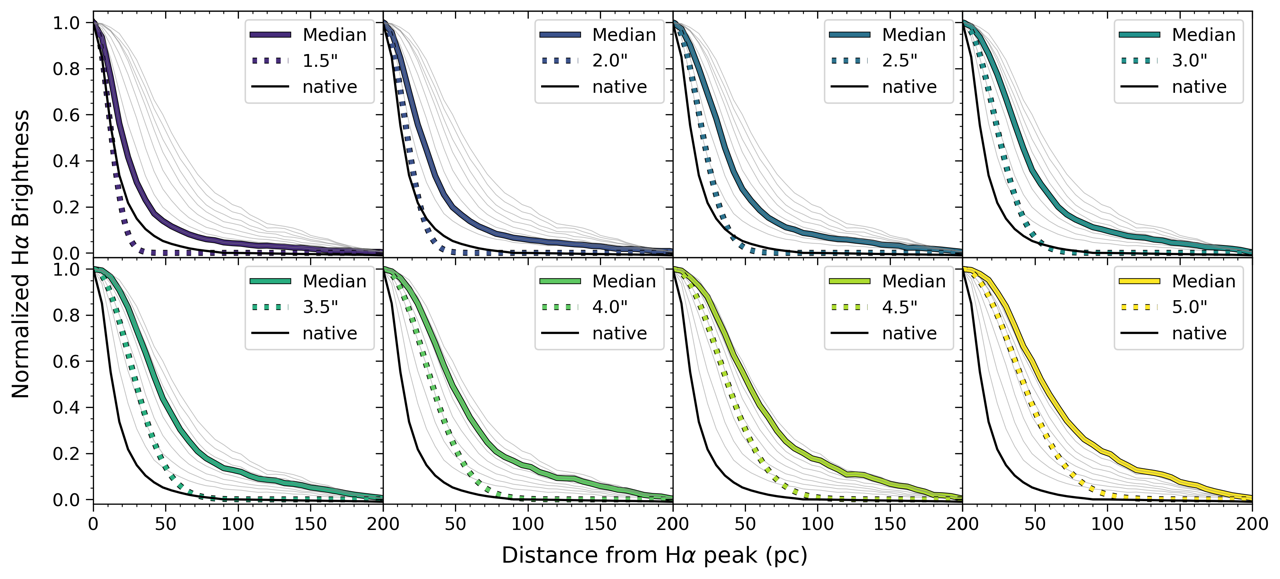}
    \includegraphics[width=\linewidth]{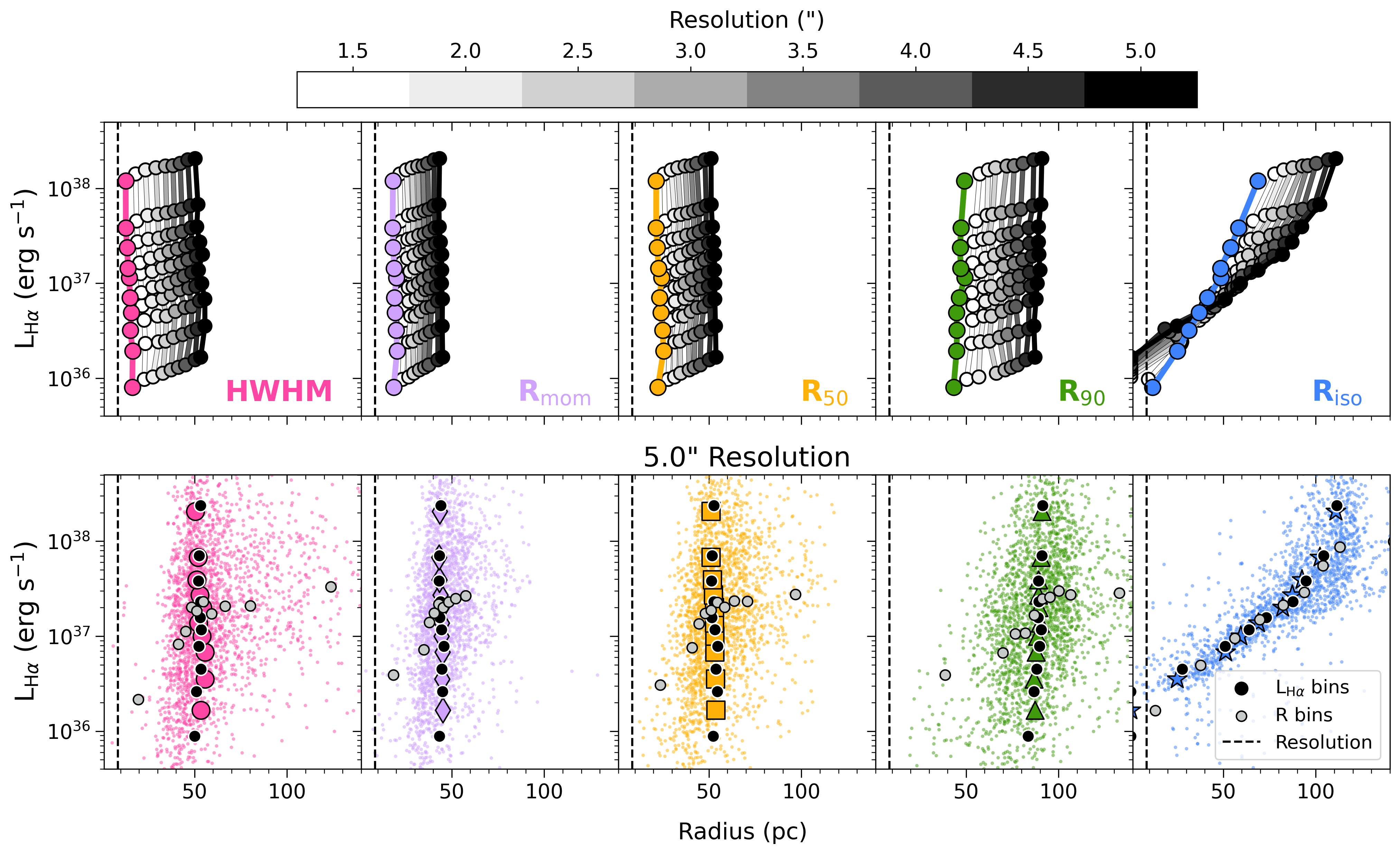}
    \caption{Effect of resolution on our measurements. Top panels: Radial profiles of the full \hii\ region catalog at varying resolutions. The median profile is shown with a solid colored line and the corresponding Gaussian PSF is denoted with a dotted colored line. The native 1'' resolution \ha profile is shown with a solid black line in all panels. Middle panels: Radius-luminosity trends for different radius measurements as the angular resolution degrades to the PHANGS-MUSE physical resolution ($\sim$85~pc). The colored points represent the \ha peak intensity bins for the native resolution map while the gray-scale points are convolved with a Gaussian kernel set by the resolution. Bottom panels: The full distribution of luminosity and radius for the region catalog convolved to 5" resolution. The different binning methods are the same as Figure~\ref{fig:RL}. We see similar trends to the native resolution for all radius measurements and binning methods, but the slope of the \riso-$L$ trend decreases with decreasing resolution. The determined region size scales up with resolution, overestimating the true region size.}
    \label{fig:convolved-profiles}    
\end{figure*}

%% file: tab-respowers.tex
\begin{table*}
    \begin{center}
    \caption{\textbf{Profile  properties as a function of resolution}}
    \begin{tabular}{ccccccccccccc}
        \toprule
        \multicolumn{13}{c}{\textbf{Double Gaussian fits}}\\
        \hline
        \hline
        \multicolumn{2}{c}{\textbf{Resolution}} & \textbf{A$_{\mathbf{1}}$} & \multicolumn{2}{c}{\textbf{HWHM$_{\mathbf{1}}$}} & \textbf{f$_{\text{P,1}}$} & \textbf{A$_{\mathbf{2}}$} & \multicolumn{2}{c}{\textbf{HWHM$_{\mathbf{2}}$}} & \textbf{f$_{\text{P,2}}$} & \multicolumn{2}{c}{\textbf{HWHM$_{\mathbf{single}}$}} & $\Delta$BIC \\
        \midrule
        \shortstack{FWHM\\ ($''$)} & \shortstack{HWHM\\ (pc)} & & \shortstack{Measured\\ (pc)} & \shortstack{Deconv.\\ (pc)} & & &  \shortstack{Measured\\ (pc)} & \shortstack{Deconv.\\ (pc)} & & \shortstack{Measured\\ (pc)} & \shortstack{Deconv.\\ (pc)} & \\
        \midrule
        1.0$''$ & 8.48 & 0.71 & 10.81 & 6.69 & 0.23 & 0.28 & 31.03 & 29.85 & 0.77 & 15.63 & 13.12 & 462 \\
        1.5$''$ & 12.73 & 0.8 & 18.35 & 13.22 & 0.24 & 0.18 & 67.4 & 66.19 & 0.76 & 24.4 & 20.82 & 392 \\
        2.0$''$ & 16.97 & 0.82 & 24.24 & 17.31 & 0.28 & 0.17 & 85.32 & 83.62 & 0.72 & 31.04 & 25.99 & 223 \\
        2.5$''$ & 21.21 & 0.83 & 30.18 & 21.47 & 0.33 & 0.16 & 97.29 & 94.95 & 0.67 & 37.47 & 34.93 & 111 \\
        3.0$''$ & 25.45 & 0.83 & 34.75 & 23.66 & 0.35 & 0.18 & 102.23 & 99.01 & 0.65 & 43.22 & 34.93 & 59 \\
        3.5$''$ & 29.69 & 0.82 & 38.36 & 24.29 & 0.35 & 0.20 & 105.57 & 101.31 & 0.65 & 48.27 & 38.05 & 30 \\
        4.0$''$ & 33.94 & 0.77 & 40.61 & 22.31 & 0.34 & 0.24 & 102.54 & 96.76 & 0.66 & 52.80 & 40.45 & 14 \\
        4.5$''$ & 38.18 & 0.72 & 42.79 & 19.33 & 0.30 & 0.29 & 101.08 & 93.60 & 0.70 & 57.98 & 43.63 & 5 \\
        5.0$''$ & 42.42 & 0.67 & 44.33 & 12.87 & 0.27 & 0.35 & 98.74 & 89.16 & 0.73 & 62.32 & 45.65 & -0.8 \\
        \midrule
        \multicolumn{13}{c}{\textbf{R-L relations}}\\
        \hline
        \hline
        \multicolumn{2}{c}{\textbf{Resolution}} & & \multicolumn{2}{c}{\textbf{HWHM}}& \multicolumn{2}{c}{\textbf{\rmom}} & \multicolumn{2}{c}{\textbf{\rf}}& \multicolumn{2}{c}{\textbf{\rn}} & \multicolumn{2}{c}{\textbf{\riso}}\\
        \midrule
        \shortstack{FWHM\\ ($''$)} & \shortstack{HWHM\\ (pc)} & & $\rho$ & \shortstack{$\alpha$\\A} & $\rho$ & \shortstack{$\alpha$\\A} & $\rho$ & \shortstack{$\alpha$\\A} & $\rho$ & \shortstack{$\alpha$\\A} & $\rho$ & \shortstack{$\alpha$\\A}\\
        \midrule
        \multirow{2}{*}{1.0"} & \multirow{2}{*}{8.48} &  & \multirow{2}{*}{0.19} & 0.93 & \multirow{2}{*}{0.39} & 1.82 & \multirow{2}{*}{0.38} & 1.56 & \multirow{2}{*}{0.43} & 1.95 & \multirow{2}{*}{0.89} & 2.1 \\
        & & & & 35.82 & & 34.62 & & 34.81 & & 33.75 & & 33.64 \\
        \multirow{2}{*}{1.5"} & \multirow{2}{*}{12.73} &  & \multirow{2}{*}{0.25} & 1.28 & \multirow{2}{*}{0.41} & 2.07 & \multirow{2}{*}{0.39} & 1.78 & \multirow{2}{*}{0.43} & 2.07 & \multirow{2}{*}{0.9} & 2.11 \\
         & & & & 35.29 & & 34.19 & & 34.43 & & 33.47 & & 33.59 \\
        \multirow{2}{*}{2.0"} & \multirow{2}{*}{16.97} &  & \multirow{2}{*}{0.3} & 1.47 & \multirow{2}{*}{0.41} & 2.25 & \multirow{2}{*}{0.39} & 1.93 & \multirow{2}{*}{0.4} & 2.19 & \multirow{2}{*}{0.91} & 2.14 \\
         & & & & 34.9 & & 33.83 & & 34.13 & & 33.18 & & 33.52 \\
        \multirow{2}{*}{2.5"} & \multirow{2}{*}{21.21} &  & \multirow{2}{*}{0.32} & 1.82 & \multirow{2}{*}{0.38} & 2.41 & \multirow{2}{*}{0.37} & 2.02 & \multirow{2}{*}{0.36} & 2.28 & \multirow{2}{*}{0.91} & 1.96 \\
         & & & & 34.29 & & 33.51 & & 33.92 & & 32.96 & & 33.81 \\
        \multirow{2}{*}{3.0"} & \multirow{2}{*}{25.45} &  & \multirow{2}{*}{0.31} & 1.77 & \multirow{2}{*}{0.37} & 2.49 & \multirow{2}{*}{0.35} & 2.11 & \multirow{2}{*}{0.37} & 2.58 & \multirow{2}{*}{0.91} & 2.08 \\
         & & & & 34.28 & & 33.31 & & 33.72 & & 32.35 & & 33.58 \\
        \multirow{2}{*}{3.5"} & \multirow{2}{*}{29.69} &  & \multirow{2}{*}{0.3} & 1.72 & \multirow{2}{*}{0.36} & 2.61 & \multirow{2}{*}{0.34} & 2.22 & \multirow{2}{*}{0.38} & 2.81 & \multirow{2}{*}{0.9} & 2.15 \\
         & & & & 34.3 & & 33.07 & & 33.48 & & 31.88 & & 33.43 \\
        \multirow{2}{*}{4.0"} & \multirow{2}{*}{33.94} &  & \multirow{2}{*}{0.29} & 1.72 & \multirow{2}{*}{0.35} & 2.47 & \multirow{2}{*}{0.33} & 2.4 & \multirow{2}{*}{0.39} & 2.94 & \multirow{2}{*}{0.89} & 2.15 \\
         & & & & 34.24 & & 33.24 & & 33.16 & & 31.58 & & 33.41 \\
        \multirow{2}{*}{4.5"} & \multirow{2}{*}{38.18} &  & \multirow{2}{*}{0.29} & 1.58 & \multirow{2}{*}{0.3} & 2.18 & \multirow{2}{*}{0.29} & 2.25 & \multirow{2}{*}{0.38} & 2.78 & \multirow{2}{*}{0.88} & 2.1 \\
         & & & & 34.48 & & 33.66 & & 33.37 & & 31.85 & & 33.5 \\
        \multirow{2}{*}{5.0"} & \multirow{2}{*}{42.42} &  & \multirow{2}{*}{0.29} & 1.52 & \multirow{2}{*}{0.29} & 2.09 & \multirow{2}{*}{0.28} & 2.2 & \multirow{2}{*}{0.38} & 2.81 & \multirow{2}{*}{0.86} & 2.1 \\
         & & & & 34.55 & & 33.79 & & 33.43 & & 31.77 & & 33.47 \\
        \bottomrule
    \label{tab:resfits}
    \end{tabular}
    \end{center}
    \tablecomments{\textbf{Properties of the radial profiles convolved to different resolutions. Top: Double and single Gaussian functions that best fit the normalized median \ha profile averaged for the whole sample at each resolution. f$_{\text{P,1}}$ and f$_{\text{P,2}}$ represent the fractional flux of the inner Gaussian (A$_1\sigma_1^2$) and the outer Gaussian (A$_2\sigma_2^2$), respectively. HWHM$_{\text{single}}$ is the HWHM measured from the best-fit single Gaussian. $\Delta$BIC$>10$ implies that a double Gaussian is preferred over a single Gaussian. Bottom: Radius-luminosity relations for each radius measurement at each resolution. We provide the Spearman rank correlation coefficient, $\rho$, and the powerlaw slope ($\alpha$) and normalization (A) of the best fit to all regions.}}
\end{table*}

%% file: fig-2dcutoutscombo.tex
\begin{figure}
    \centering
    \includegraphics[width=0.65\linewidth]{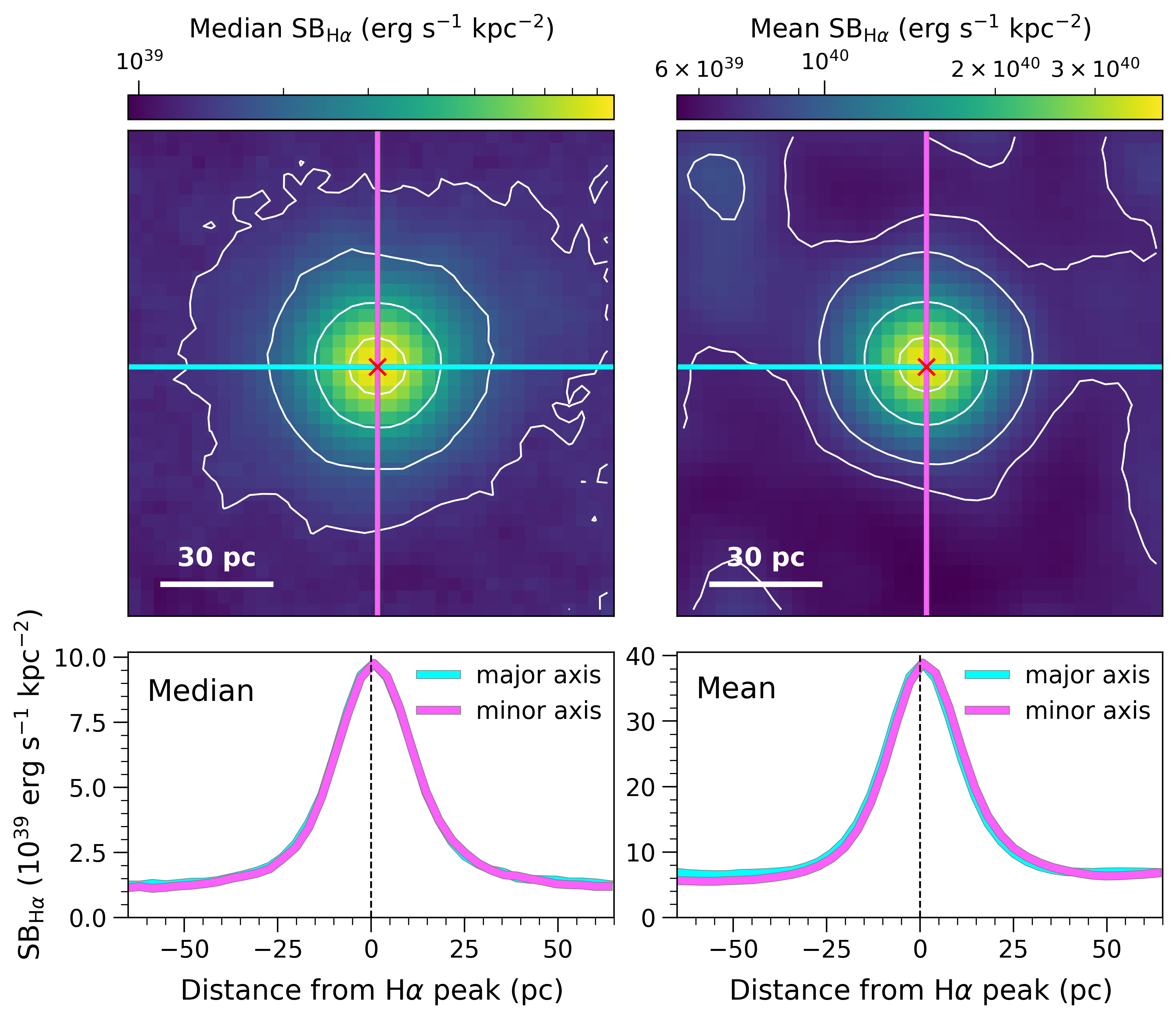}
    \caption{Two-dimensional stacks of the full sample of \hii\ regions. The left plot shows the median of the stack and the right shows the mean. Both plots have contours of the 99th, 95th, 84th, and 50th percentile of \ha brightness with a red X marking the location of the brightest pixel, used as the center of the \hii\ region. Below each stacked image are the radial profiles of \ha surface brightness along the major (cyan) and minor (magenta) axis.}
    \label{fig:2dcutouts}
\end{figure}

%% file: fig-HSTRL.tex
\begin{figure}
    \centering
    \includegraphics[width=0.75\linewidth]{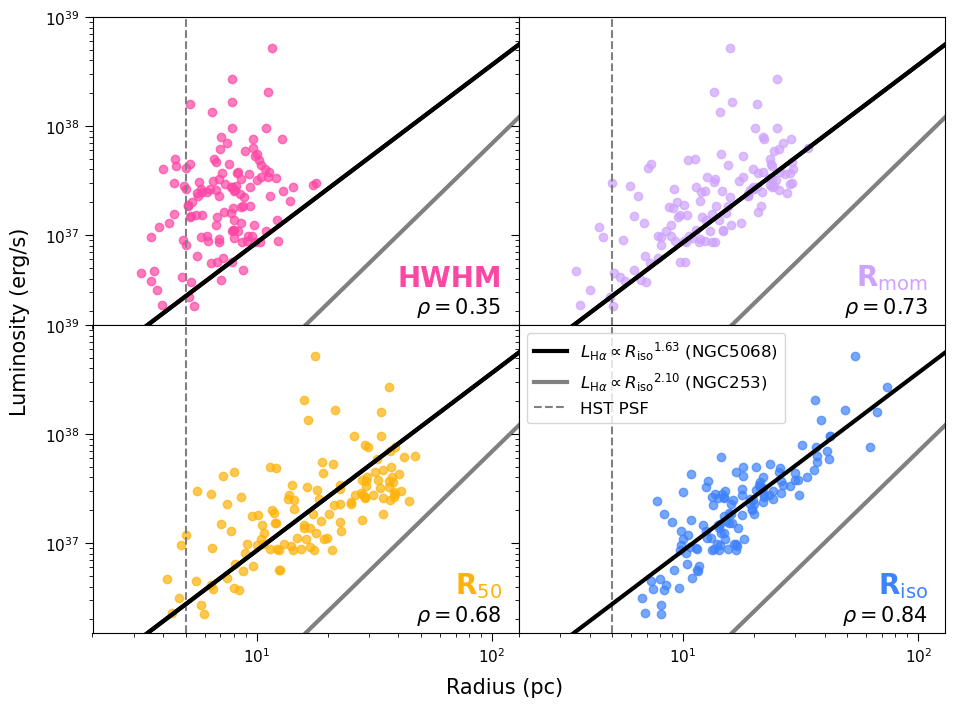}
    \caption{Luminosity-size relations measured from HST \ha images of NGC\,5068 for each radius measurement. The Spearman rank correlation coefficient, $\rho$, is shown for each radius. As in NGC\,253, the HWHM shows only a weak relationship to radius in NGC\,5068. The slope of the $L$-\riso relation in NGC\,5068 (black line) is similar to the fiducial one we measure in NGC\,253 (gray line), but the intercept is higher due to the increased isophotal intensity threshold needed to account for the sensitivity of the HST H$\alpha$ observations. The \rmom and \rf relations match the \riso relation, only with more scatter. This differs from our results for \rmom and \rf in NGC\,253, which show weaker correlation and more scatter compared to \riso (Figure \ref{fig:RL}, Table \ref{tab:RLcorrelations}).}
    \label{fig:hstrl}
\end{figure}

%% file: fig-ngc5068profile.tex
\begin{figure}
    \centering
    \includegraphics[width=0.5\linewidth]{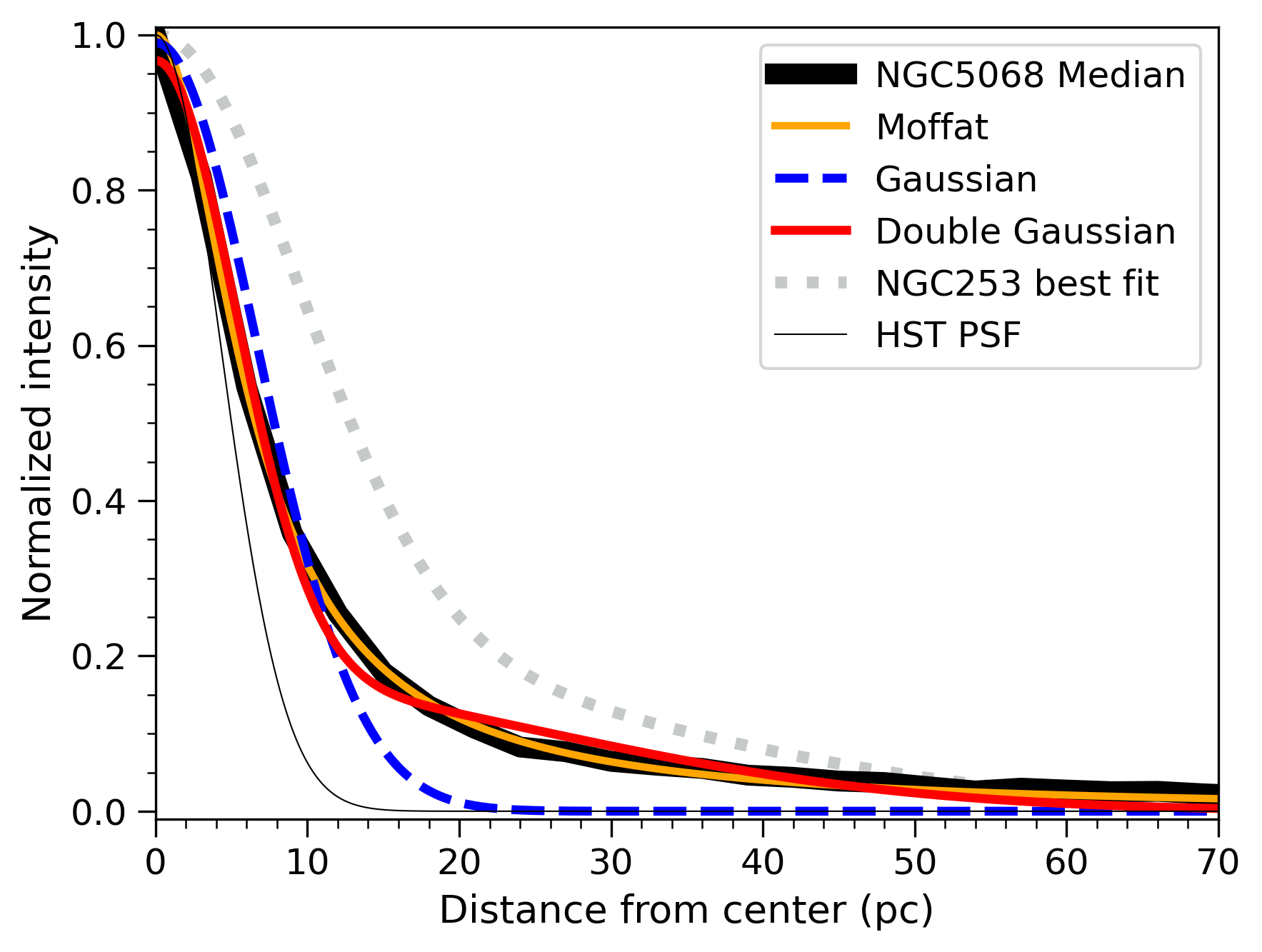}
    \caption{Median \ha profile for \hii\ regions in NGC\,5068 (black). Lines show the best fit single (blue) and double Gaussian (red) functions. As in NGC\,253 the double Gaussian is preferred because it captures the extended emission at the edge of the profile. The best-fit double Gaussian for NGC\,253 is plotted in gray for comparison. The profile shapes are very similar between the two galaxies, though the cores of the profile in NGC\,5068 appears less resolved than in NGC\,253.}
    \label{fig:ngc5068profile}
\end{figure}